\documentclass[a4paper,11pt,notitlepage,twoside]{article}
\usepackage[dvips]{graphicx}
\usepackage{amsfonts}
\usepackage{amsmath,amsthm}
\usepackage{amssymb}
\usepackage{mathrsfs}
\usepackage{bm}
\usepackage{psfrag} %Temporary, until better figures...
\usepackage{pst-node}
\usepackage{url}
\usepackage{esint}
\usepackage{epstopdf}
\usepackage{geometry}
\usepackage{hyperref}
\definecolor{linkcolour}{rgb}{0,0.2,0.6}
\hypersetup{colorlinks,breaklinks,
            linkcolor=linkcolour,citecolor=linkcolour,
            filecolor=linkcolour, urlcolor=linkcolour}
\usepackage{authblk}

\usepackage{fancyhdr}
\definecolor{myred}{rgb}{0.9,0,0}
\definecolor{mygre}{rgb}{0,0.5,0}
\definecolor{myblu}{rgb}{0.1,0.2,0.8}
\definecolor{mywhite}{rgb}{1,1,1}
\definecolor{myblack}{rgb}{0,0,0}
\definecolor{myoran}{rgb}{1,.4,.2}
\definecolor{mygray}{rgb}{.7,.7,.7}

\newcommand{\myphantom}[1]{\textcolor{mywhite}{#1}}

\def\smfrac#1#2{{\textstyle\frac{#1}{#2}}}

\newcommand{\jour}[1]{#1}

\newcommand{\be}{\begin{equation}}
\newcommand{\ee}{\end{equation}}
\newcommand{\ef}[1]{\, #1}

\newcommand{\Aut}{\mathop{\mathrm{Aut}}\nolimits}
\newcommand{\Tr}{\mathop{\mathrm{Tr}}\nolimits}

\newcommand{\Cat}{C}
\newcommand{\ud}{\,\mathrm{d}}
\newcommand{\genus}{\mathfrak{h}}

\newcommand{\atopp}[2]{\genfrac{}{}{0pt}{}{#1}{#2}}

\newcommand{\eee}{\mathbb{E}}

\newcommand{\GG}{\mathcal{G}}

\newcommand{\DD}{\mathrm{d}}
\newcommand{\dd}{\mathrm{d}}

\newcommand{\LL}{L}
\newcommand{\LxL}{\mathcal{L}}

\newcommand{\psibar}{\bar{\psi}}

\newcommand{\Zfor}{Z^{\mbox{\scriptsize{forest}}}}

\newcommand{\osp}{\mathop{\mathfrak{osp}}\nolimits}

\begin{document}

\title%[Critical Behaviour of Spanning Forests on Random Planar Graphs]
{Critical Behaviour of Spanning Forests\\
on Random Planar Graphs}
\author[1]{Roberto Bondesan}
\author[2]{Sergio Caracciolo} 
\author[3]{Andrea Sportiello}
%% \author[1]{R.~Bondesan} %\thanks{}}
%% \author[2]{S.~Caracciolo} 
%% \author[3]{A.~Sportiello}

\affil[1]{{\small Theoretical~Physics,~Oxford~University,}\myphantom{$\displaystyle{\hat{\hat{a}}}$}
{\small \myphantom{xxxxxxxxxxx}~1,~Keble~Road,~Oxford~OX1~3NP,~United~Kingdom~\myphantom{xxxxxxxxxxx}}}
\affil[2]{{\small Dip.~di~Fisica~dell'Universit\`a~degli~Studi~di~Milano~and~INFN,}\myphantom{$\displaystyle{\hat{\hat{b}}}$}
  {\small via~Celoria~16,~I-20133~Milano,~Italy}}
\affil[3]{{\small LIPN,~and~CNRS,~Universit\'e~Paris~13,~Sorbonne~Paris~Cit\'e,}\myphantom{$\displaystyle{\hat{\hat{b}}}$}
{\small 99~Av.~J.-B.~Cl\'ement,~93430~Villetaneuse,~France}}

\date{\today}

%\begin{titlepage}
\maketitle

\begin{center}
{\it Dedicated to Tony Guttmann, on the occasion of his 70th birthday.}
\end{center}

\vspace*{5mm}

\begin{abstract}
\noindent
As a follow-up of previous work of the authors, we analyse the
statistical mechanics model of random spanning forests on random
planar graphs.  Special emphasis is given to the analysis of the
critical behaviour.

Exploiting an exact relation with a model of $\mathrm{O}(-2)$-loops
and dimers, previously solved by Kostov and Staudacher, we identify
critical and multicritical loci, and find them consistent with recent
results of Bousquet-M\'elou and Courtiel. This is also consistent with
the KPZ relation, and the Berker--Kadanoff phase in the
anti-ferromagnetic regime of the Potts Model on periodic lattices,
predicted by Saleur.  To our knowledge, this is the first known
example of KPZ appearing explicitly to work within a Berker--Kadanoff
phase.

We set up equations for the generating function, at the value $t=-1$
of the fugacity, which is of combinatorial interest, and we
investigate the resulting numerical series, a favourite problem of
Tony Guttmann's.
\end{abstract}

\newpage
%%%%%%%%%%%%%%%%%%%%%%%%%%%%%%%%%%%%%%%%%%%%%%%%%%%%%%%
\section{Introduction}
\label{sec:intro}

\subsection{Spanning Forests and the Potts Model}

A \emph{spanning forest} $F$ over a graph $G$ is a spanning
subgraph\footnote{I.e., a subgraph containing all vertices.} without
cycles, thus each of its connected components is a tree.  For a graph
$G$, a set $\vec{w}=\{w_e\}$ of edge weights, and a parameter $t$,
the partition function of spanning
forests $\Zfor_G (t,\vec{w})$ on a graph $G$ is defined as
\begin{align}
  \label{eq:ZForDef}
  \Zfor_G (t,\vec{w}) 
&= 
\sum_{F \preceq G}t^{K(F)} \prod_{(ij)\in E(F)}w_{ij} \, .
\end{align}
where the sum on $F \preceq G$ means on all spanning forests
over $G$, and $K(F)$ is the number of trees in the
forest.

Models in this class are both of combinatorial interest in Graph
Theory, as first considered by the seminal work of Tutte and Whitney
before the war \cite{tuttebook}, and of interest in Critical
Phenomena, as they arise as particular limit $q\to 0$ of the
$q$-colour Potts Model in the Fortuin--Kasteleyn representation
\cite{SokTutte}.  The limit $t\to 0$ in \eqref{eq:ZForDef} corresponds
to \emph{spanning trees}, an even more studied model, whose history
dates back to the 19th century.

On a fixed (weighted) graph $G$, spanning forests can be mapped
through a fermionic Berezin integral to an $\osp(1|2)$-invariant
$\sigma$-model (or O$(n)$ $\sigma$-model analytically continued to
$n=-1$) and its critical behaviour on a square lattice~\cite{CJSSS}
and on the triangular lattice~\cite{DeGrandi} has been deduced, in the
range $t\geq 0$. In this range, RG calculations in dimension two show
the absence of critical points between $t=0$ and $t \nearrow +\infty$,
and a flow from $t=0$ towards $t = +\infty$ compatible with asymptotic
freedom.  The exact solution of the Potts Model on regular
two-dimensional lattices (which are Yang--Baxter integrable)
\cite{baxter} also suggests the absence of critical points in this
range. The picture is best understood in the case of the square
lattice. In this case, the second-order ferromagnetic transition of
the Potts Model, in the phase space parametrised by
$(q,v)$,\footnote{Notations: $v=\exp(\beta J)-1$, as e.g.\ in
  \cite{SokTutte}.} occurs on the parabola $\Gamma_{\rm F}$, defined
as $v=+\sqrt{q}$ for $0 \leq q \leq 4$. Parametrising $q = 4
\cos^2(\pi/\delta)$, the central charge of the associated conformal
field theory is
$c=1-\frac{6}{\delta (\delta-1)}$.
No other critical lines are expected in the quadrant
$q,v \geq 0$, this implying the forementioned behaviour for spanning
forests at positive fugacity, and consistently the well-known
behaviour of uniform spanning trees as a logarithmic non-unitary CFT
with $c=-2$.

On the other side, the solution of the Potts model predicts the
existence of a complex critical behaviour in the
\emph{anti-ferromagnetic regime} $-1 \leq v < 0$ (and the non-physical
one $v < -1$), i.e.\ in the quadrant $q \geq 0$ and $v < 0$, which
corresponds, for spanning forests, to finite negative values
of~$t$. The solution predicts two other dual curves 
$\Gamma_{\rm AF}^{\pm}$, namely
$v=-2 \pm \sqrt{4-q}$, where the anti-ferromagnetic criticality
occurs, and is identified with a CFT with 
$c=2-\frac{6}{\delta}$ with the peculiar presence of a non-compact
boson (thus $c=-1$ for spanning forests, obtained for $\delta=2$).  On
top of this, it predicts yet another special curve, $\Gamma'_{\rm F}$,
defined as $v=-\sqrt{q}$, all of this still in the range
$0 \leq q \leq 4$.  At $\Gamma'_{\rm F}$ there is a different
behaviour, and a CFT with $c=1- \frac{6(\delta-1)^2}{\delta}$, (i.e.,
peculiarly, yet again $c=-2$ for spanning forests).
In a series of works, Saleur and Jacobsen
\cite{saleurNPB, jacobsen2005arboreal, SJAntiferro} have tried to
clarify the nature of this transition, and its CFT content.  However,
a number of subtleties arise (role of the twist parameter in the
boundary conditions of the associated spin chain, saddle-point
approximation and ansatz in the study of the Bethe equations, special
behaviour at the Beraha numbers,~\ldots), so that the picture is
surely quite complicated, and possibly in part still conjectural.

Something that seems well-established, and was predicted already in
\cite{Sal90,saleurNPB}, is the existence of a \emph{Berker--Kadanoff phase},
i.e.\ a full interval of couplings with the critical exponents of the
self-dual curve $\Gamma'_{\rm F}$, for all points between
$\Gamma_{\rm AF}^{+}$ and $\Gamma_{\rm AF}^{-}$.
Planar duality exchanges $v=0$ with $v=\infty$, leaves fixed
$\Gamma_{\rm F}$ and $\Gamma'_{\rm F}$, and exchanges
$\Gamma^{\pm}_{\rm AF}$ (thus leaving fixed the predicted
Berker--Kadanoff interval).  The picture in the triangular lattice,
which is also Yang--Baxter solvable, seems to be analogous, but
universality is not clear and is probably absent in some respect (see
in particular the discussion in \cite{alan03} and the recent
work \cite{Vernier2015}), and it is probably
fair to say that the Potts phase diagram, for general two-dimensional
periodic lattices and out of the (simpler) ferromagnetic quadrant,
deserves investigation still nowadays.

\begin{figure}[tb]
\setlength{\unitlength}{10pt}
\begin{picture}(30,16)(-2,0)
\put(0,0){\includegraphics[scale=1]{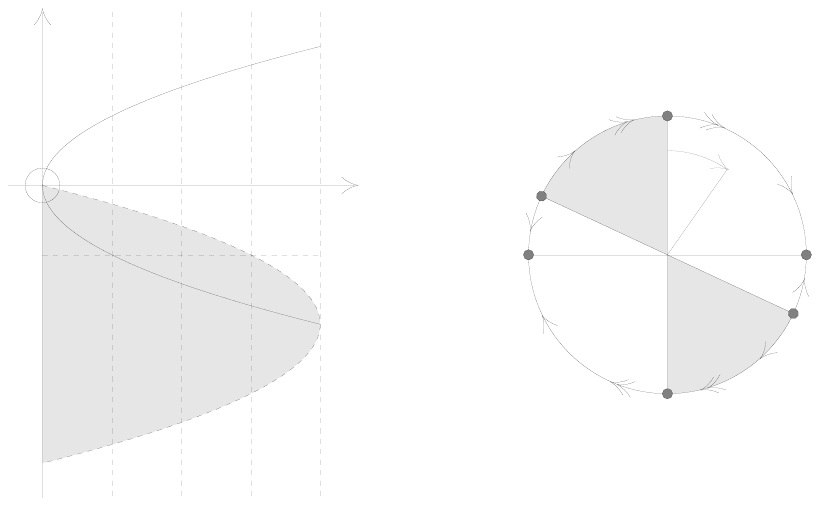}}
\put(11,9){$q$}
\put(1,14){$v$}
\put(1.7,9){\makebox[0pt][r]{$0$}}
\put(1.7,7){\makebox[0pt][r]{$-1$}}
\put(3.7,9){\makebox[0pt][r]{$1$}}
\put(5.7,9){\makebox[0pt][r]{$2$}}
\put(7.7,9){\makebox[0pt][r]{$3$}}
\put(9.7,9){\makebox[0pt][r]{$4$}}
\put(2.8,12.5){F}
\put(6.2,11){HT}
\put(8.2,7.9){HT}
\put(4.2,6){BK}
\put(6.2,8){$\Gamma^+_{\rm AF}$}
\put(6.2,3){$\Gamma^-_{\rm AF}$}
\put(6.2,6.5){$\Gamma'_{\rm F}$}
\put(6.2,12.5){$\Gamma_{\rm F}$}
\put(14.8,13){Spanning trees, $t=0$}
\put(14.8,2.5){Spanning trees, $t=0$}
\put(23,4){Berker--Kadanoff phase, $t<0$}
\put(24.3,5.8){$t=t_{\ast}$}
\put(24.7,7.8){$t=\atopp{+}{-}\infty$}
\put(24.3,9.8){$t>0$}
\put(29,9.4){High temperature}
\put(29,8.3){(massive theory)}
\put(22.5,11.7){Asymptotic freedom}
\put(19,9.5){$\arctan t$}
\end{picture}
  \caption{The phase diagram of the Potts model on the square lattice.
    There are three regions, corresponding to a ferromagnetic phase
    (F), a high-temperature phase (HT), and an
    anti-ferromagnetic/non-physical Berker--Kadanoff phase (BK).  On
    the right, we show the phase diagram for spanning forests, as a
    magnification of the neighbourhood of $(q,v)=0$. In this model we
    have a single parameter left, the ratio $t=q/v$ by which limit
    $q,v \to 0$ is taken in the Potts model. Thus, the parameter $t$
    may be represented by half of a circle, i.e.\ $t=\tan \theta$, and
    $\theta \in [-\pi/2,\pi/2]$.  Arrows on the circle denote the
    behaviour under the Renormalisation Group. The value of $t_{\ast}$
    is $-4$.  \label{fig:pottsphase}}
\end{figure}

Figure \ref{fig:pottsphase} summarises the main elements described
above, for the case of the square lattice.  For a more detailed
account of this complicated landscape, the reader may consult the
introduction of \cite{alan04}\,\footnote{Including reading the long
  footnotes there!}, and, for more details on the triangular lattice,
of \cite{alan03}.  For the behaviour of spanning forests, introduction
and conclusions of \cite{jacobsen2005arboreal} are a valuable
reference.  Some more comments, and relations with the results of this
paper, are discussed in Section~\ref{sec:flat_lattice}.

%-------------------------------------------------------
\subsection{Statistical mechanics on Random Planar Graphs}

Versions of statistical mechanics models on \emph{random planar
  graphs} are often solvable through methods of \emph{Random Matrices}
\cite{DiFrancesco1993,DiFrancesco2004,Eynard2015,bousquet2011counting},
and, mainly in the 80s and 90s, a number of well-known
statistical mechanics models have been solved in this framework
(Ising, Potts, O$(n)$ Loop Models,\ldots).  These models correspond to
the annealed average, of the fixed-graph partition function for a
graph $G$, over the ensemble of all planar graphs $G$ with the
grand-canonical measure $g^{|V(G)|}$. So, given a class $\GG=\{ G \}$ of
planar graphs (e.g., all planar graphs with vertices of prescribed
degree), and a fixed-graph partition function $Z_G(t)$ for a certain
model, possibly depending on some global parameters $t$ (like the
temperature, or, here, the component fugacity), the random planar
graph related quantity is
\be 
\label{eq.generRPG}
Z^{\rm rpg}(t,g):=
\sum_{G \in \GG}
\frac{1}{|\Aut(G)|}
g^{|V(G)|}
Z_G(t)
\ee
(here $|\Aut(G)|$, the cardinality of the group of planar
automorphisms of the graph $G$, is introduced for convenience, and is
irrelevant in the thermodynamic limit).

Besides being interesting \emph{per se}, a solution to such a model is
also relevant in understanding the phase diagram and critical
exponents in regular two-dimensional geometries, in light of the
Knizhnik, Polyakov, Zamolodchikov (KPZ) relation
\cite{KPZ,DAVID1992671,duplantier2011liouville,david2016liouville}.

In some, more rare, cases, a better analytic control on the solution
can be achieved by an alternate approach, that, instead
of using Random Matrices, makes use of combinatorial recursions
\cite{bousquet2011counting}.
% algebraic of degree 2, 
These are inspired by the Tutte solution of the original counting
problem in the 60's (the \emph{quadratic method}), but are not
confined to this, as recent years have seen a flourishing of exact
combinatorial approaches, as in \cite{schaeffer1998conjugaison,
  marcus2001bijection, chapuy2015bijection} or in
\cite{bouttier2002census, bouttier2004planar, BouGui}.  

It is thus natural to ask whether the problem of spanning forests is
also solvable on random planar graphs. The fact that this model is a
limit of the Potts Model, and that some aspects of the latter on
random planar graphs are known exactly (there is a large literature on
the subject, starting with \cite{kazakov1988exactly}), might seem to
suggest that the solution of the Potts Model already provides a
complete answer.  However, this is not the case, because of the
analytic structure of this solution (that is explicitly calculated
only on a discrete set of values for $q$, or on the self-dual critical
curve, analogue of $\Gamma_{\rm F}$), and of the peculiar difficulties
pertinent to the way in which the spanning-forest limit is taken. Even
when the approach is of combinatorial nature
\cite{bernardi2008characterization, bernardi2011counting,
  bernardi2015counting}, the present understanding of the full Potts
Model is too fragmentary for deducing the behaviour of spanning
forests just by specialising the general results, and it is instead
advisable to exploit from the beginning the simplifying properties of
the forest case.
 
Yet another ansatz could have been based on the forementioned relation
with the $\mathrm{OSP}(1|2)$ non-linear $\sigma$-model, and, by the
Parisi--Sourlas mechanism, the equivalence of the latter, at the
perturbative level, with the O$(n)$ model, analytically continued
at $n=-1$. Also this is not the case, due to the fact that the O$(n)$
model related to spanning forests, and the one which is solved on
random planar graphs, have an important difference: the former is a model
with an extra $\mathbb{Z}_2$ symmetry, i.e.\ it is a \emph{projective}
$\sigma$-model, while the latter is the O$(n)$ `Nienhuis Loop Model',
which does not have this symmetry, see the discussion
in~\cite{noiGuttAlan}.

To some surprise, instead, (see however the discussion of
Section~\ref{sec:l-d}), the solution of the O$(n)$ Loop Model at
$n=-2$ is relevant here. Indeed, a first approach has been presented
by two of the authors in \cite{CS_RMFor}\;\footnote{See also the
  master thesis of the first author, {\it Spanning trees and forests
    on genus-weighted random lattices,}
  \href{http://pcteserver.mi.infn.it/~caraccio/Lauree/Bondesan.pdf}
       {\tt
         http://pcteserver.mi.infn.it/$\sim$caraccio/Lauree/Bondesan.pdf},
       2009.}.
This approach is based on an exact combinatorial relation with a
O$(n)$ Loop Model, which, in the case corresponding to random
triangulations, had already been solved by Kostov and
Gaudin~\cite{kostov89,GAUDIN1989}.
%  and Staudacher
This led, in particular, to the first determination of the
radius of convergence for spanning forests on random planar graphs, at
fugacity $t=-1$, i.e.\ in the anti-ferromagnetic region, to be
$g_{c}=\pi/(8 \sqrt{6})$, remarkably a value whose algebraic nature is
incompatible with the $D$-finiteness of the corresponding generating
function (see the pertinent discussion in~\cite{BMCourt2015}).

Our approach was not completely satisfactory. The combinatorial part
is exact, and performed for arbitrary $p$-angulations, but the
solution is not self-contained, as it is based on the forementioned
former solution of Kostov. The original specialisation to the cubic
case can be extended, in part, by use of later results by Kostov and
Staudacher \cite{KoStau} (a task done here), however an issue remains
unsolvable: the solution by Kostov and Staudacher is done at the level
of the saddle-point approximation of the Random-Matrix integral, which
is the solution method with the smallest amount of mathematical
control, e.g.\ on finite-size corrections (w.r.t., e.g., the use of
orthogonal polynomials, and, no telling, w.r.t.\ explicit bijective
approaches).

A quite satisfactory full solution has been presented by
Bousquet-M\'elou and Courtiel in \cite{BMCourt2015} (see also
\cite{CourtTh}). It turns out that the simplest case corresponds to
$p$-angulation with $p$ even (and, among these, quadrangulations are
the best example). Nonetheless, the authors manage to solve also the
case of triangulations. Their approach relies strongly on a general
recursive decomposition invented by Bouttier and Guittier
\cite{BouGui}. Said in a few words, there exist explicit bivariate
hypergeometric series, $\theta$ and $\phi_{1,2}$, such that (the
derivative of) the generating function $F(z,u)$, satisfies a system
of three equations, of the form\footnote{The variables $(z,u)$ in
  \cite{BMCourt2015} correspond to $(g^2,t)$ of our paper, and
  $F(z,u)$ is given by with 
 the grand canonical partition function of Eq.~\eqref{eq.generRPG}
 without the factor $1/|\Aut(G)|$.}
\begin{align}
F'(g,t) &= \theta(R,S)
\ef,
&
R &= g + t\, \phi_1(R,S)
\ef,
&
S &= t\, \phi_2(R,S)
\ef.
\end{align}
A remarkable and surprising result of \cite{BMCourt2015} is that, in a
full range of fugacities, including $-1 \leq t < 0$ and possibly
extending below $t=-1$, the generating function has a string
susceptibility \emph{distinct} from the one of pure gravity,
\emph{and} shows peculiar inverse-logarithmic corrections.

Note that such a behaviour appears to be new within models of random
planar graphs, which are exactly-solved through methods of bijective
combinatorics and recursive decompositions. This is at no surprise, as
such an `exotic' behaviour would be impossible in a `simple'
random-matrix model, with a single matrix and a polynomial
potential. In this case, it emerges in connection to the structure of
the system of equations of Bouttier and Guittier, in vicinity of a
singularity in the complex plane.

This is thus in contradiction with the simplest picture of models of
Quantum Gravity with matter, which, by KPZ, are supposed to be
analogous to their Euclidean counterparts, that, in turns, under the
assumption of generality of the critical manifold in the phase space,
shall show criticality on a manifold of co-dimension one in the phase
space.

However, as we have already discussed above, the modern comprehension
of the phase diagram of the Potts model does \emph{not} follow this
simple pattern. In particular, in the anti-ferromagnetic region we
have a Berker--Kadanoff phase, with the same exponents as the theory
at its edges, with the exception of one special point (and, in our
case, the same exponents of spanning trees). Thus, the fact that the
string susceptibility takes a constant non-trivial value in a whole
interval all along $t \geq 0$ is \emph{in perfect agreement} with
Saleur's prediction of the Berker--Kadanoff phase, and the KPZ
correspondence. On the contrary, the presence of inverse powers of
logarithms, to our knowledge, couldn't be predicted in advance by
general arguments (and may be present only as a speciality of the $q=0$
case, where the critical behaviour is a
\emph{logarithmic CFT}, but we cannot conclude on this point).

To our knowledge, in retrospect, the result of \cite{BMCourt2015}
is the first case in which the KPZ counterpart of a Berker--Kadanoff
phase has been observed to be valid in a mathematically rigorous way,
thus producing a highly non-trivial check of the KPZ correspondence,
which still nowadays is considered quite mysterious.

%-------------------------------------------------------
\subsection{Plan of the paper}

This paper is a natural continuation of the investigation of the model
started in~\cite{CS_RMFor}. On one side, it uses essentially the same
main tool, namely the correspondence with the O$(n)$ loop model and
the saddle-point solution of the latter by Kostov and Staudacher
\cite{KoStau}, either directly, or as a guideline for performing a
more extensive analysis. On the other side, this calculation is now
performed while having in mind the procedure in \cite{BMCourt2015}. We
will see that interesting connections between the two derivations will
emerge in the treatment.

As we said, the approach of \cite{BMCourt2015} is completely
rigorous. Which is good, of course, but, as a side effect, the authors
refrain from extending their analysis to situations in which it shall
be supplemented by widely believed (but not rigorously proven) further
hypotheses. An example is the fact that, in the cubic case, the
authors claim the appropriate behaviour only for Ces\`aro means of the
coefficients, see \cite[Remark 1, page 46]{BMCourt2015}. Another
example is that the authors do not try to extend the analysis to
$t<-1$, where a certain positivity property (discussed later on in
Section \ref{sec:comb_crit}) is lost.

We make here a different choice.  We believe that the phase diagram of
the model under consideration is still puzzling, with questions which
are genuinely open. So we find it useful to establish reasonable
derivations, even, if necessary, at the expenses of mathematical
control.  We rederive all the main results of \cite{BMCourt2015},
extend the analysis to all the range of $t$, and determine the
presence of a critical value $t_{\ast} < -1$, which we identify with
the other endpoint of the Berker--Kadanoff phase. Along the
derivation, we determine an analytic expression for the spectral
density, a result partially contained in \cite{KoStau}, but clearly
not in \cite{BMCourt2015} (where there is no interpretation of the
results in terms of random matrices).

Our methods are intrinsically less rigorous than the ones in
\cite{BMCourt2015}. Nonetheless, given the increase of combinatorial
methods for dealing with O$(n)$ models in full rigour, also from the
point of view of random matrices, \cite{borot2011recursive,
  borot2012more, borot2012loop}, it is conceivable that, in future,
our approach can also be improved to the appropriate level of
mathematical rigour.

In Section~\ref{sec:rm_sf} we will review the derivation of a matrix
integral for the spanning-forest model.  In
Section~\ref{sec:sf_to_O-2}, we establish the mapping of the model
onto the O$(n=-2)$ gravitational loop model. This corrects a mistyped
formula in~\cite{CS_RMFor}.  In Section~\ref{sec.mainana}, which is
the main part of the paper, we complete the analysis of the model,
started in~\cite{CS_RMFor}. In passing, we see the emergence of
connections with the quantities which are relevant in the (apparently
orthogonal) approach of \cite{BMCourt2015}.
Sections \ref{sec.zeri} and \ref{sec:comb_crit} present mostly
numerical data. In the first of these sections, we study the zeroes in
the complex plane (for $t$) of the generating functions of spanning
forests with fugacity $t$, on random planar graphs of fixed volume. In
the second one we produce recursions for the series in the cubic case,
and at fugacity $t=-1$, which is of special combinatorial interest.
Finally, in Section \ref{sec:flat_lattice} we investigate the
consistency of our findings with the known properties of the Potts
Model on flat lattices, and the KPZ relation, and conclude with a
number of open questions.

%%%%%%%%%%%%%%%%%%%%%%%%%%%%%%%%%%%%%%%%%%%%%%%%%%%%%%%%%
\section{A Random Matrix Model for Spanning Forests}
\label{sec:rm_sf}
Here we derive the matrix integral for the random planar graph
partition function (\ref{eq.generRPG}), in the case of spanning
forests, (\ref{eq:ZForDef}), reviewing the approach described in
detail in~\cite{CS_RMFor}.

The derivation is composed of three steps. First, we perform a
resummation of trees into effective vertices. Second, we perform a
change of variables which transforms the resulting series into a
polynomial, thus obtaining a one-matrix model with a polynomial
potential, and a peculiar structure of Vandermonde--like factors.
Finally, we review the basic manipulations customary in the O$(n)$
Loop Model, and match the resulting matrix integral to the one
obtained at the previous step.

Given a graph $G$ and a spanning forest $F$ over it, define the
\emph{contraction} $G/F$ as the graph in which the components of the
forest are shrunk to points, by identifying vertices pertaining to
edges of the same tree.  Again, let $|\Aut(G)|$ be the cardinality of
the automorphisms of the graph $G$ (respecting the embedding).  Then,
at slight difference with (\ref{eq:ZForDef}), we define the partition
function for spanning forests over a fixed graph $G$ as
\begin{equation}
  \label{eq:ZSpForFixG}
  Z_G(t) =  \sum_{F \preceq G} t^{K(F)} \frac{|\Aut(G)|}{|\Aut(G/F)|}
\end{equation}
We have introduce for later convenience the additional ratio of the
cardinality of automorphisms. This is normally done with not much
hassle, as it is expected that, in the limit of large random graphs,
this does not affect the leading behaviour and so the analysis of
criticality.\footnote{In fact, that's not only the leading behaviour
  which is unaffected. The whole contribution to the partition
  function of graphs with non-trivial automorphisms shall be
  exponentially small.}  An alternate approach could consist in
working with a slight different ensemble of random graphs, in which
there is a marked local structure (an appropriate choice in this case
would be a marked half-edge not in the forest), which breaks all
possible automorphisms and is easily related to the previous
expression, by differentiation w.r.t.\ a parameter associated to the
volume of the graph. The advantage of this second approach, besides
mathematical elegance, is the fact that the generating series is
guaranteed to be integer-valued.

Consider a generic graph $G$, equipped with an embedding (i.e., with a
cyclic ordering of the half-edges incident to each given vertex).
Denote by $\genus(G)$ the genus of the surface isomorphic to the
resulting 2-dimensional cell complex.  Let us consider $\GG$, the
ensemble of such graphs, with vertices of homogeneous degree 
\begin{align}
  k = h+2 \, ,
\end{align}
and consider the formal power series in which these graphs
are weighted by the customary `random matrix' measure 
$N^{-2\genus(G)} g^{|V(G)|}/|\Aut(G)|$:
\begin{align}
  \label{eq:ZSpForRanG}
  Z(t,g,N) &=  \sum_{G}N^{-2\genus(G)} g^{|V(G)|}
  \sum_{F \preceq G} t^{K(F)} \frac{1}{|\Aut(G/F)|} \nonumber \\
  &= \sum_{F}t^{K(F)}g^{|V(F)|} 
  \sum_{G \succeq F}  \frac{N^{-2\genus(G/F)}}{|\Aut(G/F)|} 
\ef.
\end{align}
In the manipulations above we have crucially used, besides the trivial
fact $|V(F)| = |V(G)|$, also the slightly more subtle fact
$\genus(G/F) = \genus(G)$ (the contraction of trees, i.e.\ of
subgraphs with no cycles, is the \emph{only} contraction operation
that does not erase cycles in the graph. As a corollary, it is in turns
guaranteed not to decrease the genus).

We will investigate directly the planar limit $N\to \infty$, and we
will be mainly interested in the \emph{thermodynamic limit}, that is
when the moments of expected size of the graph, $\eee(|V(G)|^n)$,
diverge for $n \geq n_0$ (for some $n_0$).  This occurs when $g$ tends
to its critical value, i.e.\ the radius of convergence $g_c(t)$ of the
series above (with $N \to +\infty$). As notoriously the case in
similar circumstances, the series at finite $N$ can be formally
written as a random matrix integral, with matrices of size $N$,
although in fact this series is not convergent, but rather an
asymptotic one. 

\begin{figure}[tb]
 \centering
  \includegraphics[scale=1.5]{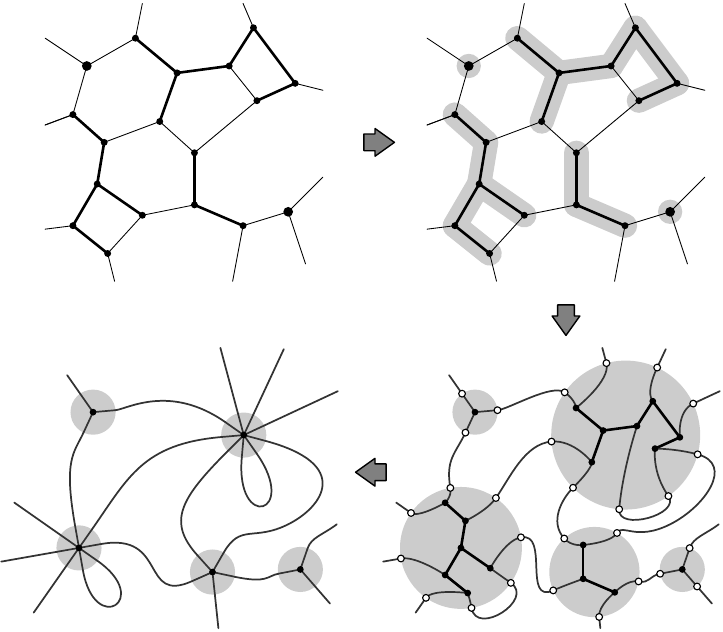}
  \caption{The procedure for shrinking the trees. Given a spanning
    forest $F$ over a graph $G$, we mark the components of the forest
    (here in bold) , and we shrink the trees with their leaves to
    single vertices respecting the ordering of the the border legs.
    Then we obtain $G/F$ (graph on the bottom left).  The images on
    the right side represent intermediate steps, where we deform the
    graph and color in grey the region to be shrunk, to better
    visualize the procedure.}
  \label{fig:treeshrink}
\end{figure}

In order to represent \eqref{eq:ZSpForRanG} as a matrix model, we
shall first contract the trees to points, and then represent the
combinatorics of the resulting effective vertices by a suitable
integral.  This procedure, described in the following lines, is also
illustrated by Figure \ref{fig:treeshrink}.  Let $F \preceq G$ be a
spanning forest, with components $T_a$. We represent its edges as
bold, and the edges in the complement as thin. Now, subdivide each
thin edge into three edges in series. The two external ones are
incident to original vertices, and thus to $F$, while the internal one
is not (these are the edges not contained in the gray-shadowed
region). Now, the edges in the shadowed region, thick and thin
altogether, make a new forest $F'$, with components $T'_a$, and the
vertices of $T_a$ are exactly the vertices of $T'_a$ which are not a
leaf (if there are $n$ of them, there are $hn+2$ leaves in $T'_a$). It
is convenient to imagine each of these trees $T'$ as embedded in a
disk, with the leaves on the boundary.

The number of trees of the like of $(T'_a)$s, with $n$ vertices, and
counted with a factor $1/|\Aut(T')|$ for the automorphisms under
cyclic rotation, can be derived from a recurrence relation
\cite{CS_RMFor}, and is given by
\begin{equation}
  \label{eq:Ahn'}
  A'_{h,n} = \frac{((h+1)n)!}{n! (hn+2)!} \, .
\end{equation}
For $h=1$, this corresponds to $C_n/(n+2)$, where $C_n$ is the $n$-th
Catalan number,
and the factor $n+2$ accounts for the cyclic symmetry among the $n+2$
leaves. For higher values of $h$, they are related to the appropriate
generalisation, that goes under the name of 
\emph{Fuss--Catalan numbers} (see e.g.\ \cite[Sec.~7.5]{GKP})..

Now contract every $T'_a$ to a single vertex, respecting the
ordering of the border legs. The resulting graph is $G / F$.
% The whole procedure is depicted in figure \ref{fig:treeshrink}.
To keep track of the graph parameters all along this shrinking
procedure, we assign to every effective vertex, obtained from a tree
$T'$ with $n$ vertices, an effective coupling $g_n = t g^n A'_{h,n}$.
These vertices have degree $hn+2$.

The generating function of graphs $G/F$, counted with the associated
multiplicities, symmetries, and factor for the number of vertices, can
thus be represented by the following one-Hermitian-matrix integral:
\begin{equation}
  \label{eq:MatIntSpFor}
  Z(t,g,N) = \frac{1}{N^2} \log \displaystyle{\frac{
      \int \DD M\, e^{-N\, \Tr V(M)}}
    {\int \DD M\, e^{-N \Tr \frac{1}{2} M^2} }}
\end{equation}
where the potential is 
\begin{align}
  \label{eq:PotSpFor}
  V(z) &= \frac{1}{2} z^2 - \sum_{n\ge 1}g_n z^{hn+2}
= \frac{1}{2} z^2(1 - t A'_h(gz^h)) \, ,
\end{align}
and we have called $A'_h(x)$, the generating function $A'_h(x) =
\sum_{n\ge 1} A'_{h,n} x^n$.

At this point we perform the integration over the angular degrees of
freedom, and reduce to the eigenvalues basis, i.e., we write $M =
U\,\Lambda\,U^{-1}$ and integrate over $U$. This produces the square
of the Vandermonde determinant $\Delta(\{\lambda_i\}) = \prod_{i<j}
(\lambda_i-\lambda_j)$ as Jacobian of this transformation:
\begin{equation}
  \label{eq:MatIntSpForEigen}
  Z(t,g,N) = \frac{1}{N^2} \log \displaystyle{\frac{
      \int \prod_{i=1}^N \ud \lambda_i \,\Delta^2(\{\lambda_i\}) \,
      e^{-N \sum_i V(\lambda_i)}}
    {\int \prod_{i=1}^N \ud \lambda_i \, \Delta^2(\{\lambda_i\}) \,
      e^{-N \sum_i \frac{1}{2} \lambda_i^2}}}
\end{equation}
and the eigenvalues are confined to the analyticity region of
$A'_h(x)$, namely $|x| < \frac{h^h}{(h+1)^{h+1}}$.

We have a single random matrix, which is convenient for the solution,
but unfortunately the potential is not polynomial, this being in
general a crucial obstacle.  Nonetheless, there exists a suitable
change of variables that leads to a polynomial potential, at the
price of an additional Vandermonde--like term in the
integrand. Indeed, calling $z = x(1-gx^h)$, we have
\begin{equation}
  \label{eq:A'hxParam}
  z^2 A'_h(g z^h) = \frac{x^2}{2}\left( \frac{2}{h+2} g x^{h} - 
    (g x^{h})^2 \right) \, .
\end{equation}
so that, if we perform the change of variables $\lambda_i(x_i) = x_i
(1-gx_i^h)$, we end up with the following expression:
\begin{equation}
  \label{eq:MatIntSpForx}
\begin{split}
  Z(t,g,N) 
&=
\frac{1}{N^2} \log \displaystyle{\frac{
      \int \prod_{i=1}^N \ud x_i \, \Delta^2(\{x_i\}) \,
      \hat{\Delta}_g(\{x_i\}) \, e^{-N \sum_i \tilde{V}(x_i)}}
    {\int \prod_{i=1}^N \ud x_i \, \Delta^2(\{x_i\}) \, e^{-N\sum_i
        \frac{1}{2} x_i^2}}}
\\
&=
\frac{1}{N^2} \log \displaystyle{\frac{
      \int \prod_{i=1}^N \ud x_i \, \Delta^2(\{x_i\}) \,
      \hat{\Delta}_g(\{x_i\}) 
\, e^{-N \sum_i \tilde{V}(x_i)}}
    {\int \prod_{i=1}^N \ud x_i \, \Delta^2(\{x_i\}) \, 
      \hat{\Delta}_g(\{x_i\}) 
e^{-N\sum_i \tilde{V}_0(x_i)}}}
\end{split}
\end{equation}
where the potentials $\tilde{V}$ and $\tilde{V}_0$
are now both polynomial
\begin{align}
  \label{eq:PotPolySpFor}
\tilde{V}(x) 
&=
\frac{x^2}{2} \left(1 -2 g\left(1+\frac{t}{h+2}\right)x^h
    + g^2(1+t)x^{2h}\right) 
\\
\tilde{V}_0(x) 
&=
\frac{x^2}{2} \left(1 -2 g x^h
    + g^2 x^{2h}\right) 
\end{align}
and the new factor $\hat{\Delta}_g(\{x_i\})$ comes from the combination
of the Jacobian and the change of the Vandermonde determinant in the new
coordinates, and is
\begin{equation}
  \label{eq:DeltaHatgx}
  \hat{\Delta}_g(\{x_i\}) = \prod_{i,j} (1 - g(x_i^h + x_i^{h-1}x_j + \dots
  + x_j^h)) \, .
\end{equation}

%%%%%%%%%%%%%%%%%%%%%%%%%%%%%%%%%%%%%%%%%%%%%%%%%%%%%%%
\section[Mapping to the $\mathrm{O}(n)$ Gravitational Loop Model at $n=-2$]
{Mapping to the \textbf{O(\textit{n})}
Gravitational Loop Model at \hbox{\textbf{\textit{n}\,=\,-\!-2}}}
\label{sec:sf_to_O-2}

%{\raisebox{3.2pt}{\rule{7pt}{1.2pt}}}

%-------------------------------------------------------
\subsection{Loop-dimer gas coupled to gravity}
\label{sec:l-d}

The O$(n)$ loop model has been proposed by Nienhuis
\cite{nienhuis1982exact} as a model with $n$-dimensional vector fields
that, on cubic lattices, allowed for a simple combinatorial
description in terms of self-avoiding cycles. The idea of universality
in critical phenomena is that the critical exponents are related to
the symmetry group of the system, including the target space of the
elementary degrees of freedom, and in the Nienhuis model the
underlying symmetry under (analytically-continued in $n$) O$(n)$
global rotations appears as a ``weight factor $n$ per loop'', in a way
similar to how a ``weight $q$ per component'' appears in the
Fortuin--Kasteleyn formulation of the Potts Model, related to the
(analytically-continued in $q$) permutational symmetry group
$\mathfrak{S}_q$ for the Potts colours.  We also remark that however
there exist other spin models (in particular integrable vertex models)
which can be mapped onto loop models of O$(n)$ and $q$-state Potts
type and whose symmetry is different, see \cite{Yung1995} for a systematic
discussion.

Because of its combinatorial simplicity, its definition on arbitrary
cubic graphs, and of its formulation well-adapted to the `model
building' of random matrices (see e.g.\
\cite[sec.~2.5]{DiFrancesco2004}), the O$(n)$ loop model had a
prominent role in the development of statistical mechanics on random
geometries, starting from the first seminal
results \cite{kostov89, GAUDIN1989, KoStau}, and is still considered a
major realisation of the KPZ counterpart of the family of minimal
CFT's, and the most promising for a rigorous formalisation of this
relation \cite{1742-5468-2011-01-P01010, Borot2013, BouGui}.

Therefore, and this will be crucial in this paper, once we have
established the announced correspondence with the model of spanning
forests, we have at disposal a large amount of established results and
techniques.

Dimer models are another statistical-mechanics model which, besides a
long tradition on ordinary regular graphs, is well adapted to the
formalism of random planar graphs \cite{stauDimers}
%  [ANCHE GINSPARG MOORE ?]  
(see also \cite{DiFrancesco2004}, pages 12-13 and 33). It is in
fact one of the simplest cases of solvable such models, and the
criticality has a simple pattern, which makes this model a candidate
pedagogical example to illustrate the emerging features of Random
Matrix theory.

In this section we consider a more general model, in which we have
underlying graphs of fixed degree, $k=h+2$, and, as matter fields,
both dimers and non-intersecting loops.\footnote{Versions of the
  Nienhuis Model lead to non-intersecting loops, i.e.,
  vertex-disjoint collections of cycle subgraphs, when defined on
  cubic graphs, \emph{and} through a fine-tuning of either the
  two-body interaction, or of the vertex measure, or both. The
  first choice is more customarily investigated. In that case, the
  extension to graphs of higher degree leads to more general
  edge-disjoint collections of walks on the graph, with suitable
  vertex weights. However, on arbitrary graphs, under the second
  choice one can construct from the O$(n)$-invariant Nienhuis field
  theory, a combinatorial model of dimers and non-intersecting loops
  like that described here.}
The loops have fugacity $n$.  This model, that we shall call
\emph{loop-dimer gas model} (we will write \emph{O$(n)$-loop--dimer
  gas model} when we need to emphasize the value of $n$), has been
studied by Kostov and Staudacher \cite{KoStau}, as a continuation of
the work of Kostov and Gaudin \cite{kostov89,GAUDIN1989} on the O$(n)$
model (with no dimers). As we will see, it is this model that is
exactly related to spanning forests.

Such a hybrid model may appear baroque at a first look. However, at a
deeper inspection, some justifications can be identified. First, the
O$(n)$ loop model configurations are defined as a collection of cycles
in the underlying graph which are vertex-disjoint. On a cubic graph,
this coincides with cycles which are edge-disjoint. Edge-disjoint
paths are in particular non-backtracking. Vertex-disjoint paths are
also non-backtracking, \emph{unless} they are cycles of length 2, in
which case they form a structure isomorphic to a dimer. As,
intrinsically, for cubic graphs and `long' cycles, we do not see a
difference between the constraints of being vertex- or edge-disjoint,
it makes sense to consider the only possible local discrepancy --
namely dimers -- as included in the theory with an arbitrary weight.

A second, deeper reason is the fact that, as we have mentioned in
the introduction (see \cite{noiJPhysA} for a full account), on an
\emph{arbitrary} graph (even without averaging on the ensemble of
planar graphs) there is a formulation of the spanning-forest partition
function in terms of a Grassmann action which has a quadratic and a
quartic term. The quadratic part is of the form
`mass$\;+\;$Laplacian', and the quartic part is associated to the
marking of a dimer, i.e.,
% if $d_v$ is the degree of vertex $v$, 
the action is
\be
\exp(S_{\rm forest}) = 
\exp \bigg(
\sum_v (t + 
\sum_{u \sim v} w_{uv}
) \psibar_v \psi_v
- \sum_{(uv)} w_{uv} (\psibar_v \psi_u + \psibar_u \psi_v)
- t
\sum_{(uv)} w_{uv} \psibar_u \psi_u \psibar_v \psi_v
\bigg)
\ee
On the other side, following the combinatorial rules of Grassmann
Algebra, we can investigate the meaning of an action
\be
\exp(S) = 
\exp \bigg(
\sum_v m_v \psibar_v \psi_v \bigg)
\prod_{(uv)} (1 - a_{uv} (\psibar_v \psi_u + \psibar_u \psi_v) 
% ) \prod_{(uv)} (1 
+ b_{uv} \psibar_u \psi_u \psibar_v \psi_v )
\ee
If we expand this polynomial, and keep only the term $\prod_v
\psibar_v \psi_v$ (up to reordering), as appropriate for
Grassmann--Berezin integration, we see that the contributions can be
identified to configurations of edge-disjoint oriented cycles and
dimers. Edges $(uv)$ in the cycles take a factor $a_{uv}$.  Edges
$(uv)$ covered by a dimer take a factor $b_{uv}$. Vertices $v$ not
covered neither by a cycle, nor by a dimer, take a factor $m_v$.
(Unoriented) cycles come with a `topological' weight, which has a
factor 2, accounting for the two possible orientations, and a factor
$-1$, coming from the reordering of the Grassmann variables along the
cycle.  Thus, this is exactly a model of O$(-2)$ loops and dimers
(this calculation is also sketched at the very end
of~\cite{jacobsen2005arboreal}).

Now, taking care of sign factors coming from the reordering of
fermions, we can easily exponentiate the edge factors in the action, to get
\be
1 - a_{uv} (\psibar_v \psi_u + \psibar_u \psi_v) 
+ b_{uv} \psibar_u \psi_u \psibar_v \psi_v 
=
\exp \big[
- a_{uv} (\psibar_v \psi_u + \psibar_u \psi_v) 
+ (b_{uv}+a_{uv}^2) \psibar_u \psi_u \psibar_v \psi_v 
\big]
\ee
that is, exactly the spanning-forest action, provided we match the
parameters as
\be
\left\{
\begin{array}{l}
m_v = t + \sum_{u \sim v} w_{uv} \\
a_{uv} = w_{uv} \\
b_{uv}+a_{uv}^2 = -t w_{uv}
\end{array}
\right.
\ee
If the underlying graph has uniform degree $k$, and uniform weights,
this reduces to
\be
\left\{
\begin{array}{l}
m = t + k w \\
a = w \\
b+a^2 = -t w
\end{array}
\right.
\ee
that is, the partition function of loops and dimers coincides with the
one for spanning forests, under the identification above, on the
subspace of parameters
\be
\label{eq.476564}
b=(k-1)a^2-am
\ef.
\ee
Later on, for conforming to the notation of Kostov and Staudacher
(which in turns is fixed by the form of a certain functional
equation), we will parametrise the weight of a loop--dimer
configuration as
\be
\label{eq.confwei}
(-g_1)^{V-V_{\rm L}-2d}
n^\ell (4 b_0)^{(-V_{\rm L})}
(g_{\rm D} g_1^2)^d
\ee
where $V_{\rm L}$ is the number of vertices covered by loops, $\ell$
is the number of loops, and $d$ is the number of dimers. We will also
set $g_2=-(k-1)g_{\rm D} g_1^2$.
Thus we can identify
\be
\left\{
\begin{array}{l}
m = -g_1 \\
a = 1/(4 b_0) \\
b = -g_2/(k-1)
\end{array}
\right.
\ee
and for spanning forests, requiring $n=-2$, we can reformulate the
relation (\ref{eq.476564}) above as
\be
\label{eq.zerodive}
1 + \frac{4 b_0}{k-1} g_1 + \left(\frac{4 b_0}{k-1}\right)^2 g_2 = 0
\ef.
\ee
Curiously enough, as we will see, the case of spanning forests with
fugacity $-1$, which has a nice interpretation as trees with no
internal activity, is also a special point in the theory of loops and
dimers, as it corresponds to the point in the subspace of parameters
in which dimers have zero fugacity. (We do not know if this is an
accident or has a deeper reason.)

%-------------------------------------------------------
\subsection{Reduction to a one-matrix theory}
\label{ssec:goto1mat}

Let us now study the loop-dimer gas model on random planar graphs,
with uniform degree $k=h+2$.  As anticipated in (\ref{eq.confwei}), we
weight configurations of dimers and loops by the factor
\[
(-g_1)^{V-V_{\rm L}} 
n^\ell (4 b_0)^{-V_{\rm L}}
g_{\rm D}^d
\]
where $V$ is the total number of vertices, $\ell$ is the number of
loops, $d$ is the number of dimers, and $V_{\rm L}$ is the number of
vertices covered by loops.

Let us start by assuming that $n$ is a non-negative integer (we will
perform an analytic continuation afterwards). As customary for the
construction of Random Matrix theories associated to
statistical-mechanics models, following the rules of Wick theorem (see
e.g.\ \cite[sect.\ 2]{DiFrancesco2004}), we shall introduce matrices $B$ for
unoccupied edges, $D$ for dimers, and $\{ A_c \}_{c=1,\dots,n}$ for
marked edges, coloured in one of $n$ colours (this reproduces the
factor $n^\ell$).

As a result, this model is described by the following matrix integral:
\begin{align}
  \label{eq:OnRMGeneral}
  Z^{l-d}_N(g_1,b_0,\gamma) 
  &= 
  \log \displaystyle{\frac{\int \DD A_1
      \dots\DD A_n \DD B \DD D \,
      e^{-N\Tr W(A_1,...,A_n,B,D)}}
    {\int \DD A_1\dots\DD A_n\DD B \DD D \, e^{-N\Tr W_0(A_1,...,A_n,B,D)}}}\\
  W(A_1,...,A_n,B,D)
  &= 
  \frac{1}{2}B^2 + \frac{1}{2} \sum^n_{c=1} A_c^2 + \frac{1}{2}D^2
  + \frac{g_1}{k} B^k + \sqrt{g_{\rm D}} g_1 B^{k-1} D +\nonumber \\
  &\quad - \frac{1}{8b_0} \sum_{c=1}^n\sum_{h'=0}^h B^{h'}A_cB^{h-h'}A_c 
  \nonumber \\
  W_0(A_1,...,A_n,B,D)
  &= 
  \frac{1}{2} B^2 + \frac{1}{2} \sum^n_{c=1} A_c^2 + \frac{1}{2} D^2\, .
  \nonumber
\end{align}
We have made the choice of having the quadratic part trivial. As a
result, in the model construction, we have weights associated to
``vertices'' of interaction, and no weights associated to edges. A
monomial of the form
$-\frac{1}{|\Aut|} g \prod_i M_{c(i)}$ corresponds to a vertex with edge-colour
content $\{c(i)\}$, weighted with a factor $g$.\footnote{Here $|\Aut|$
  is the number of automorphisms of the string $\{c(i)\}$
  w.r.t.\ cyclic shifts.  For example, a vertex $-g/3 \;ABBABBABB$ makes
  a vertex of degree 9, in which one in three edges are of type $A$,
  the others being of type $B$.}
As a result, it is easily recognised that, as desired, vertices of the
graph not in any dimer or loop, associated to the term $g_1/k B^k$,
come with a weight $-g_1$. Vertices in a loop, associated to the last
term of the potential, come with weight $1/(4 b_0)$, and each of the
two vertices of a dimer come with weight $-\sqrt{g_{\rm D}} g_1$.

If we shrink dimers to points, those can be seen as vertices of
degree $2k-2$ with a coupling $g_{\rm D} g_1^2$. The algebraic
counterpart of this is the Gaussian integral over matrix $D$. As a
result, the action becomes
\begin{equation}
  \label{eq:action_no_dimers}
  W
  = 
  \frac{1}{2}B^2 + \frac{1}{2} \sum^n_{c=1} A_c^2 
  + \frac{g_1}{k} B^k + \frac{g_2}{2k-2} B^{2k-2} 
  - \frac{1}{8b_0} \sum_{c=1}^n\sum_{h'=0}^h B^{h'}A_cB^{h-h'}A_c \, ,
\end{equation}
where we have introduced $g_2 := -(k-1)g_{\rm D} g_1^2$ (again, with
the aim of exposing a $1/|\Aut|$ factor).  After passing to the
eigenvalue basis for matrix $B$
\begin{equation}
  \label{eq:change_coord}
  B \longrightarrow U \mbox{diag}(x_1,\dots, x_N) U^{-1}, 
  \qquad A_c \longrightarrow U A_c U^{-1}\, ,
\end{equation}
the part of the action depending on
the matrices $A_c$ reads
\begin{equation}
  \label{eq:Ac_term}
  \frac{1}{2} \sum_{ij} (A_c)_{ij} (A_c)_{ji} 
  \left(1 - \frac{1}{4b_0}(x_i^h + x_i^{h-1}x_j+\dots+x_j^h)\right)\, ,
\end{equation}
and if we integrate out the matrices $A_c$ we end up with the following
one-matrix integral:
\begin{equation}
  \label{eq:OnAcIntegrated}
  Z^{l-d}_N(g_1,g_2,b_0) 
  = 
  \log \displaystyle{\frac{
      \int \prod_{i=1}^N \ud x_i \,
      \Delta^2(\{x_i\}) \, \left(\hat{\Delta}_{1/(4b_0)}(\{x_i\})\right)^{-n/2}\,
      e^{-N\sum_i V(x_i)}}
    {\int \prod_{i=1}^N \ud x_i\,  \Delta^2(\{x_i\}) \, e^{-N/2 \sum_i x_i^2}}}\, .
\end{equation}
where $\hat{\Delta}$ is the same quantity defined in \eqref{eq:DeltaHatgx}, and
\begin{equation}
  \label{eq:V_l-d}
  V(x) = \frac{x^2}{2} + \frac{g_1}{k}x^k + \frac{g_2}{2k-2} x^{2k-2} \, .
\end{equation}

Now we compare the partition functions for spanning forests
\eqref{eq:MatIntSpForx} and for the O$(n)$ loop-dimer gas
\eqref{eq:OnAcIntegrated}. These are identical if we set $n=-2$ and
identify the parameters of the two models in the following way:
\begin{equation}
  \label{eq:ParsMatching}
  \left\{ 
    \begin{array}{l} 
      \frac{1}{4b_0} = g\\ 
      g_1 = -g (k+t)\\ 
      g_2 = (k-1) g^2 (1+t)
    \end{array} \right. \, .
\end{equation}
Note that this identification corrects the derivation in
\cite{CS_RMFor}, and will allow us to draw conclusions on the critical
behaviour of spanning forests using the available results for the
O$(n)$ model. Also note that, not surprisingly, this substitution
satisfies the condition~(\ref{eq.zerodive}). That is, if we substitute
(\ref{eq:ParsMatching}) into 
$1 + \frac{4 b_0}{k-1} g_1 + \big(\frac{4 b_0}{k-1}\big)^2 g_2$
we get
$1 - \frac{k+t}{k-1} + \frac{1+t}{k-1} = 0$.
In turns, the procedure of this section, of averaging over a certain
statistical ensemble of (planar) graphs with fixed degree $k$, has
obviously implemented the same condition that holds separately for
each and every graph of degree $k$.

%%%%%%%%%%%%%%%%%%%%%%%%%%%%%%%%%%%%%%%%%%%%%%%%%%%%%%%
\section{Analysis of the Random Matrix Model at genus zero}
%in the planar limit}
\label{sec.mainana}

In this section we will discuss the critical behavior of spanning
forests on random planar graphs by studying the large-$N$ limit of the
matrix integral \eqref{eq:MatIntSpFor}.  We refer the reader to the
reviews \cite{DiFrancesco1993,DiFrancesco2004,Eynard2015} for
background material on techniques for matrix models.  We will try to
give a self--consistent presentation, mainly following
\cite{DiFrancesco1993} and adapting it to our case.  An ingredient of
our problem, with respect to the (nowadays standard) treatments
presented in the forementioned reviews, is that we have to trade one
or the other complication: either we have an extra factor
$\hat{\Delta}$ beside the customary Vandermonde factors (in variables
$x$, w.r.t.\ notations of Section \ref{sec:rm_sf}), or we deal with a
potential of the one-matrix model which is not polynomial (in
variables $z$). We adopt this second approach. As will be discussed,
the standard formulas for a polynomial potential go through with
little modification, but obtaining explicit solutions poses new
technical challenges.

%-------------------------------------------------------
\subsection{Saddle-point equation}

To facilitate the singularity analysis it is convenient to perform the
change of variables $M\to g^{-1/h} M$ which, apart from an irrelevant
prefactor $g^{-N^2/h}$, leaves us with the following matrix integral
replacing the numerator of \eqref{eq:MatIntSpFor}:
\begin{align} 
  \label{eq:35}
  \int \DD M\,
  \exp\left(-\frac{N}{v}\, \Tr W(M)\right)\, ,
\end{align}
where $W$ coincides with $V$ with $g$ set to one:
\begin{align}
  \label{eq:V_new}
  W(z) &= \frac{1}{2} z^2 - t \sum_{n\ge 1}A'_{h,n} z^{hn+2}\, ,
\end{align}
and for easiness of notation we introduced
\begin{align}
  v = g^{2/h} \, .
\end{align}
We recall that the parameter $h+2$ is the degree of the graph,
and that the series in $n$, $A'_{h,n}$, has region of convergence
\begin{align}
  \label{eq:def_domain}
  |z| < r_{h} \equiv
  \frac{h}{(h+1)^{(h+1)/h}}
  \, .
\end{align}
The potential can be analytically continued to a larger region in the
complex plane, as will be discussed below for the first few values of
$h$.

In the eigenvalue basis, the integrand is $e^{-N^2
  S(\{\lambda_i\})}$, with the action
\begin{equation}
  \label{eq:action}
    S(\{\lambda_i\}) = \frac{1}{v}\frac{1}{N} \sum_{i=1}^N W(\lambda_i)
    - \frac{1}{N^2}\sum_{1\le i\neq j\le N}\log|\lambda_i-\lambda_j|
    \, ,
\end{equation}
and the saddle-point equation $\partial_{\lambda_\ell} S = 0$ 
results in
\begin{equation}
  \frac{1}{v}W'(\lambda_\ell)
  =
  \frac{2}{N}\sum_{i(\neq \ell)}\frac{1}{\lambda_\ell-\lambda_i}
  \, .
\end{equation}
In the large-$N$ limit the action \eqref{eq:action} becomes a functional
of the spectral density
\begin{align}
  \rho(\lambda) = \lim_{N\to\infty}\frac{1}{N}\sum_{i=1}^N \delta(\lambda
  -\lambda_i)\, ,
\end{align}
and reads
\begin{align}
  \label{eq:S_limit}
  S(\lambda) 
  = 
  \frac{1}{v}\int \ud \lambda \rho(\lambda) W(\lambda)
  - 
  \fint\ud \lambda\ud \lambda'  \rho(\lambda)\rho(\lambda')
  \log |\lambda-\lambda'|\, ,
\end{align}
where the double integral in the second summand is regularized by its
Cauchy principal value.  It is understood that the support of $\rho$
shall be contained within the forementioned domain of analyticity of
$W$.  In the planar limit the saddle-point equation becomes the
following integral equation
\begin{align}
  \frac{1}{v}
  W'(\lambda) 
  &=
  2\fint \ud \lambda' \frac{\rho(\lambda')}{\lambda-\lambda'}\, ,\quad
    \lambda \in \text{supp}(\rho)\, ,
\label{eq.r654e564}
\end{align}
which computes the extrema of the functional $S(\rho)$.
This equation has to be solved together with the normalization
condition
\begin{align}
\label{eq.r654e564b}
  \int \ud \lambda \, \rho(\lambda) = 1\, .
\end{align}
Given a solution of the saddle-point and normalization equations,
the generating function of spanning forests for random planar graphs
is then given by
\begin{align}
  \label{eq:ZNinfty}
  \lim_{N\to \infty} Z(t,g,N) = S_0(\rho_0) - S(\rho)\, ,
\end{align}
where $Z(t,g,N)$ was defined in \eqref{eq:MatIntSpFor}. The action $S_0$
is given by \eqref{eq:S_limit} with $W(\lambda)/v$ replaced by
$V_0(\lambda) =\lambda^2/2$, and thus
$\rho_0$ is given by the celebrated Wigner semicircle formula,
%$\rho_0(\lambda) = \frac{1}{2\pi}\sqrt{4-\lambda^2}$
%with support $[-2,2]$, and 
$S_0(\rho_0)$ is just a constant, which can be ignored
as far as the singularity analysis is concerned.

%-------------------------------------------------------
\subsection{Solution of the saddle-point equation}

As standard, the singular integral equation representing the saddle
point is solved by rewriting it as a Riemann--Hilbert problem for the
resolvent $\omega$. Introducing
\begin{align}
  \omega(z) = \frac{1}{N}\sum_{i=1}^N \frac{1}{z-\lambda_i}\, ,
\end{align}
few lines of calculation, starting from the saddle-point
equation, show that in the large $N$ limit $\omega$ satisfies
\begin{align}
  \omega^2(z) = \frac{1}{v}W'(z)\omega(z) - \frac{1}{4v^2}P(z)\, ,
  \quad
  P(z) = 4v\int \ud \lambda \rho(\lambda)\frac{W'(z)-W'(\lambda)}
  {z-\lambda}\, .
\end{align}
The quantity $P(z)$ is yet to be determined. As apparent from its
definition, it is guaranteed to be analytic only in the domain where
$W(z)$ is.  At large $N$, the relation between $\omega$ and $\rho$ is
\begin{align}
  \omega(z) 
  &=
  \int \ud \lambda \frac{\rho(\lambda)}{z-\lambda} \, ,
\end{align}
which shows that $\omega$ is an analytic function in 
$\mathbb{C} \setminus \text{supp}(\rho)$, this despite $W$ being
possibly not analytic in all $\mathbb{C}$ (as is the case
here). Furthermore, Eq.~(\ref{eq.r654e564b}) implies the asymptotic
behavior
\begin{align}
  \label{eq:omega_zinfty}
  \omega(z)\sim \frac{1}{z}\, ,\quad z\to \infty
\ef .
\end{align}
Define $\omega^\pm(\lambda)=\omega(\lambda\pm i 0)$ as the values
right above and below the support of $\rho$. One has the following
equations
\begin{align}
&
\left\{
\begin{array}{l}
  \omega^-(\lambda) - \omega^+(\lambda) = 2\pi i \rho(\lambda)\, ;
\\  
\rule{0pt}{13pt}%
  \omega^-(\lambda) + \omega^+(\lambda) = \frac{1}{v}W'(\lambda)\, ;
\end{array}
\right.
&&
\lambda \in \text{supp}(\rho)\, .
\label{eq:RHP}
\end{align}
The latter corresponds to the aforementioned Riemann--Hilbert problem.
Roughly speaking, its solvability requires that $W'$ is continuous on
any open set within $\text{supp}(\rho)$,
% with the edges removed, 
a condition certainly met in our case.  (See e.g.~\cite{Gakhov1990}
for a more precise treatment.)

We now discuss the solution of the second of the equations above.
% \eqref{eq:RHP}. 
We make the customary \emph{one--cut assumption} that the support of
$\rho$ consists of a single segment, that we parametrise as
$[a_1,a_2]$.  This means that the problem has to be solved with the
further requirement that $\rho > 0$ on $[a_1,a_2]$, where $a_1,a_2$
will be determined as functions of the parameters in the problem. If
positivity fails for some parameter range, then the assumption is
invalidated, and, most probably, multiple cuts need to be considered.
Let us define
\begin{align}
  \sigma(z) = (z-a_1)(z-a_2)\, ,
\end{align}
we can write
\begin{align}
  \omega(z) = \frac{1}{2v}\left( W'(z)-\sqrt{W'(z)^2-4 P(z)}\right)
    = \frac{1}{2v}\left( W'(z)- M(z) \sqrt{\sigma(z)} \right)\, ,
\end{align}
where $M$ is analytic in some open set containing the interval
$[a_1,a_2]$.  These prescriptions are sufficient to determine $M$ and
$\omega$ uniquely. We report here the derivation of \cite{Eynard2015},
while paying attention to analyticity properties, which are slightly
different in the case at hand. We recall that $\mathbb{C}\setminus
[a_1,a_2]$ is the domain of analyticity of $\omega/\sqrt{\sigma}$, and
consider in it a point $z$ and a cycle $C_z$ around $z$.  Further, we
denote by $C_{[a_1,a_2]}$ a cycle around $[a_1,a_2]$ which does not
enclose $z$, and assume that it is contained in the domain of
analyticity of $W$.  Then, one has
\begin{align}
  \frac{\omega(z)}{\sqrt{\sigma(z)}}
  =
  \frac{1}{2\pi i}
  \oint_{C_z} \frac{\ud w }{w-z}  \frac{\omega(w)}{\sqrt{\sigma(w)}}
  =
  -\frac{1}{2\pi i}
  \oint_{C_{[a_1,a_2]}} 
  \frac{ \ud w }{w-z}  \frac{\omega(w)}{\sqrt{\sigma(w)}}\\
  =
  -
  \frac{1}{2\pi i} \frac{1}{2v}
  \oint_{C_{[a_1,a_2]}} 
  \frac{\ud w }{w-z}  \frac{W'(w)}{\sqrt{\sigma(w)}}
  =
  \frac{1}{2\pi i} \frac{1}{v}
  \int_{a_1}^{a_2} 
  \frac{\ud \mu }{\mu-z}  \frac{W'(\mu)}
  {\sqrt{\sigma(\mu)}}\, .
\end{align}
Correspondingly, the formula for the spectral density is
\begin{align}
  \label{eq:rhoW}
  \rho(\lambda) = \frac{1}{2\pi^2 v}\sqrt{-\sigma(\lambda)}
  \fint_{a_1}^{a_2} 
  \frac{\ud \mu }{\mu-\lambda}  \frac{W'(\mu)}
  {\sqrt{-\sigma(\mu)}}\, .
\end{align}
The spectral edges of the support are chosen to ensure the correct
large-$z$ behavior of $\omega$, equation~\eqref{eq:omega_zinfty}, and are
determined by the following equations:
\begin{subequations}
  \label{eq:intVp}
\begin{align}
  \label{eq:intVp1}
  \int_{a_1}^{a_2} 
  \ud \mu \frac{W'(\mu)}{\sqrt{-\sigma(\mu)}} 
&= 0\, ,
\\
  \label{eq:intVp2}
  \int_{a_1}^{a_2} \ud \mu 
  \frac{\mu W'(\mu)}{\sqrt{-\sigma(\mu)}} 
&= 2\pi v\, .
\end{align}
\end{subequations}
The first one is a necessary condition for the solvability of the
Riemann--Hilbert problem, while the second one ensures normalization
of $\rho$.  By passing to contour integrals around the support, and
changing variables via the Joukowski map,
\begin{align}
  \label{eq:Joukowski}
  \mu &= S + z + \frac{R}{z}\, ,
&
  S &= \frac{a_1+a_2}{2}\, ,
&
  R &= \left(\frac{a_1-a_2}{4}\right)^2 \, ,
\end{align}
one can rewrite these conditions as
\begin{subequations}
  \label{eq:ointVp}
\begin{align}
  \oint \frac{\ud z}{2\pi i z} W'(\mu(z)) &= 0\, ,
\\
  \oint \frac{\ud z}{2\pi i} W'(\mu(z)) &= v\, .
\end{align}
\end{subequations}
We now specify these formulas to our potential \eqref{eq:V_new}.
Its derivative is
\begin{align}
  \label{eq:Vp_new}
  W'(z) &= z - t  \sum_{n\ge 1}A_{h,n} z^{hn+1}\, ,
\end{align}
where the $A_{h,n}$ numbers are Fuss--Catalan numbers:
\begin{align}
  \label{eq:Ahn}
  A_{h,n} = \frac{((h+1)n)!}{n! (hn+1)!} \, .
\end{align}
The conditions on $a_1$, $a_2$, or equivalently $R$, $S$ (through
(\ref{eq:Joukowski})), then read as follows:
\begin{subequations}
  \label{eq:RS}
\begin{align}
  S &= t\, \sum_{n\ge 1}\sum_{q\ge 0}
  A_{h,n} 
  \frac{(hn+1)!}{(hn-q+1)!^2(hn-2q+1)}
  R^{hn-q+1} S^{2q-hn-1}
  % \frac{R}{S}
  % \left(\frac{R}{S}\right)^{h n}
  % \left(\frac{S^2}{R}\right)^q
  \, ,
\\
  \label{eq:vR}
  v &= R - t
  \sum_{n\ge 1} \sum_{q\ge 0}
  A_{h,n} 
  \frac{(hn+1)!}{(hn-q+1)!(hn-q)!(hn-2q)!} 
  R^{hn-q+1}S^{2q-hn}
  % R
  % \left(\frac{R}{S}\right)^{h n}
  % \left(\frac{S^2}{R}\right)^q
  \, .
\end{align}
\end{subequations}
These equations coincide with those appearing in 
\cite[Thm 3.1]{BMCourt2015} (they use $z$ and $u$ in place of $v$ and
$t$).  Remarkably, they were derived with a purely combinatorial
method, quite different from the one presented.  The fact that both
approaches lead to the same equations is rooted in previous work
\cite{BouGui} on which \cite{BMCourt2015} is based, and the connection
between combinatorial methods and matrix
models~\cite{bouttier2011matrix, DiFrancesco2004}.

%-------------------------------------------------------
\subsection[Saddle-point equation for the 
$\mathrm{O}(n=-2)$ 
model]
{Saddle-point equation for the 
% O$(n=-2)$ 
\hbox{\textbf{O(\textit{n}\,=\,-\!-2)}}
model}

We present now the reformulation of the saddle-point equation from the
point of view of the O$(n)$ model. As already mentioned, the resulting
potential will be polynomial, at the expenses of a more complicated
form of the equation.  In order to make contact with the formulas of
Section \ref{sec:l-d}, we will work here with the original $V$ which
contains $g$ explicitly.  We remind the reader that the connection
with the O$(n)$ model emerges upon the change of variables 
$z \to x(z)$, whose inverse is $f(x) := z(x) = x(1-g x^h)$ (note that $z'(x)= 1-g(h+1)x^h$),
and that this leads to a polynomial potential:
\begin{align}
\tilde{V}(x) 
&=
  V(x(z)) = \frac{x^2}{2}+\frac{g_1}{h+2}x^{h+2}+
  \frac{g_2}{2h+2}x^{2h+2}\, ,
\\
\frac{d\tilde{V}(x)}{dx}
& =
\frac{dV(z)}{dz} \frac{dz(x)}{dx}
= V'(f(x)) f'(x) 
= x
  + g_1 x^{h+1}
  + g_2 x^{2h+1}
  \, .
\end{align}
We define $\tilde{\rho}$ as the distribution density in the new
variable
\begin{align}
  \tilde{\rho}(y)
  := 
  f'(y)
  \rho(f(y)) \, ,
\end{align}
the factor $f'(y)$ arises from the Jacobian of the transformation, i.e.
\begin{align}
  \ud y \tilde{\rho}(y) &= \ud \mu \rho(\mu)\, ,
&
\mu&=f(y)\, .
\end{align}
In terms of $\tilde{\rho}$, the large-$N$ action is
\begin{align}
  \tilde{S}(\tilde{\rho}) 
  = 
  \int \ud x \tilde{\rho}(x) \tilde{V}(x)
  - 
  \fint\ud x \ud y
  \tilde{\rho}(x)\tilde{\rho}(y)
  \log |f(x)-f(y)| \, ,
\end{align}
and the saddle-point equation becomes:
\begin{align}
\label{eq.76365356}
  x
  + g_1 x^{h+1}
  + g_2 x^{2h+1}
  =
  2 f'(x)
  \fint \ud y
  \frac{\tilde{\rho}(y)}
  {f(x)-f(y)}
\, .
\end{align}
Noting that $(f(x)-f(y))/(x-y)$ is a polynomial in $x$ and $y$,
and calling $\xi_1^{(h)}(y)$, \ldots, $\xi_h^{(h)}(y)$ its roots
as a polynomial in $x$, it is in
fact the case that\footnote{More generally, if $P(x)$ is a polynomial
  with distinct roots $\xi_i$, then 
  $\sum_i \frac{1}{x-\xi_i} = \frac{P'(x)}{P(x)}$.}
\be
\frac{1}{x-y} + \sum_{j=1}^h
\frac{1}{x-\xi_j^{(h)}(y)}
=
\frac{ f'(x) }
  {f(x) - f(y)}
\ee
so that (\ref{eq.76365356}) can be reformulated as
\begin{align}
\label{eq.76365356b}
  x
  + g_1 x^{h+1}
  + g_2 x^{2h+1}
  =
  2 
  \fint \ud y
  \tilde{\rho}(y)
\bigg(
\frac{1}{x-y} + \sum_{j=1}^h
\frac{1}{x-\xi_j^{(h)}(y)}
\bigg)
\ef.
\end{align}
It is easy to see that, for small $g$, these roots converge to the
roots of unity, multiplied by $g^{-1/h}$, for example, at $h=1$ and 2
we have
\begin{align}
  &h=1: \quad&
\xi_1^{(1)}(y) &= (g^{-1}-y)\\
  &h=2: \quad&
\xi_{1,2}^{(2)}(y) &= -\frac{y}{2} \pm \sqrt{g^{-1}-\smfrac{3}{4} y^2}
\end{align}
and thus, for $x$ and $y$ small w.r.t.\ such a quantity, the
denominators $x-\xi_j^{(h)}(y)$ do not vanish. Let $b$ be the largest
value in the support of $\tilde{\rho}$. It will turn out that we have
an interesting behavior when $b$ is large enough for one of these
denominators to vanish. If we imagine to increase the value of $b$,
this is first possible when both $x$ and $y$ are near $b$, i.e.\ when
$\lim_{x,y \to b} \frac{f(x)-f(y)}{x-y}= \lim_{x \to b} f'(x)=0$,
which corresponds to the condition
$(h+1) g b^h=1$. In terms of the original variable $\lambda$, this singularity
occurs when the right spectral edge reaches the critical radius of $W$.

On the other hand, Eq.~\eqref{eq.76365356} can also be rewritten as
\begin{align}
  x
  + g_1 x^{h+1}
  + g_2 x^{2h+1}
  =
  2 f'(x) 
  \fint \ud y
  \frac{\tilde{\rho}(y)}{f'(y)}
  \left(
    \frac{1}{x-y}
    +
    \sum_{i=1}^h
    \frac{1}{\xi_i^h(x)-y}
  \right)
  \ef.  
\end{align}
In terms of the resolvent
\begin{align}
  \tilde{\omega}(z) = f'(x) \int \ud y \frac{\tilde{\rho}(y)}{f'(y)}\frac{1}{z-y}
%  =
% f'(x) \int \ud y \frac{\rho(f(y))}{z-y}
\, ,
\end{align}
this equation reads as:
\begin{align}
  x
  + g_1 x^{h+1}
  + g_2 x^{2h+1}
  =
  \tilde{\omega}^+(x)
  +
  \tilde{\omega}^-(x)
  +
  2
  \sum_{i=1}^h\tilde{\omega}(\xi^h_i(x))
\, , \quad x\in\text{supp}(\tilde{\rho})
\end{align}
since $f'(x)$ is a polynomial and therefore continuous, and 
$\xi_i(x)$ are outside the support of $\tilde{\rho}$.
We remark that similar equations appear also in other statistical 
mechanical problems, see
e.g.~\cite{Kostov2002245,KOSTOV2000513,PZinn2000},
and general techniques for dealing with these equations have been
developed in \cite{Borot2013}.

To the best of our knowledge the critical behavior of the O$(n)$ loop
gas on a graph of arbitrary degree, as the one appearing in
section \ref{ssec:goto1mat}, has not been discussed in the
literature. The only explicit results relevant for us are for the case
$h=1$, and are discussed in several works, among which the seminal
paper \cite{KoStau}.  The method employed in \cite{KoStau} is to start
from our equation \eqref{eq.76365356} and map it back to the standard
singular integral equation appearing in \eqref{eq.r654e564}, at the
expenses of a non polynomial potential.  The singular equation is then
solved along the lines discussed above.  The authors of \cite{KoStau}
discussed however only isolated critical points corresponding to $t=0$
and $t=-1$ in our notation, and not the presence of a massless phase,
the quantum gravity counterpart of the Berker--Kadanoff phase, which
to our knowledge has been found only much later, in
\cite{BMCourt2015}. In section \ref{ssec.anacubi} we will present a
self--contained discussion of the solution for $h=1$, which agrees
with the one presented by \cite{KoStau} at the specific values of the
couplings mentioned above.

Interestingly, the criticality of one-matrix models with
non-polynomial potentials has been investigated in recent work
\cite{ambjorn2016generalized}, whose findings are possibly of
relevance also for the model at hand.

%-------------------------------------------------------
\subsection{Singular behavior of the partition function}

Knowledge of the spectral density determines the partition
function and all the correlators of a matrix model. 
Recall Eq.~\eqref{eq:ZNinfty} and the formula:
\begin{align}
  \left< \Tr f_1(M)\cdots \Tr f_n(M) \right>
  =
  N^{n}
  \int \ud \lambda_1\cdots \ud \lambda_n
  f_1(\lambda_1)\cdots f_n(\lambda_n)
  \left< \rho(\lambda_1)\cdots \rho(\lambda_n)
  \right>\, .
\end{align}
Here $\left< \cdot \right>$ stands for the average w.r.t~the measure
defined by \eqref{eq:35}.  Further, in one-matrix models with a
connected spectral support, the critical behavior is determined only
by the spectral edges, and this result is universal in the sense that
it holds for any potential \cite{DiFrancesco1993}. For the reader's
convenience, we re-derive this result in this section.  The reason for
this universality can be traced back to equation~\eqref{eq:RHP} and
the way $v$ enters in it.  Indeed the function
\begin{align}
  \Omega(z) = \frac{\partial (v \omega(z))}{\partial v}\, ,
\end{align}
satisfies 
\begin{align}
  \Omega^+(\lambda)+  \Omega^-(\lambda)= 0\, ,
\end{align}
independently of $W$. From the behavior $\Omega(z) \sim z^{-1}$ as
$z\to\infty$, and the one-cut assumption one obtains
$\Omega(z)=1/\sqrt{\sigma(z)}$, and therefore
\begin{align}
  \frac{\partial (v \rho(\lambda))}{\partial v} =
  \frac{1}{\pi}\frac{1}{\sqrt{-\sigma(\lambda)}}\, .
\end{align}
One then notices that in the large-$N$ limit the derivative of $Z$
is
\begin{align}
  v^3 \frac{\partial Z}{\partial v}
  =
  v N \left< \Tr W \right>
  \sim
  N^2
  \int_{a_1}^{a_2} \ud \lambda \,(v \rho(\lambda))\, W(\lambda)\, ,
\end{align}
and therefore
\begin{align}
\nonumber
  \frac{\partial^2}{\partial v^2}
  \left(
    v^3 \frac{\partial Z}{\partial v}
  \right)
  &\sim
  \frac{N^2 }{4\pi i}
  \Bigg(
    \frac{\partial a_1}{\partial v}
    \oint \ud \lambda 
    \frac{W(\lambda)}{(\lambda-a_1)^{\frac{3}{2}}(\lambda-a_2)^{\frac{1}{2}}}
% \\    &\qquad
+
    \frac{\partial a_2}{\partial v}
    \oint \ud \lambda 
    \frac{W(\lambda)}{(\lambda-a_2)^{\frac{3}{2}}(\lambda-a_1)^{\frac{1}{2}}}
  \Bigg)\\
  &=
  N^2 v \frac{\partial }{\partial v} \log(16 R)
  \, ,
\end{align}
where the second equality follows from integrating by parts the
constraints (\ref{eq:intVp}). Furthermore, near to the critical point,
as $\epsilon = 1 - v/v_c \to 0$, after replacing powers of $v$ with
$v_c$ and integrating, one obtains
\begin{align}
  \label{eq:ZlogR}
  Z'' \sim \frac{N^2}{v_c^2} \log(16 R)\, .
\end{align}
Therefore the singular behavior of $Z$ can be obtained just 
by knowing the behavior of $R$ close to a critical point,
and this can be inferred from equations~\eqref{eq:RS}.
By the transfer principle,
% \cite{flajodl},
knowing the leading dependence of $Z$ in $\epsilon$ dictates the
asymptotic behavior of $Z_n$, where $Z=\sum_{n\ge 1}Z_n g^n$:
\begin{align}
	\label{eq:Zn}
   Z = \epsilon^{-\alpha} \log^\beta(\epsilon)
   \Rightarrow
	Z_n = g_c^{-n} n^{\alpha-1} \log^\beta(n)
	\left(\frac{1}{\Gamma(\alpha)}+\frac{\beta}{\log(n)}
	\frac{\ud}{\ud s}\frac{1}{\Gamma(s)}\Big|_{s=\alpha}
	+\dots\right)\, .
\end{align}
For a proof of this result, see \cite[Thm.~6.2]{Flajolet}.
From the asymptotic behavior of $Z_n$ we define
the critical exponents
\begin{align}
  Z_n \sim n^{\gamma-3} \log(n)^{\gamma'}\, ,
\end{align}
where, at least when $\gamma'=0$, $\gamma$ is interpreted as the
string susceptibility exponent and is related to the central charge of
the model coupled to gravity by the KPZ
formula~\eqref{eq:KPZ}~\cite{DiFrancesco1993}.

Let us pause for a moment our calculations, and make a digression on
the subtleties of using KPZ here. The original KPZ formula~\cite{KPZ}
reads
\begin{equation}
  \label{eq:KPZ}
  \gamma = 2 - \frac{1}{12}\left(25 - c + \sqrt{(1-c)(25-c)}\right)
\ef.
\end{equation}
From this, and $\gamma = -1$, it is easily recovered the well-known
value $c=-2$ for the central charge of uniform spanning trees in
dimension two. The situation is in fact more complex, although the
conclusions are unchanged. The formula above holds, in principle, only
for unitary CFTs. In non-unitary theories, the formula is modified
into
\begin{equation}
  \label{eq:KPZ}
\begin{split}
  \gamma 
&= 2 - \frac{2}{24+c_{\rm eff}-c}\left(25 - c +
  \sqrt{(1-c_{\rm eff})(25-c)}\right)
\\
&=
2 - \frac{1}{12(1-h_{\rm min})} 
\left(25 - c + \sqrt{(1+24 h_{\rm min}-c)(25-c)}\right)
\ef,
\end{split}
\end{equation}
where $c_{\rm eff}=c-24 h_{\rm min}$ is the `effective' central
charge, and $h_{\rm min}$ is the lowest conformal dimension in the
theory \cite{stauDimers, brezin1990ising}.  In the CFT for spanning
trees, there is a single field with negative dimension, $h=-1/8$ (see
\cite{SALEUR1992, Ivashkevich1999}), so that we shall obtain
$\gamma=0$. However, this field is associated to a twist operator for
the boundary conditions, which is absent from the setting in which we
constructed our theory, so that the formula with $h_{\rm min}=0$ shall
be used, and $\gamma=-1$ is obtained accordingly. We conjecture that,
in genus 1, there might be a way of constructing a model in which the
twist operator is present (besides the obvious generalisation of the
treatment here to higher genus, in which it would be absent), but we
do not investigate this aspect here.

To the best of our knowledge, the exponent $\gamma'$ has not been
discussed in the context of Liouville theory coupled to the present
CFT.  Nonetheless, logarithmic corrections in lattice discretizations
of $c=1$ CFTs are well known, both for flat \cite{affleck1989critical}
and fluctuating graphs \cite{BREZIN1990}, and their interpretation
here may also be accessible at the light
of~\cite{jacobsen2005arboreal}.

Below we will discuss the critical behaviour in several cases. Some of
these cases were already studied in \cite{BMCourt2015}, and for these
we re-derive and extend their results from a random matrix
perspective. The generating function for spanning forests $F(z,u)$
considered in \cite{BMCourt2015} is expressed in terms of $R,S$ and,
since the definition of $F$ lacks the factor $|\Aut(G/F)|$ present in
\eqref{eq:ZSpForRanG}, it is expected to have the same singular
behavior as $\partial Z/\partial g$. We remark that the random matrix
derivation provides a direct connection between the singularity of $Z$
and that of $R$ via \eqref{eq:ZlogR} (in particular, there is no need
of usig $S$, even in the case of odd degree), and this fact, not
used in \cite{BMCourt2015}, simplifies the singularity analysis.

Let us conclude this section with a comment on the case of spanning
trees, which is arguably the simplest case.  As is apparent from
formula \eqref{eq:ZSpForRanG}, spanning trees emerge as
\begin{align}
  Z_{\text{tree}}(g,N)
  =
  \lim_{t\to 0}\frac{\partial}{\partial t} Z(t,g,N)
  =
  -\frac{1}{N}
  \sum_{n\ge 1} g^n A'_{h,n}\left<  M^{hn+2}\right>_0\, ,
\end{align}
where $\left< \cdot \right>_0$ is the average w.r.t.~the Gaussian
measure. In the large $N$ limit the result is well known
\cite{DiFrancesco2004} and given by the Catalan numbers:
\begin{align}
  \lim_{N\to \infty}\frac{1}{N} \left<  M^{r}\right>_0 = 
  \begin{cases}
    \frac{1}{p+1}\binom{2p}{p} & \text{if } r=2p\, ,\\
    0 & \text{otherwise,}
  \end{cases}
\end{align}
(and recall that $A'_{h,n}$ is itself a Fuss--Catalan number).  One
then recovers the expression for the generating function of spanning
trees on random planar graphs exactly from the matrix integral (see
\cite{CS_RMFor} for further details and references).  Here we just
recall that the asymptotics $Z_{n}\sim g_c^{-n}n^{-4}$, with $g_c$
determined by the radius of convergence of the series,
\begin{align}
  g_c = 
  2^{-h}\frac{h^h}{(h+1)^{(h+1)}}\, .
\end{align}
This behavior corresponds via the KPZ formula to $c=-2$, the known
central charge for the problem on flat lattices (see although the
\emph{caveat} mentioned a few lines above).

%-------------------------------------------------------
\subsection{The case of quartic graphs}
\label{ssec.anaquart}

We start by discussing the case of quartic graphs, i.e.~$4$-regular
graphs. Since the potential is even, the support will be of the form
$[-a,a]$, and therefore $S=0$, $R=a^2/4$.  Equations~\eqref{eq:RS}
then reduce~to
\begin{align}
  \label{eq:R_4}
& g = R - t \Phi(R) \, ,
&
\Phi(R)
&=
  R\left({}_2F_1(\tfrac{1}{3},\tfrac{2}{3};2;27 R)-1\right)
  \, .
\end{align}
Note that the radius of convergence of $\Phi$ is 
\begin{align}
  \label{eq:r_2^2/4}
  \tau = \frac{r_{2}^2}{4}= \frac{1}{27}\, ,
\end{align}
where $r_2$ is the radius of convergence of $W$.  The singularity
analysis of this implicit equation for $t\ge -1$ was carried out in
\cite[Sec.~8]{BMCourt2015}. Here we will recall the main steps, and
extend the analysis to all values of $t$. (See \cite{Flajolet} for a
reference on singularity analysis. The words singular, non--regular
and non--analytic are used as synonyms.)  The central result is an
instance of the implicit function theorem, which asserts that given a
function $\Omega(R)$ analytic at $R_0$, and $\Omega'(R_0)\neq 0$,
there exists an inverse $R(g)$, such that $\Omega(R)=g$, and $R(g)$ is
analytic in a neighbourhood of $g_0=\Omega(R_0)$. The proof is well
known and relies on analyticity of $\Omega$ at $R_0$ to expand it in
Taylor series.  There are thus two competing sources for the failure
of analyticity of an implicit function: (1)\ $\Omega$ is singular at
$R$; (2) $\Omega$ is regular at $R$ but $\Omega'(R) = 0$.  If $R(g)$
is analytic at the origin, the singular behavior of $R$ and the
exponential growth of its coefficients are dictated by $\rho$, the
radius of the \emph{nearest} singular point.
% encountered.

In our case $\Omega(R) = R - t\, \Phi(R)$, and $\Omega'(R)= 1-t\,\Phi'(R)$.
We note that our $\Omega$ can be analytically continued to $\mathbb{C}
\setminus [\tau,\infty)$.  Our goal being the singularity analysis for
arbitrary real $t$, we will not restrict the values of $R,g$ to their
initial range of definition, namely $R\in (0,\tau), g>0$, but consider
them as complex parameters, and postpone the interpretation of the
results, and possible `physical restrictions', to the end of the calculation.
% We will comment later on the interpretation of the results.  

In light of the discussion above, the two
kinds of singular points are:
\begin{enumerate}
\item triples $(t,R_1,g_1)$ such that $R_1 = \tau$ and 
  $g_1(t) = \tau-t\, \Phi(\tau) = \frac{1}{27} (t+1) - \frac{\sqrt{3}}{12\pi} t$;
\item triples $(t,R_2,g_2)$ such that 
  $t^{-1} = \Phi'(R_2)$ and
  $g_2(t(R_2)) = R_2 - \frac{\Phi(R_2)}{\Phi'(R_2)}$ (as we substitute
  $t^{-1}=\Phi'(R_2)$ in (\ref{eq:R_4})).
\end{enumerate}
Once we fix a value of $t$, among all triples $(t,R,g)$ realising one
of the two mechanisms above, we have to identify the one with
smallest value of $|g|$.\footnote{We expect this be unique for almost
  all values of $t$, and that the special points in which we do not
  have unicity shall be interesting, and studied separately.}

For this identification, it is useful to observe that $\Phi'(R)$ is an
increasing function of $R$, on the portion of the real axis
complementary to the cut, and has values at the endpoints
$\Phi'(-\infty)=-1$ and $\Phi'(\tau)=+\infty$.

We will identify different behaviours, for $t$ being valued in
different ranges. Let us start with $t>0$, which is the easiest case.
One has $R_2\in (0,\tau) < R_1$, and, since
\begin{align}
  \frac{\partial R}{\partial g} = \frac{1}{1-t\Phi'(R)} > 0\, 
  \, ,\quad \text{for } t>0\, ,
\end{align}
we deduce $g_2 < g_1$. Furthermore, as $R\to 0$, we have 
$g_1\to -\infty$ and $g_2\to 0$, thus $g_1$ shall be
discarded. Therefore one has the critical values:
\begin{align}
  \label{eq:Rcgct>0_4}
  \Phi'(R_c)=t^{-1}\, ,\quad
  g_c = R_c - t \Phi(R_c)\, ,\quad \text{for } t>0\, .
\end{align}
Close to the critical point, we set 
\begin{align}
  \label{eq:Rc_gc}
  R=R_c(1-\delta)\, , \quad g=g_c(1-\epsilon) \, ,
\end{align}
with $\delta,\epsilon\to 0^+$. Expanding equation \eqref{eq:R_4} to second
order allows to derive the leading dependence of $\delta$ on $\epsilon$:
\begin{align}
  \delta &\sim C\sqrt{\epsilon}\, ,
&
  C&= \sqrt{\frac{2g_c}{R_c^2 \,\Phi''(R_c)}\frac{1}{t}}\, .
\end{align}
We can now immediately deduce, from \eqref{eq:ZlogR}, the critical
behavior of $Z$ as $\epsilon\to 0$
\begin{align}
  \label{eq:Zsing_4_t>0}
  Z_{\text{sing}} \sim 
  -\frac{N^2}{g_c^2}\frac{4}{15}C \, \epsilon^{5/2}  \, ,
  \quad \text{for } t>0\, .
\end{align}
In turns, this implies $Z_n\sim g_c^{-n} n^{-7/2}$, and $\gamma=-1/2$,
so that $c=0$.  This behaviour is that of the so-called 
\emph{pure gravity}, which indicates that the spanning forest model is
massive in this regime, analogously to what happens on the regular
square lattice, for which there are no critical points in the forest
model in the ferromagnetic phase.

We now turn our attention to the more interesting case of $t<0$.  As
already noted, the equation $t^{-1}=\Phi'(R)$ has no solution for
$t\in [-1,0)$.  As a result the only singularity must be attained at
$R=R_c$, the radius of convergence of $\Omega$:
\begin{align}
  \label{eq:Rcgct<0_4}
  R_c &= \frac{1}{27}\, ,
&
  g_c &= \frac{t+1}{27} - t \frac{\sqrt{3}}{12\pi}
%  \text{for } 
% &(t<0)
  \, .
\end{align}
Using the same notation as in \eqref{eq:Rc_gc}, we expand
around $R_c$ equation \eqref{eq:R_4} to have at leading order:
\begin{align}
  \label{eq:eps_delta_4}
  \epsilon 
&\sim D \,t\, \log(\delta) \delta\, ,
&
  D
&=
  \frac{\sqrt{3}}{2\pi}\frac{R_c}{g_c}\, .
\end{align}
Note that both $\delta$ and $\epsilon$ are on a neighbourhood of the
origin, and positive, the function $g_c(t)$ is positive for all $t<0$,
and $\log(\delta) \delta < 0$, thus this equation is consistent for
$t<0$. The inversion of this equation has been discussed at length in
\cite[Sec.~7]{BMCourt2015}.  If we denote the inverse as
$\delta=\Upsilon(\epsilon)$, $\Upsilon(\epsilon)\to 0$ as $\epsilon\to
0$ and further:
\begin{align}
  \epsilon \sim D\, t\, \log(\Upsilon(\epsilon))\Upsilon(\epsilon)\, .
\end{align}
Taking the logarithm, one has $
\log(\Upsilon(\epsilon))\sim \log(\epsilon)$, which finally implies
\begin{align}
  \label{eq:invzlogz}
  \delta = \Upsilon(\epsilon) \sim \frac{1}{D\, t} \frac{\epsilon}
  {\log(\epsilon)}\, .
\end{align}
Plugging this into \eqref{eq:ZlogR} gives
the following critical behavior:
\begin{align}
  Z_{\text{sing}} \sim 
  -\frac{N^2}{6 g_c^2 D}\frac{1}{t}\frac{\epsilon^3}{\log(\epsilon)}
  \, .
%  \quad \text{for } t<0\, .
\end{align}
In turns, equation \eqref{eq:Zn} implies $Z_n\sim
g_c^{-n}n^{-4}\log^{-2}(n)$.  Then, we find $\gamma=-1$ for all $-1\le
t<0$, which corresponds to a massless phase. According to the KPZ
formula, and neglecting the logarithms, this value of $\gamma$
corresponds to $c=-2$.

\begin{figure}[tb]
  \centering
  \includegraphics[scale=.5]{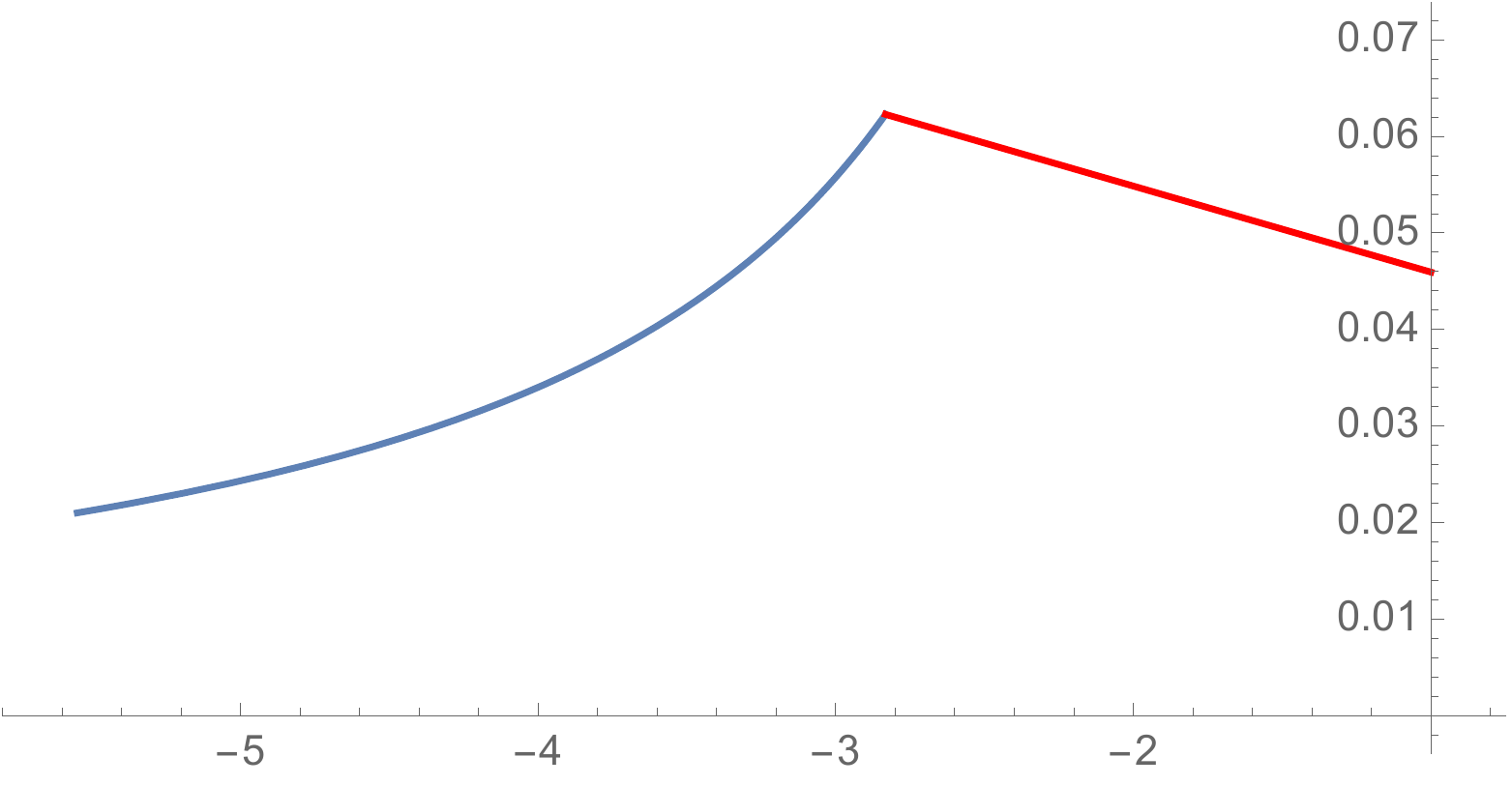}
  \caption{The crossing between $|g_1(t)|$ (red) and $|g_2(t)|$ (blue) 
  at $t=t_\ast\simeq -2.83240$. Only the minimum between the two
  curves is shown.}
  \label{fig:2}
\end{figure}

We now discuss the regime $t<-1$.  In this case positivity arguments
on $Z$ cannot be used, and in fact a branch of $g_2$, negative and of
smaller absolute value than $g_1$, emerges for $|t|$ sufficiently large.
We present in figure \ref{fig:2} the plot of $|g_1|$ and
$|g_2|$, whose expressions are easily obtained.\footnote{For $g_2(t)$,
  the curve is more conveniently found parametrically in terms of $R$.}
% We remark that $g_1>0$ and $g_2<0$ for $t<-1$. 
After a further interval, where the singularity at $g_1$ is
encountered first and the behavior is the same as in $-1\le t<0$, at a
value $t_\ast$ we have $|g_1(t_\ast)| = |g_2(t_\ast)|$, then the
solution $g_2$ wins.  The numerical value of $t_\ast$ is
\begin{align}
  t_\ast &= -2.83240\dots
\label{eq.valtast4}
\end{align}
which corresponds to $R = -0.18470\dots$ via the relation
$t^{-1}=\Phi'(R)$.\footnote{At higher precision, 
  $t_\ast = -2.832397443908381\ldots$ and $R=-0.18469834952811662\ldots$} 
The value of $R$ satisfies the relatively simple equation
\be
\Phi(R) - \big( R + \tfrac{1}{27} \big) \Phi'(R)
=
-\Phi(\tfrac{1}{27}) = 
\tfrac{1}{27} - \tfrac{\sqrt{3}}{12 \pi}
\ef,
\ee
but we are not aware of any more explicit expression (e.g., as a root
of a polynomial in $\mathbb{Q}[\sqrt{3}/\pi]$).

For $t\le t_\ast$ (note, $t_\ast$ included) $R$ has a square-root
singularity as in the case $t>0$. However, in this regime the
resulting values of $R$ and $g$ are both negative. While analytic
continuation for negative $g$ of a large-$N$ result in a matrix model
is a well-understood and fundamental mechanism, see the example of the
`basic' quartic potential $V(M) = \tfrac{1}{2}M^2+\tfrac{1}{4}M^4$
\cite{DiFrancesco1993}, a negative value of $R$ is hard to interpret
in light of its role in the Joukowski map.

In any case, the implication of this finding on the enumerative
results seem clear, and state that, for $t\le t_\ast$, the model exits
the massless phase and is characterized by pure-gravity exponents. A
very similar picture, and with a better analytical control, is
presented in the case of cubic graphs, discussed in Section
\ref{ssec.anacubi} below.

In the remainder of this section we will investigate the spectral
density, and, because of the issue mentioned above, we restrict to 
$t> t_\ast$. The derivative of the potential is
\begin{align}
  W'(z) = (1+t)z-\frac{2}{\sqrt{3}} \,t \sin\left( \tfrac{1}{3}
  \arcsin\left(\tfrac{3\sqrt{3}}{2} z \right)\right)\, ,
\end{align}
and upon rescaling variables as
\begin{align}
  \lambda = a x
  =
  2\sqrt{R} x
  \, ,
\end{align}
and denoting $\hat{\rho}(x)=\rho( a x)$, 
one gets from formula \eqref{eq:rhoW}:
\begin{align}
  \hat{\rho}(x) = \frac{\sqrt{1-x^2}}{2\pi^2}\frac{1}{g}\fint_{-1}^{1}\ud y
  \frac{W'(2\sqrt{R} y)}{\sqrt{1-y^2}}\frac{1}{y-x}\, ,\quad
  \text{supp}(\hat{\rho}) = [-1,1]\, .
\end{align}
Using the integral
\begin{align}
  \label{eq:97}
  \fint_{-1}^{1} 
  \frac{\ud y }{y-x}  \frac{\alpha + \beta y}
  {\sqrt{1-y^2}}
  =
  \beta \pi
  \, , \quad \text{for any } \alpha,\beta\, .
\end{align}
we get:
\begin{align}
  \hat{\rho}(x) = \frac{\sqrt{1-x^2}}{2\pi^2}\frac{1}{g}
  \left(
  2\pi(1+t)\sqrt{R}
  -
  \frac{2}{\sqrt{3}} \,t 
  \fint_{-1}^{1}\ud y
  \frac{
  \sin \big( \tfrac{1}{3}
  \arcsin \big(\sqrt{27 R} y \big) \big)
  }{\sqrt{1-y^2}}\frac{1}{y-x}
  \right)
  \, .
\end{align}
In particular the critical density for $t_\ast<t<0$, using
\eqref{eq:Rcgct<0_4}, reads as:
\begin{align}
  \hat{\rho}_c(x) = \frac{\sqrt{1-x^2}}{\sqrt{3}\pi^2}\frac{1}{g_c(t)}
  \left(
  \frac{\pi}{3}(1+t)
  -
  t 
  \fint_{-1}^{1}\ud y
  \frac{
  \sin\left( \tfrac{1}{3}
  \arcsin\left( y \right)\right)
  }{\sqrt{1-y^2}}\frac{1}{y-x}
  \right)
  \, . 
\label{eq.655564}
\end{align}
We were not able to compute analytically the remaining
integral. However, due to the simple form of (\ref{eq.655564}), is is
easily verified that $\hat{\rho}_c(x)$ is positive in its interval of
definition for all values $t_\ast<t<0$.\footnote{Because the integral attains
  its minimum value for $x=0$, which is $1.20057\ldots\,$, this being
  larger than $\pi/3 = 1.04719\ldots$}

\begin{figure}[tb!]
\[
\begin{array}{ccc}
%    \begin{subfigure}[b]{0.3\textwidth}
%        \includegraphics[width=.3\textwidth]{rhoc_quartic_t-10.jpg}&
        \includegraphics[width=.4\textwidth]{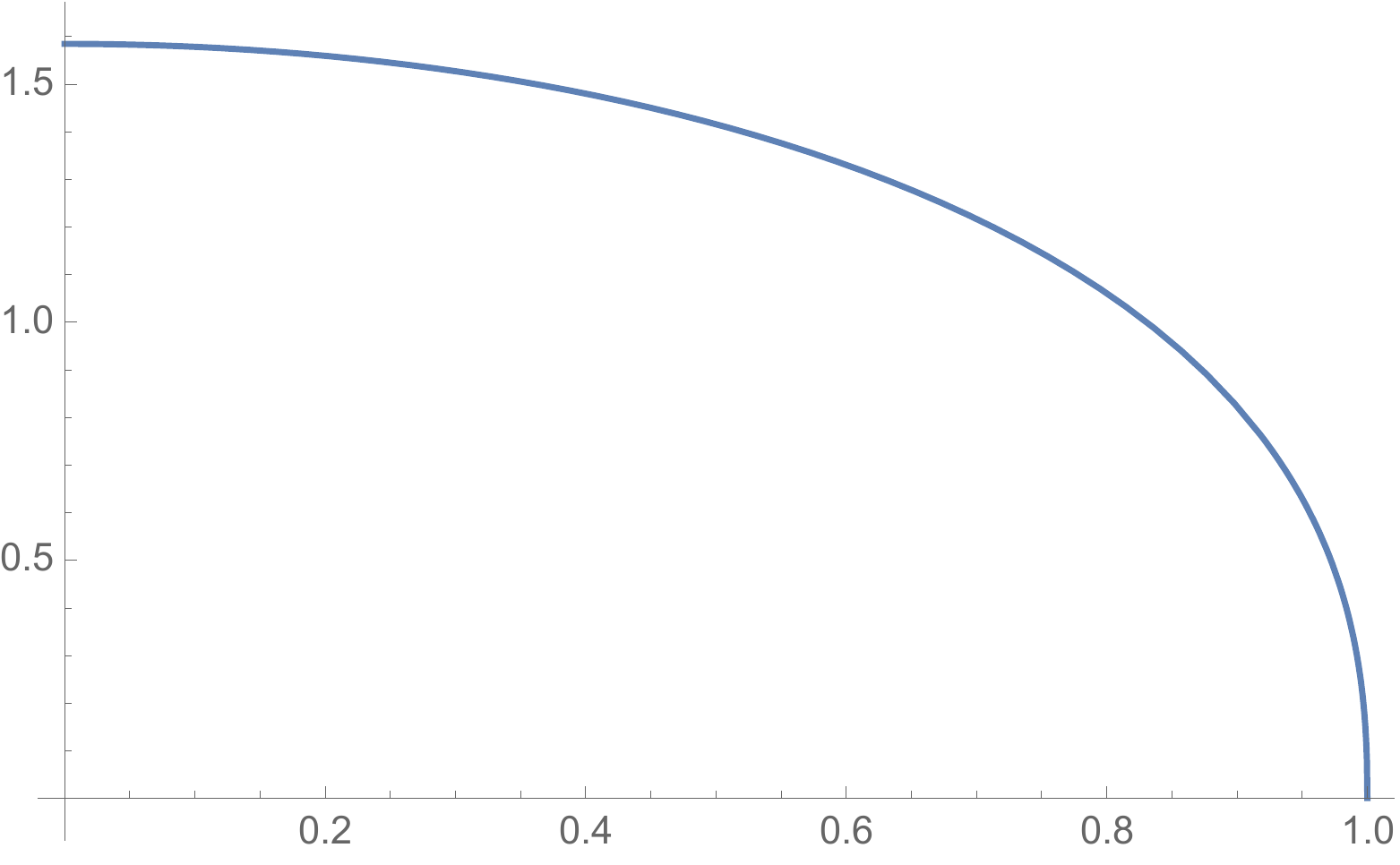}&&
        \includegraphics[width=.4\textwidth]{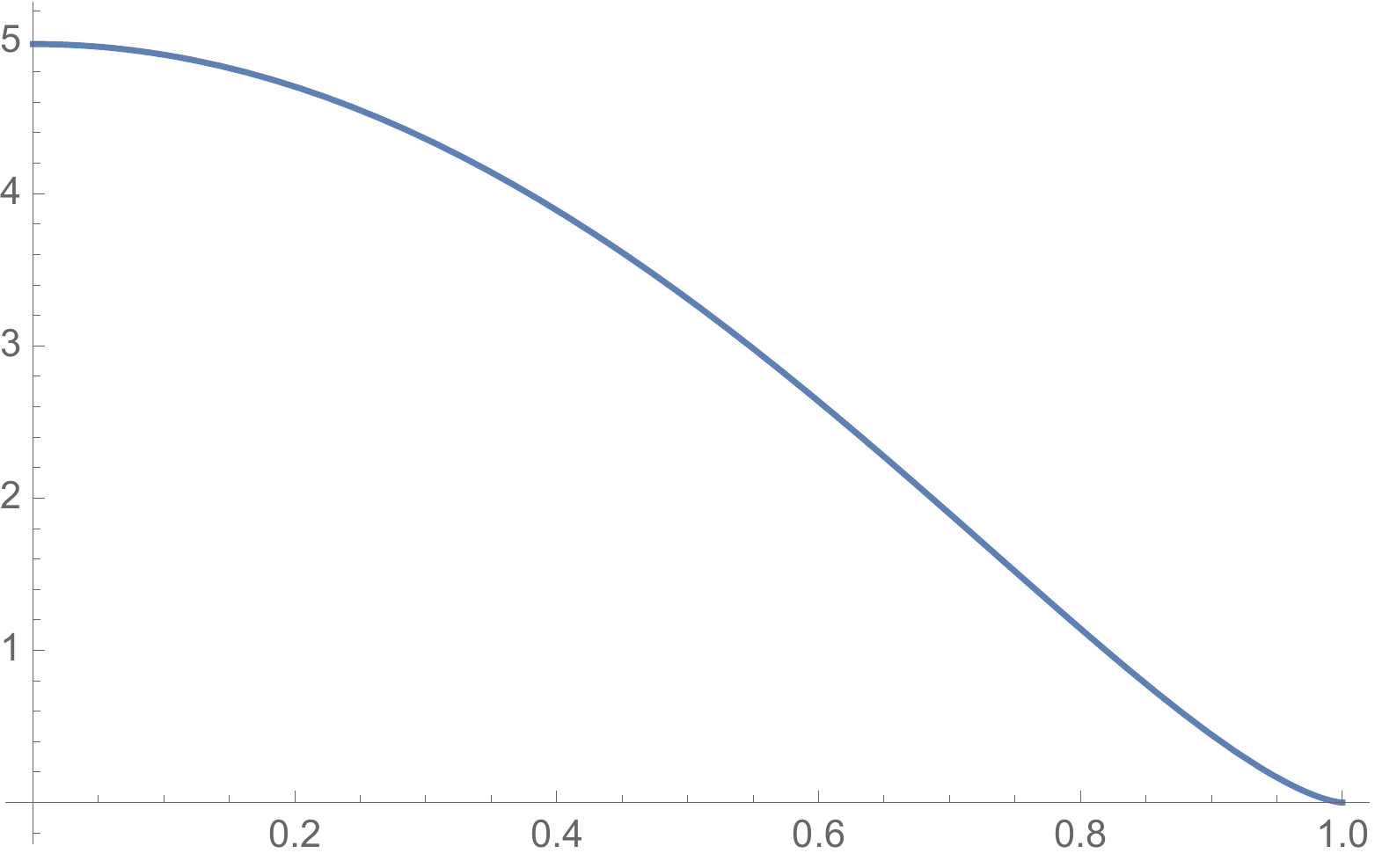}
\\
%        { t=-10 }&
        { t=-0.5\  (R=\frac{1}{27})}&&
        { t\simeq 3.23\  (R=\frac{1}{27}-0.01) }
\end{array}
\]
    \caption{Critical spectral density for quartic graphs, rescaled to
      the interval $[-1,1]$ (of which, in light of symmetry, we show
      only the right half).}
  \label{fig:density_4}
\end{figure}

We present in figure \ref{fig:density_4} the critical density for
two different values of $t$. We note clearly the different slope at
the spectral edge for $t_\ast<t<0$ and $t>0$ which entails the different
critical exponents. 

%-------------------------------------------------------
\subsection{Generalization to arbitrary even degree}

The quartic case shows the main features of all the cases of
% is instrumental for all 
even degree. For arbitrary 
$h = 2 p$, with $p\in \mathbb{N}$,
%% \begin{align}
%%   h = 2 p\, , \quad p\in \mathbb{N}\, ,
%% \end{align}
$S=0$ because of symmetry, and \eqref{eq:RS} reduce to 
\begin{align}
\label{eq.47682654}
  v &= R - t \,\Phi(R)\, ,
&
  \Phi(R) &= \sum_{n\ge 1}\frac{((2p+1)n)!}{(pn+1)!(pn)!}\frac{R^{pn+1}}{n!}\, .
\end{align}
Denoted the radius of convergence of $\Phi$ as
\begin{align}
  \label{eq:tau}
  \tau_h &= \frac{r_{h}^2}{4} = \frac{1}{4}\frac{h^2}{(h+1)^{2(h+1)/h}}
  \, ,
%% \quad
%%   \tau_h^h = 
%%   \frac{1}{64},\frac{1}{729},\frac{729}{4194304},\frac{256}{9765625},\dots
\end{align}
(the first few values are
$  \tau_h^h = 
  \frac{1}{64},\frac{1}{729},\frac{729}{4194304},\frac{256}{9765625},\ldots$),
the series $\Phi(R)$ is the following hypergeometric series 
\begin{align}
  \Phi(R) + R = R \; {}_{2p}F_{2p-1}
  \left(
  \tfrac{1}{2p+1},\tfrac{2}{2p+1},\dots,\tfrac{2p}{2p+1};
  \tfrac{1}{p},\tfrac{2}{p},\tfrac{2}{p},
  \tfrac{3}{p},\tfrac{3}{p},
  \dots,\tfrac{p}{p},\tfrac{p+1}{p};
%   \left(\tfrac{R}{\tau_{2p}}\right)^{p}
  (R/\tau_{2p})^{p}
  \right)\, .
\end{align}
Apparently, this is an overwhelmingly complicated expression.
Nonetheless, as we annotate later on, certain known properties of
hypergeometric series are instrumental to establish the main features
of the model.

We proceed as in the quartic case by computing the functions $v_1(t)$
and $v_2(t)$, for the singular triples $(t,R,v)$ satisfying one of the
two singularity mechanisms, i.e.\ corresponding to the singularity of
$\Phi$ at $R=\tau_{2p}$, and to $t^{-1} = \Phi'(R)$, respectively.
\begin{align}
\label{eq.287364573}
  v_1(t) &= g_1^{1/p}(t) = (1+t) \tau_{2p}
  - \tau_{2p} C_p\, t \, ,\\
  v_2(t) &= g_2^{1/p}(t) = 
  R - \frac{\Phi(R)}{\Phi'(R)}\Big|_{t^{-1}=\Phi'(R)} \, ,
\end{align}
where $C_p$ is defined as
\begin{align}
  C_p = \frac{\Phi(\tau_{2p})}{\tau_{2p}}+1\, .
\end{align}
Even though a closed form for $C_p$ is not easily available, it is
remarkable that (\ref{eq.287364573}) is an affine function for all
$p$'s.

The results that follow have been established numerically for the
cases $p=1,2,3,4$, and conjectured to hold generally. One can extend
the numerical analysis to bigger $p$ with no difficulties, and it
should also be possible to prove these results (although we do not do
this here). First of all, from the definition (\ref{eq.47682654}) it
is easily evinced that $\Phi'(R)$ is in fact a function of $R^p$, and
in particular it has a useful symmetry under $\mathbb{Z}_p$ rotations
of its argument
\begin{align}
  \Phi'( z ) = \Phi'( \omega^n z )\, ,
  \quad \omega = e^{2\pi i /p}\, , \quad n=0,1,\dots,p \,.
\end{align}
Given this symmetry, the analysis can be restricted to the cone
\begin{align}
  \left\{R = \rho e^{i\theta} \,\big|\, 
  \rho \ge 0\, ,\theta\in [0,\tfrac{\pi}{p}] 
  \right\} \, .
\end{align}
We determined numerically the values of $R$ which give
$\Im(\Phi'(R)) = 0$, so that the equation $t^{-1}=\Phi'(R)$ has a
solution for real $t$. These lie on the boundary of the cone, 
in particular one has:
\begin{align} 
  \label{eq:110}
  \Phi' :   (0,\tau_{2p}) \cup   \left\{\rho e^{i\pi/p} \, , 
  \rho \in (0,\infty)  \right\}  
  \mapsto 
  (0,+\infty)
  \cup
  (-1,0)
  \, ,
\end{align}
and the region $(-\infty,-1)$ has no preimage under $\Phi'$.
(The point $t=-1$ corresponds to $R=\infty \times e^{i\pi/p}$.)
Then, just as already observed in the $p=1$ case, the equation
$\Phi'(R)=t^{-1}$ has no solution  for $t\in [-1,0]$.

% t>0
For what concerns the regime $t>0$, it is then governed by exactly the
same formulas (\ref{eq:Rcgct>0_4}--\ref{eq:Zsing_4_t>0}), with $\Phi$
replaced by the appropriate series (\ref{eq.47682654}) and $g$
replaced by $v = g^{1/p}$, so the generating function for $t>0$ has
the behaviour of pure gravity for any even degree.  This holds
for the same reasonings as in the quartic case, and doesn't require
any subtle control on the relevant expressions.

% -1<=t<0
For $-1\le t<0$, the singularity occurs at the critical radius
of $\Phi$, since $t^{-1}=\Phi'(R)$ has no solution.
So $R_c = \tau_{2p}$ and the critical curve is:
\begin{align}
  v_c(t) = g_c^{1/p}(t) = (1+t) \tau_{2p}
 - \tau_{2p} C_p\, t \, .
\end{align}
The expression for $\tau_{2p}$ generalises (\ref{eq:r_2^2/4}), and the
one above for $v_c(t)=(g_c(t))^{1/p}$ generalises
(\ref{eq:Rcgct<0_4}). In the case $p=1$, we have 
$v_c(t) \geq 0$ for all $t<0$ as a result of the inequality 
$C_1 - 1 = \frac{9 \sqrt{3}}{4 \pi} \geq 0$. The analogous statement
here reads $C_p -1 = \frac{\Phi(\tau_{2p})}{\tau_{2p}} \geq 0$, which
holds at sight from the positivity of the series coefficients.
More remarkably, since
\begin{align}
  \psi_{2p-1} = 
  \tfrac{1}{p}+\tfrac{2}{p}+\tfrac{2}{p}+
  \tfrac{3}{p}+\tfrac{3}{p}+
  \dots+\tfrac{p}{p}+\tfrac{p+1}{p}
  -
  \left(\tfrac{1}{2p+1}+\tfrac{2}{2p+1} + \dots + \tfrac{2p}{2p+1}\right)
  = 1\, ,
\end{align}
general results on hypergeometric series \footnote{See
  \url{http://functions.wolfram.com/HypergeometricFunctions/HypergeometricPFQ/06/01/04/02/0002/}.}
imply that for any $p$ the leading singular term in the expansion of
$\Phi$ around $\tau_p$ is $\log(\delta)\delta$, where we defined
$\delta$ as before, $R = \tau_{2p}(1-\delta)$.  For example, in the case
$p=2$ (i.e., graphs of degree six), equation~\eqref{eq:eps_delta_4}
is modified into
\begin{align}
\epsilon &\sim D\, t\, \log(\delta) \delta\, ,
&
  D
&=
  \frac{\sqrt{5}}{4\pi}\frac{R_c}{\sqrt{g_c}}\, . 
\end{align}

We now move to the region $t<-1$. According to formula \eqref{eq:110},
the preimage of this region under $\Phi'$ is the ray
$\rho e^{i\pi/p} \, , \rho \in (0,\infty)$. The phenomenology is
exactly the same as in the quartic case. By comparing
$|v_1|$ and $|v_2|$ one establishes the existence a critical value
$t_\ast$ such that the singularity given by $t^{-1} = \Phi'(R)$ takes
over that due the lack of analyticity at $\tau_{2p}$, and
implies the pure gravity scaling exponent.
The values of $t_\ast$ determined are reported in the following
table together with
$R_\ast$, $t_\ast^{-1} = \phi'(R_\ast)$:
\begin{center}
  \begin{tabular}[h]{cccc}
    $h = 2p$ & $t_\ast$ & $R_\ast e^{\pi i (n+1/p)}$ & $|g_{\ast}|$
% \quad (0 \leq n < p)$
\\
    \hline
\rule{0pt}{13pt}%
    2 & $-2.83240 \dots$ & $0.184609\dots$ & $0.0622653\dots$   \\
    4 & $-3.44339 \dots$ & $0.189499\dots$ & $0.00920182\dots$  \\
    6 & $-3.95672 \dots$ & $0.194037\dots$ & $0.00163489\dots$  \\
    8 & $-4.40930 \dots$ & $0.197990\dots$ & $0.000317077\dots$ \\
  \end{tabular}
\end{center}
The multiplicity of choices for the argument of 
$R_\ast$ ($0 \leq n < p$ in the notation above) reflects the
aforementioned $\mathbb{Z}_p$ symmetry of $\Phi'(R)$. We note that the
absolute values of both $t_\ast$ and $R_\ast$ slightly increase with
$p$. Even worse than in the case $p=1$, for $t\le t_\ast$ the
interpretation of $R$ in terms of the support of the (supposedly
one-cut) spectral density seems to breaks down. 

Finally, the above findings imply that all values of $p$ show the same
universal critical behavior, while (as expected) the critical value
$t_\ast$ is not universal.

%-------------------------------------------------------
\subsection{The case of cubic graphs}
\label{ssec.anacubi}

In the case of cubic graphs, the potential has a more explicit
expression, and is thus of a more tractable form that in case of
higher degree. In particular, we have for the derivative
\begin{align}
  W'(z) = (1+t)z + \frac{t}{2}(-1+\sqrt{1-4 z})\, .
\end{align}
On the other side, equations \eqref{eq:RS} are now more complicated
than in the cases analysed in the previous sections, since $S\neq
0$. For these reasons, we analyse this case starting from the integral
constraints \eqref{eq:intVp}.  This approach is different from the
analysis of the cubic case in \cite{BMCourt2015}, while it essentially
coincides with the one of~\cite{KoStau}.

Instead of dealing with $R$ and $S$, we work here directly with $a_1$
and $a_2$ (recall that the two pairs of parameters are related by
\eqref{eq:Joukowski}).  In order to proceed, we shall need the
following integrals (valid for $a_2 < 1/4$, as is the case here)
\begin{align}
  &\int_{a_1}^{a_2} \ud \mu \frac{A + B\mu}{\sqrt{(\mu-a_1)(a_2-\mu)}}
  =
  \pi \left( A + B \frac{a_1+a_2}{2} \right)\, ,\\
  \label{eq:int_diff_1}
  &\int_{a_1}^{a_2} \ud \mu \sqrt{\frac{ 1-4 \mu }
  {(\mu-a_1)(a_2-\mu)}}
  =
  2 \sqrt{1-4 a_1} E\left(
    4\frac{a_2-a_1}{1-4 a_1}\right)
  \, ,\\
  &\int_{a_1}^{a_2} \ud \mu \sqrt{\frac{\mu-a_1}{a_2-\mu}}
    (A+B \mu)
  =
  \pi  \frac{a_2-a_1}{2} \left( A + B
  \frac{a_1 + 3 a_2}{4} \right)
  \, ,\\
  \label{eq:int_diff_2}
\begin{split}
  &\int_{a_1}^{a_2} \ud \mu 
  \sqrt{\frac{(\mu-a_1)(1-4 \mu)}{a_2-\mu}}
  =\\
  &\quad
  \tfrac{4}{3} \sqrt{\tfrac{1}{4}-a_1} 
  \left(\left(\tfrac{1}{4}-a_2\right)
    K\left(4\frac{a_2-a_1}{1-4 a_1}\right)+\left(2
      a_2 -a_1 -\tfrac{1}{4}\right) 
    E\left(4\frac{a_2-a_1}{1-4 a_1}\right)\right)
  \, ,
\end{split}
\end{align}
where $K(m),E(m)$ are the complete elliptic integrals of the first and
second kind with parameter (square of elliptic modulus) $m$.  The less
trivial equations (\ref{eq:int_diff_1}) and (\ref{eq:int_diff_2}) can
be found in \cite[eqs.~(3.141.2) and (3.141.26)]{gradshtein2014}.
Equations~\eqref{eq:intVp} then read as
\begin{align}
  \label{eq:121}
  (1+t) \pi \frac{a_1+a_2}{2} -\frac{t}{2} \pi
  +t \sqrt{1-4 a_1}  E\left(
    4\frac{a_2-a_1}{1-4 a_1}\right)  &= 0\, ,\\
  \label{eq:122}
  \begin{split}
   -t\pi  \frac{a_2-a_1}{4} 
    + (1+t) \pi
    \frac{(a_2-a_1) (a_1 + 3 a_2)}{8}
%    +   \tfrac{2t}{3} \sqrt{\tfrac{1}{4}-a_1} 
\qquad  \qquad &
\\
%\times 
   + \tfrac{t}{12} \sqrt{1-4a_1} 
\left(\left(1-4a_2\right)
      K\left(4\tfrac{a_2-a_1}{1-4 a_1}\right)+\left(
        8 a_2 - 4 a_1 - 1\right) 
      E\left(4\tfrac{a_2-a_1}{1-4 a_1}\right)\right)
    &=
    2\pi g^2
  \, .
  \end{split}
\end{align}
The spectral density can be derived from equation~\eqref{eq:rhoW}.  At
this aim it is convenient to perform an affine change of coordinates
as to set the support of the density on $[-1,1]$: let us change
coordinates as\footnote{These parameters are quite similar to $R$ and
  $S$ in (\ref{eq:Joukowski}), it is just $A=\sqrt{R}$ and $B=S$.}
\begin{align}
 \lambda &:= f(x) = A x + B
  \, ,
&
  A &= \frac{a_2-a_1}{2}\, ,
&
  B &= \frac{a_1+a_2}{2}\, ,
%  \text{supp}(\hat{\rho})=[-1,1]\, ,
\end{align}
and let us call $\hat{\rho}(x) \equiv \rho(f(x))$ the
resulting distribution. Then we have
\begin{align}
  \hat{\rho}(x) 
  &= 
  \frac{\sqrt{1-x^2}}{2\pi^2 g^2}
  \fint_{-1}^{1} 
  \frac{\ud y }{y-x}  \frac{W'(f(y))}
  {\sqrt{1-y^2}}\\
  &=
  \frac{\sqrt{1-x^2}}{2\pi^2 g^2}
  \left(
    (1+t) A \pi
    +
    t
    \sqrt{A}
    \fint_{-1}^{1} 
    \frac{\ud y }{y-x}  
    \sqrt{\frac{\frac{1-4B}{4A}- y}{1-y^2}}
  \right)
  \, .  
\label{eq.rhocubigene}
\end{align}
where we used the integral~\eqref{eq:97}. The parameters $A$ and $B$ as
a function of $t$ are to be determined in what follows.

The analysis of \cite{BMCourt2015} shows that, in the regime $t>0$,
cubic and quartic graphs behave similarly, and in both cases this
corresponds to a massive phase on the flat lattice. This could be
rederived here, but the procedure and conclusions are similar to the
ones already depicted for the case of even degree, and are
comparatively less interesting than the behaviour at negative values
of the fugacity.

Thus, we will focus our attention directly on the regime $t<0$,
where, yet again, at least in an interval containing $t \in [-1,0]$,
and possibly extending below $t=-1$, criticality is expected to appear
as the support of the spectral density touches the boundary of the
analyticity region of $W$. The function $W$ can be analytically
continued to the complex plane minus the cut on the real axis $z \geq
1/4$, so in this interval the critical value is $a_{2,c} = 1/4$, while
$a_{2,c} < 1/4$ if the criticality mechanism is different. For
convenience, in the following we will trade $a_1$ for an equivalent
variable defined as
\begin{align}
    s = \sqrt{1-4 a_1}\, .
\end{align}
As $a_{2} \leq 1/4$, and 
$a_1 < a_2$, we have that $s>0$.

At a critical point with $a_2=1/4$, the above equations reduce to algebraic ones:
\begin{align}
  s^2 (t+1)+2 (t-1)&=\frac{8}{\pi} s t\\
  \label{eq:eq2_crit}
  s^4 (t+1)+4 s^2 (t-1)-\frac{32}{3\pi} 
  s^3 t &= -256 g^2\, .
\end{align}
Solving the first for $s$, with the condition $s>0$, gives:
\begin{align}
  s_c&=
  \frac{1}{\pi}
  \frac{4 t + \sqrt{2} \sqrt{\pi ^2-\left(\pi ^2-8\right) t^2}}
  {1+t}\, , 
\end{align}
valid within the interval $t_\ast\le t<0$, with $t_\ast$ defined as
\begin{align}
  \label{eq:t_ast}
  t_\ast \equiv -\frac{\pi}{\sqrt{\pi^2-8}} = -2.2976\dots
\end{align}
Therefore, plugging $s_c$ in \eqref{eq:eq2_crit}, we have that,
provided that the leading criticality mechanism comes from the
singularity of $W$ at $1/4$, the critical values of the parameters as
functions of $t$ are
\begin{align}
  \label{eq:126}
  a_{2,c} &= \frac{1}{4}\, ,
\\
  a_{1,c} &= \frac{1}{4}-
    \frac{1}{4\pi^2}
    \frac{\big(4 t + \sqrt{2 \pi ^2-2 \left(\pi ^2-8\right) t^2}\big)^2}
    {(1+t)^2}\, ,\\
  \label{eq:127}
  g_c^2
  &=
  \frac{
(4 t+\sqrt{2} \sqrt{8 t^2+\pi ^2(1-t^2)})^3
(-4 t+3\sqrt{2} \sqrt{8 t^2+\pi ^2(1-t^2)}) 
% \left(\sqrt{2 \pi ^2-2 \left(\pi ^2-8\right) t^2}+4 t\right)^3 \left(3
%    \sqrt{2 \pi ^2-2 \left(\pi ^2-8\right) t^2}-4 t\right)
}{768 \pi ^4
   (t+1)^3}\\
 \label{eq.75467565}
  &=
  \frac{512 t^4+96 \pi ^2 \left(1-t^2\right) t^2+16 \sqrt{2} \left(8 t^2+\pi ^2
      \left(1-t^2\right)\right)^{3/2} t+3 \pi ^4
    \left(1-t^2\right)^2}{192 \pi ^4 (t+1)^3}\, .
\end{align}
We have also determined that this criticality pattern may occur at
most on the interval $t_\ast\le t<0$ (and we know in advance, from
\cite{BMCourt2015}, that it occurs at least on the interval $-1 \le
t<0$). Later on, we determine that the interval in which this
criticality pattern is dominant is $t_\ast < t<0$.

Note that the formula (\ref{eq.75467565}) above coincides with the
expression for the critical radius of the series first obtained in
\cite[eq.~(77)]{BMCourt2015}, of which it thus provides an alternate
derivation.
At $t=t_{\ast}$, the expressions for $g^2_c$ and $a_{1,c}$ simplify
considerably:
\begin{align}
g^2_c(t_\ast) 
&= \frac{1}{3 (\pi^2-8)^2 (\frac{\pi}{\sqrt{\pi^2-8}}-1)^3}
\label{eq.gctc}
\\
a_{1,c}(t_\ast) &= 
\frac{1}{4}-\frac{4}{(\pi -\sqrt{\pi ^2-8})^2}\, .
\end{align}
To compute the critical exponents in the range $t_\ast\le t<0$, we
investigate the behaviour close to the critical point.  We set
$g^2=g_c^2(1-\epsilon)$, $a_{2}=a_{2,c}(1-\delta)$, $s=s_{c}(1-\eta)$
and consider the limits
$\epsilon$, $\delta$ and $\eta\to 0^+$. The goal is to compute the singularity
of the partition function from \eqref{eq:ZlogR}, which is
\begin{align}
  Z'' \sim 2\frac{N^2}{g_c^4} 
  \log(\tfrac{1}{4}s_c^2 + \tfrac{1}{4}
  \delta - \tfrac{1}{2}s_c^2\, \eta)\, ,
\end{align}
and requires computing the dependence of $\eta$ and $\delta$ on
$\epsilon$.  If we write equation~\eqref{eq:121} in terms of $\eta$
and $\delta$, and expand around $\delta= 0$, we have
\begin{align}
  \label{eq:135}
  \left(2 s_c (-4 t + \pi s_c (1 + t)) \eta 
  - \pi s_c^2 (1 + t) \eta^2\right)s_c(1-\eta)
  =
  2 t \delta \log(\delta)
  + 
  o(\delta \log(\delta))
  \, .
\end{align}
Note that $o(\delta \log(\delta))$ contains also terms of the type
$\eta \delta \log(\delta)$, and that
the coefficient of $\eta$ on the l.h.s.~is positive
for $t_\ast<t<0$, while it vanishes at $t_\ast$:
\begin{align}
  (-4 t + \pi s_c (1 + t)) |_{t=t_\ast} = 0\, .
\end{align}
We start by analysing the case $t > t_\ast$.  In this case we can
neglect the order $\eta^2$ on the l.h.s., and the subleading terms on
the r.h.s., and the equation above gives:
\begin{align}
  \label{eq:129}
  \eta (1+ \mathcal{O}(\eta))
  =
  \frac{t}{s_c^2 (-4 t + \pi s_c (1 + t)) } \delta\log(\delta)
%%   \, ,
%% \quad
%%   \text{for } t_\ast<t<0
  \, .
\end{align}
We can now look at equation~\eqref{eq:122}, which, after expansion, reads
\begin{align}
  \label{eq:139}
  \begin{split}
    128 g^2_c \pi \epsilon + o(\epsilon)
    =
    \eta^2 
    s_c^2 
    \left(\pi  \left(3 s_c^2 (t+1)+2 (t-1)\right)-16 s_c t\right) +o(\eta^3) \\
    +4 s_c\, t \,\delta \log(\delta) 
    +
    o(\delta\log(\delta))\, ,
  \end{split}
\end{align}
Note that the coefficient of order $\eta$ is identically zero for any $t$.
Using \eqref{eq:129}, the terms of order $\eta^2$ and  higher 
can again be neglected, and $\delta$ can be expressed in terms of $\epsilon$
as in \eqref{eq:invzlogz}.  In conclusion, for $t_\ast<t<0$ we have
\begin{align}
  \eta 
&\sim 
% \frac{32 g_c^2\pi}{s_c^3(-4 t + \pi s_c (1 + t))}
-2 \alpha \,
  \epsilon \, ,
&
\alpha &= -\frac{64 g_c^2\pi}{s_c^3(-4 t + \pi s_c (1 + t))}\, ,
\\
  \delta 
&\sim 
s_c \beta \,
% \frac{32 g_c^2}{s_c t}
\frac{\epsilon}{\log(\epsilon)} \, ,
&
  \beta & = \frac{32 g_c^2}{s_c^2 t}\, .
\end{align}
The $\epsilon$--dependent part of the partition function is, with
$\alpha$ and $\beta$ as above,
\begin{align}
  Z'' 
  \sim 
  \frac{2 N^2}{g_c^4} 
  \left(
    \alpha \epsilon
    +
    \beta \frac{\epsilon}{\log(\epsilon)}
  \right)
  \, ,\quad
  \text{for } t_\ast<t<0\, ,
\end{align}
This coincides with the singular behaviour found in
\cite[Eq.~78]{BMCourt2015}, where $Z''$ is there called $F'$.
%  and obtained there using different methods. 
The term proportional to $\alpha$ is non--singular and does not affect
the large-$n$ behavior of the coefficients $Z_n$, which thus have the same
form as in the quartic case: $Z_n\sim g_c^{-n}n^{-4}\log^{-2}(n)$.

We now turn our attention to the case $t=t_\ast$ (which was not
discussed in \cite{BMCourt2015}). Then, as already remarked, the
coefficient of $\eta$ in \eqref{eq:135} is zero.  On the l.h.s.~of
\eqref{eq:135} the leading term is of order $\eta^2$, while the
r.h.s.~is unchanged, since terms of the form $\eta \delta
\log(\delta)$ are still suppressed compared to $\delta
\log(\delta)$. Therefore one obtains:
\begin{align}
  \eta = \frac{\pi-\sqrt{\pi^2-8}}{4\sqrt{2}}
  \sqrt{-\delta\log(\delta)}\, .
\end{align}
Now we have to take into account the $\eta^2$ term in
Eq.~\eqref{eq:139}, which is of the same order as the summands
involving $\delta\log(\delta)$.  However, the coefficient multiplying
$\eta^2$ is exactly zero at $t=t_\ast$,
\begin{align}
  \left(\pi  \left(3 s_c^2 (t+1)+2 (t-1)\right)-16 s_c t\right)|_{t=t_\ast}
  =
  0\, ,
\end{align}
the first non-vanishing term in $\eta$ is of order $\eta^3\sim
(\delta\log(\delta))^{3/2}$, and therefore negligible.
Thus,  for $t=t_\ast$ we find
\begin{align}
  \eta &\sim 
  \frac{1}{2\sqrt{3}} \epsilon^{1/2}
  \, ,
&
  \delta &\sim -\frac{8}{3\pi (\pi-\sqrt{\pi^2-8})^2}
  \frac{\epsilon}
  {\log(\epsilon)} \, ,
\end{align}
and the partition function is:
\begin{align}
    Z_{\text{sing}}'' 
  \sim 
  -
  \frac{2 N^2}{g_c^4} 
  \left(
    \frac{1}{\sqrt{3}} \epsilon^{1/2}
    +
    \frac{1}{6\pi} \frac{\epsilon}{\log(\epsilon)}
  \right)
  \, .
% \quad \text{for } t=t_\ast\, .
\end{align}
The leading singular behavior is thus $\epsilon^{1/2}$, which
corresponds to pure gravity, even though it comes along with peculiar
subleading corrections, with the exponents (and logarithms) of the
Berker--Kadanoff phase.

Thus, we have determined that the massless phase, which is the quantum
gravity counterpart of the Berker--Kadanoff phase and which is
characterized by $\gamma=-1$, $\gamma'=-2$ leading singularity, occurs
for cubic graphs in the region $t_\ast < t < 0$, while at the point
$t=t_\ast$ the spanning forest model has a critical point
characterized by the pure gravity critical exponent.

%%%%%%%%%%%%%%
%density
We shall now discuss the region $t<t_\ast$. Before doing this, let us
come back to the expression (\ref{eq.rhocubigene}) for the spectral
density. Let us introduce a shortcut $d=\frac{1-4B}{4A} \geq 1$ for
this recurring combination.  The integration in principal value on the
variable $y$ can be performed.  When $d>1$, the remaining integral is
\cite[3.131.3,3.137.3]{gradshtein2014}:
\begin{align}
  \begin{split}
    \fint_{-1}^{1} 
    &\frac{\ud y }{y-x}  
    \sqrt{\frac{d - y}{1-y^2}}
    =-
  \frac{2 }{\sqrt{d+1}}K\left(\frac{2}{d+1}\right)\\
    &-\frac{2}{\sqrt{d+1}}\frac{ d-x}{x+1}
 \left(\Pi \left(\frac{2}{x+1}\Big|\frac{2}{d+1}\right)
   -\frac{\pi}{2}
 \sqrt{\frac{(x+1)(d+1)}{(x-1)(d-x)}}
%   -\frac{\pi }{2 \sqrt{1-\frac{2}{x+1}}
%   \sqrt{1-\frac{x+1}{d+1}}}
\right)
\, ,
  \end{split}
\label{eq.resudens3}
\end{align}
where 
% $d\ge 1$ and 
$\Pi(n|k^2)$ is the complete elliptic integral of
third kind with elliptic characteristic $n$ and modulus $k$.  When
$a_2=a_{2,c}=1/4$, corresponding to $d=\frac{1-4B}{4A}=1$, the
integral simplifies considerably, and just gives
% it simplies to:
\begin{align}
  \fint_{-1}^{1} 
  \frac{\ud y }{y-x}  
  \frac{1}{\sqrt{1+y}}
  =
  \frac{2}{\sqrt{x+1}}
  \log \left(
  \frac{\sqrt{2} - \sqrt{x+1}}{\sqrt{1-x}}\right)\, .
\end{align}
Thus, for $t_\ast \le t < 0$ the critical density is:
\begin{align}
  \begin{split}
      \hat{\rho}_c(x) &= 
  \frac{\sqrt{1-x^2}}{2\pi^2 g^2_c}
  \left(
    (1+t) A_c \pi
    +
   \frac{2\,t \sqrt{A_c}}{\sqrt{x+1}}
  \log \left(
  \frac{\sqrt{2} - \sqrt{x+1}}{\sqrt{1-x}}\right)
  \right)\, ,\\
  A_c &= \tfrac{1}{2}(\tfrac{1}{4}-a_{1,c})
  \,.
  \end{split}
\end{align}
where $a_{1,c}$ and $g_c$ are as in equations~(\ref{eq:126}) and
(\ref{eq:127}). It is easily verified that the density is positive,
and that at the edges it behaves as
\begin{align}
  \hat{\rho}_c(x) &= 
  \frac{t \sqrt{A_c}}{2\pi^2 g^2_c}
  \sqrt{1-x}
 \;\big( \log(1-x) + \mathcal{O}(1) \big)
%  +  o(\sqrt{1-x}\log(1-x)) 
\\
  &=
    \frac{\sqrt{2} \pi  (t+1) A_c-2 t \sqrt{A_c}}{2 \pi ^2 g_c^2}
  \sqrt{1+x}
 \; \big( 1 + \mathcal{O}(1+x) \big)
%  o(\sqrt{x+1})\, .
\end{align}
At the right edge there is the logarithmic singularity responsible for
the peculiar critical behaviour of the model,
% logarithmic behavior of the partition function, and 
while at the left edge the behavior is instead the `ordinary'
non-critical one, i.e.\ a square-root singularity. We note that $g_c$
and $A_c$ are both strictly positive in the range $t\in [t_\ast,0]$,
while the coefficient of $\sqrt{1+x}$ in the expansion around $x=-1$
is positive only for $t\in (t_\ast,0]$ and is zero exactly at
$t=t_\ast$. Therefore, at $t=t_\ast$ we see the occurrence of the
phase transition discussed above, and the behavior at the left
spectral edge becomes $\sim (x+1)^{3/2}$.  This is similar in form to
what happens in the simplest criticality mechanism, leading to the
pure-gravity exponent in one-matrix models of random matrix theory
\cite{DiFrancesco1993}, and it is consistent with the pertinent
singularity of the partition function discussed above. In particular,
it proves in retrospective that in the range $t\in [t_\ast,0]$ we had
identified the leading criticality mechanism.

We will now identify the critical behaviour of pure gravity in all the
regime $t\leq t_\ast$.  At this aim we show that, in this regime,
there exist critical values $a_1(t)$, $a_2(t)$ such that the spectral
density has a $(x+1)^{3/2}$ singular behaviour at $x=-1$.  We start by
expanding the density (\ref{eq.resudens3}) around $x=-1$, using
well-known properties of elliptic functions:
\begin{align}
  \hat{\rho}(x) 
  &=
\frac{t \sqrt{A} \sqrt{1+x}}{\sqrt{2}\pi^2 g^2}
\,
\big( 1-\tfrac{1}{4}(x+1) + O((x+1)^2) \big)
\,
\big( R_0 + R_1 (x+1) + O((x+1)^2) \big)
\ef;
    \\
R_0 &=
    \frac{1+t}{t} \sqrt{A} \pi
    + 
    \frac{(d-1)}{\sqrt{d+1}} K\left(\tfrac{2}{d+1}\right)-
     \sqrt{d+1} E\left(\tfrac{2}{d+1}\right)
\ef;
% \frac{\pi  \, _2F_1\left(\frac{1}{2},\frac{3}{2};2;\frac{2}{d+1}\right)-4
%    K\left(\frac{2}{d+1}\right)}{2 \sqrt{d+1}}
 \\
R_1 &=
\frac{1}{3}
 \left(\frac{d-1}{\sqrt{d+1}} 
    K\left(\tfrac{2}{d+1}\right)
    -
    \frac{d}{\sqrt{d+1}} E\left(\tfrac{2}{d+1}\right)\right)
\ef.
\end{align}
In order for the density to have a $(x+1)^{3/2}$ singularity,
we need to impose that $R_0=0$ (and verify that $R_1 \neq
0$\;\footnote{This being a consequence of the fact that
$(2-2m) E(m) - (2-m) K(m) \in [-1,0)$ for $m \in (0,1]$.}).
This fixes a relation among $t$, $A$ and $d$, that,
together with \eqref{eq:121}, 
gives the critical values for $A$ and $d$ as functions of $t$.
From 
\begin{align}
  a_1 &= \frac{1}{4}- A(d+1)\, ,
&
  a_2 &= \frac{1}{4}- A(d-1)\, ,
\end{align}
we get the system:
\begin{subequations}
  \label{eq:156}
\begin{align}
  &\frac{t+1}{t}
  =
  \frac{2}{\pi  (1-4 A d)}
  \left(\pi -4 \sqrt{A (d+1)} E\left(\tfrac{2}{d+1}\right)\right)\, ,
  \\
  &\frac{t+1}{t}
  =
  \frac{1}{\sqrt{A}\pi}
  \left(
    \sqrt{d+1} 
    E\left(\tfrac{2}{d+1}\right)
    -
    \frac{d-1}{\sqrt{d+1}} 
    K\left(\tfrac{2}{d+1}\right)
\right)
  \, .
\end{align}
\end{subequations}
We satisfy the conditions $a_1<a_2 \leq 1/4$ if and only if 
$A>0$ and $d \geq 1$.  As a first check we discuss again $d=1$. In
this case, we shall find that $t=t_\ast$ is a solution.  The
system above simplifies to
\begin{align}
%  d = 1 \, ,\qquad
  \frac{t+1}{t} &= 
  -\frac{8 \sqrt{2 A}-2 \pi }{\pi(1 -4  A)}\, ,
&
    \frac{1+t}{t} 
    &=\frac{\sqrt{2}}{\sqrt{A} \pi}\, .
\end{align}
There are two solutions:
\begin{align}
  A_{\pm} (d=1)
  = \frac{1}{16} \left(-4+\pi ^2 \pm \pi  \sqrt{\pi ^2-8}\right)
  = \{0.635326 , 0.0983747\}
  \, ,\quad
  t = \pm t_\ast\, .
\end{align}
This confirms our previous result that at $t=t_\ast$ the left edge has
a $(x+1)^{3/2}$ singularity.  Conversely, the $(x+1)^{3/2}$
singularity at $t=-t_{\ast}$ is
% a criticality mechanism which is
shadowed by the $(1-x)^{3/2}$ singularity, that occurs before.

For $d>1$ the picture remains similar, although the expressions are
less explicit. The triple $(t,A,d)$ with appropriate singularity, at
$t=t_\ast$ and $d=1$, is deformed continuously, and monotonically in
the three parameters, as $d$ grows, the limit of $t$ for
$d \to \infty$ being $t=-\infty$.  The spurious branch at $t=-t_\ast$
and $d=1$, similarly, is also deformed monotonically (this time, with
$t$ increasing, instead that decreasing), so it is always shadowed by
the criticality at the other endpoint of the support.

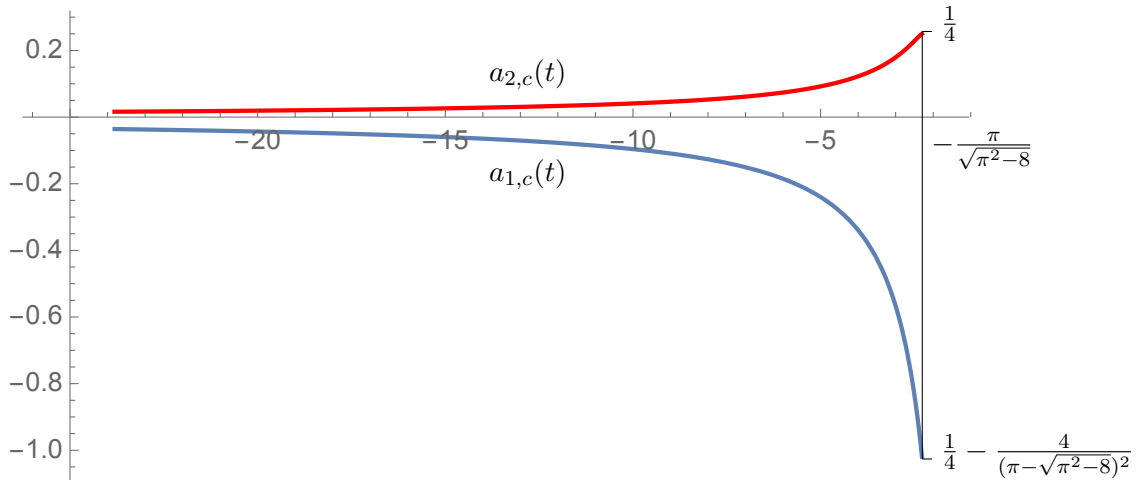
\begin{figure}[tb]
\setlength{\unitlength}{180pt}
\begin{picture}(2,1)
\put(0,0){\includegraphics[height=180pt]{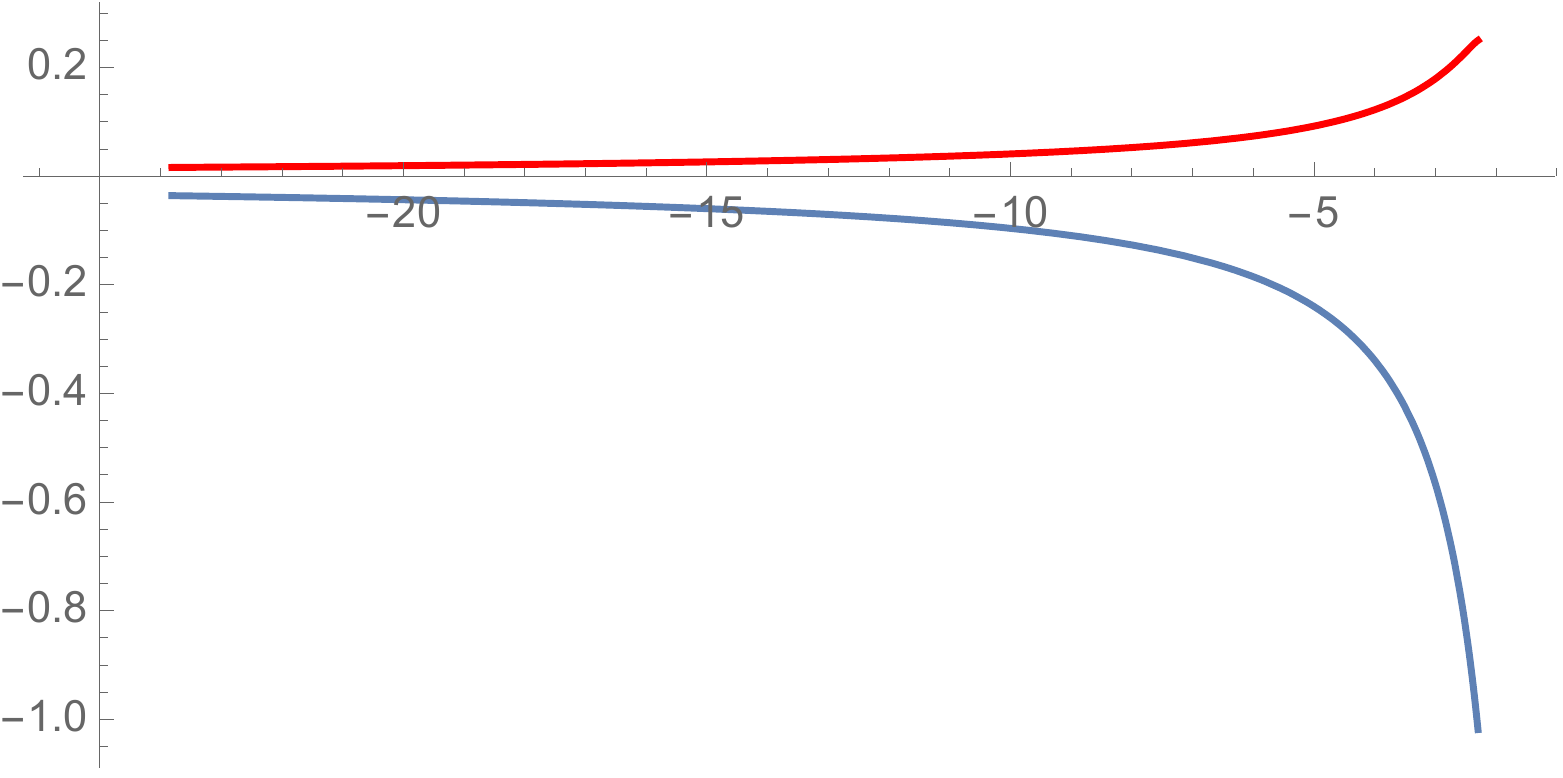}}
\put(1.901,0.057){\line(0,1){.896}}
\put(1.901,0.057){\line(1,0){.02}}
\put(1.901,0.953){\line(1,0){.02}}
\put(1.94,0.953){$\frac{1}{4}$}
\put(1.94,0.057){$\frac{1}{4}-\frac{4}{(\pi -\sqrt{\pi ^2-8})^2}$}
\put(1.92,0.7){$-\frac{\pi}{\sqrt{\pi ^2-8}}$}
\put(1,.85){$a_{2,c}(t)$}
\put(1,.64){$a_{1,c}(t)$}
\end{picture}
  \caption{\label{fig.a1a2t}The support of the spectral density for 
$t \leq t_{\ast}$.}
\end{figure}

A convenient way of expressing the general result is to derive, by
mean of \eqref{eq:156}, parametric expressions $A(d)$ and $t(d)$, to
be analysed in the regime $d \geq 1$. We obtain two branches for the
function $A(d)$, which in turn give two branches for $t(d)$. For the
reasons discussed above, we have to keep only the branch with negative
values of $t$ (which, indeed, is a monotonic function with image
$(-\infty,t_{\ast}]$). These expressions can be used, for example, to
plot $a_1(t)$ and $a_2(t)$ in this range of parameters, as shown in
Figure~\ref{fig.a1a2t}.

%%%%%%%%%%%%%%%%%%%%%%%%%%%%%%%%%%%%%%%%%%%%%%%%%%%%%%%
\section{Complex zeroes of $R_k(t)$ and $Z_k(t)$}
\label{sec.zeri}

Both in the case of $h$ even and odd, the behaviour at and below
$t_\ast$, where the Berker--Kadanoff phase has its lower endpoint and
the system is again massive, is an interesting new element w.r.t.\ the
treatment of \cite{BMCourt2015}, and deserves some deeper investigation.

In this short section we inspect more closely the numerical series
$R(g,t)$ and $Z(g,t)$. Let us call $R_k(t)$ and $Z_k(t)$\footnote{In
  fact, for the second series, the natural quantity in the system of
  equations is a \emph{derivative} of $Z$, but this changes the
  polynomials $Z_k(t)$ by an overall factor, and thus does not affect
  the zeroes.}
the polynomials in $t$ associated to these series
\begin{align}
R(g,t) &= g + \sum_{k \geq 2} g^k t \; R_k(t)
&
Z(g,t) &= \sum_{k \geq 2} g^k \; Z_k(t)
\end{align}
We know for sure that the $Z_k(t)$'s are real-positive if 
$t \geq -1$. This is a consequence of the fact that these quantities
can be written as a sum over trees, with non-negative weights (this is
detailed in Section \ref{sec:comb_crit}).

The quantities $Z_k(t)$ may or may not become negative for values
$t<-1$. Our numerical observations are compatible with the following
conjectures.

For $h=1$, at $t > t_\ast$ there exists a value $n_0(t)$ such that
$Z_k(t) < 0$ if $k < n_0$ and $Z_k(t) \geq 0$ if $k \geq n_0$, while
at $t \leq t_\ast$ all $Z_k(t)$ are negative.  In other words, all the
polynomials $Z_k(t)$ have a unique real root $t_0(k)$, these roots
form a monotonically decreasing sequence, with limit $t_\ast^{(h=1)}$.

For $h=2$, $Z_{2k}(t)>0$ for all values of $t$, while $Z_{2k+1}(t)$
have a pattern similar to the one of the $Z_k$'s for $h=1$, i.e., each
polynomial has a unique real root $t_0(k)$, and these roots form a
monotonically decreasing sequence, with limit $t_\ast^{(h=2)}$.

The fact that, for $h=2$ and $t<t_\ast$, the coefficients of the series
alternate in sign is compatible with the finding of Section
\ref{ssec.anaquart}, where it is evinced that at $t \leq t_c$ the main
singularity is present for a negative value of $g$.
% {sec.mainana}
Similar patterns emerge for the polynomials $R_k(t)$.

More generically, it is interesting to investigate the roots in the
complex plane of the polynomials $R_k(t)$ and $Z_k(t)$. Indeed, it
turns out that these roots have a simple behaviour, and seem to
accumulate on some `C-shaped' curve in the complex plane. It is the
symmetry of this curve that determines the sign behaviour depicted
above: as the roots come in complex-conjugate pairs except for the
ones on the real axis, if we assume that all the roots are aligned,
roughly along this curve, then it follows that there is exactly one
real root, or no real root at all, if the polynomial has odd or even
degree, respectively.

We will present data for the polynomials $Z_k(t)$ in the case $h=1$
and $h=2$ (data for the $R_k$'s, and, at $h=1$, for $S_k$'s, are
qualitatively similar). Before doing this, we solve exactly a `toy
version' of the system of equations relating $R$, $S$ and $Z$, which
shows explicitly the features depicted above.

%-------------------------------------------------------
\subsection{The toy model}

Consider the equation
\be
R = g + t \, \Phi(R)
\ee
with the choice
\be
\Phi(R) = \frac{R (1-\sqrt{1-4R})}{2}
= R^2 \big( 1 + R \; {}_{2}F_1(1,\tfrac{3}{2};3;4R) \big)
\ef.
\ee
% (this can be restated in terms of ${}_{2}F_1(1,\frac{3}{2};3;4R)$, which 
The second equality shows the analogy with the equations appearing in
our model, see e.g.\ equation (\ref{eq:R_4}).

This equation is sufficiently simple that it can be solved for
$R=R(g,t)$
\be
R(g,t) = g + 
\sum_{i=1}^{\infty}
\sum_{j=1}^i
\frac{(i+j)! (2i-j-1)!}{i! (i+1)! (i-j)! (j-1)!}
g^{i+1} t^j
\ee
The polynomials $R_k(t) = t^{-1} [g^{k+1}] R$ are thus, up to an overall
factor,
\be
\label{eq.rktoy}
R_k(t) \propto
\sum_{j=0}^k t^j 
\binom{k}{j} \bigg/
\binom{3k+2}{2k-j}
%(k+j+2)! (2k-j)!
\ee
(curiously enough, up to a factor $1/((k + j + 1)(k + j + 2))$, these
polynomials are proportional to the celebrated \emph{refined
  enumerations of Alternating Sign
  Matrices}~\cite{mills1983alternating}).

Now, for $t$ to be a zero of such a polynomial, we shall have
\be
\sum_j \exp
%\Big[
% k 
\Big(
% \frac{j}{k} 
j \ln t +
\ln (2k-j)! + \ln (k+j+2)! - \ln j! - \ln (k-j)!
\Big) 
%\Big]
= 0
\ef.
\ee
In the limit of large $k$, calling $x=j/k$ and $\ell(x) = x \ln(x)$,
we shall have 
\begin{align}
\int_0^1 dx
\exp ( k \, S(x,t) )
&= 0
\end{align}
with
\begin{align}
S(x,t) &= 
x \ln t + \ell(2-x) + \ell(1+x) - \ell(x) - \ell(1-x)
\end{align}
The integral is dominated by (one or more of) the saddle points, which
in this case are at the positions
\be
x_{\pm}(t) = \frac{-1\pm \sqrt{1-t+t^2}}{-1+t}
\ee
The two saddle points coincide when $t= t_{\pm} := \exp(\pm i \pi/3)$.
The integrand is not zero at none of the saddle points, so the only
possibility for the integral to vanish is that the contributions at
the two saddle points are equal in absolute value, and of opposite
phase. The phase varies rapidly, and shall be discarded (its leading
variation rate would determine the asymptotic \emph{density} of the
roots of the polynomials on the limit curve, but we do not perform
this calculation here). The necessary condition on the absolute value,
namely $Re(S(x_+(t),t)-S(x_-(t),t)) = 0$, is verified on four arcs:
\begin{itemize}
\item the arc of the circle of radius 1 and center 0, at the left of
  $t_{\pm}$;
\item the arc of the circle of radius 1 and center 1, at the right of
  $t_{\pm}$;
\item the straight segment connecting $t_{\pm}$;
\item the two straight segments, connecting $t_{+}$ with $+i \infty$
  and $t_{-}$ with $-i \infty$ (this is a single arc in the Riemann
  sphere).
\end{itemize}
It turns out that the roots of the polynomials are asymptotically
supported on the first of the four arcs mentioned above.

\begin{figure}[tb]
\[
\includegraphics[scale=.5]{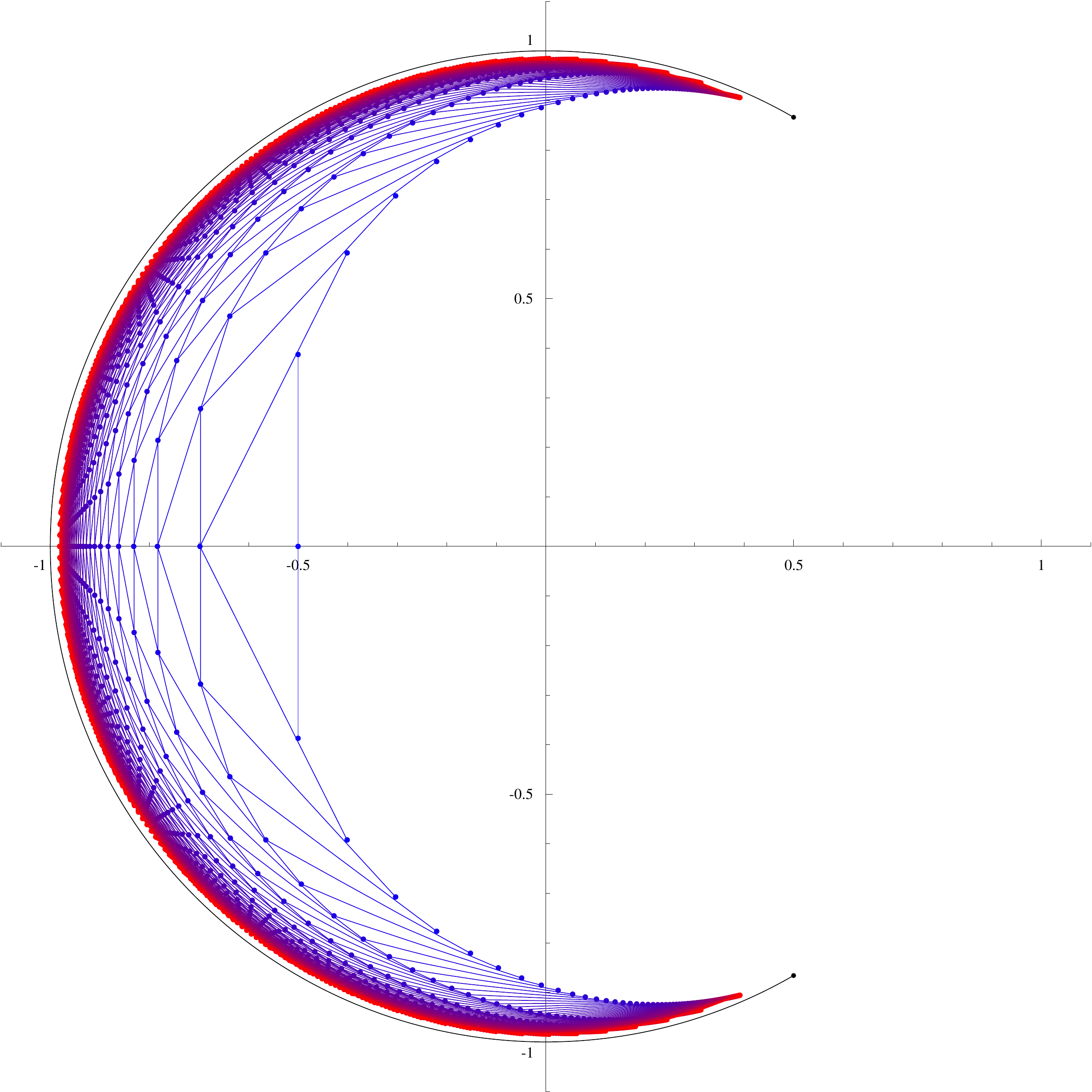}
\]
\caption{\label{fig.roottoy}Roots of the polynomials $R_k(t)$ defined
  in (\ref{eq.rktoy}), for $k \leq 80$, and their limit curve.}
\end{figure}

See Figure \ref{fig.roottoy} for an illustration of the properties
discussed above.

\subsection{The true model}

\begin{figure}[tb]
\[
\includegraphics[scale=.4]{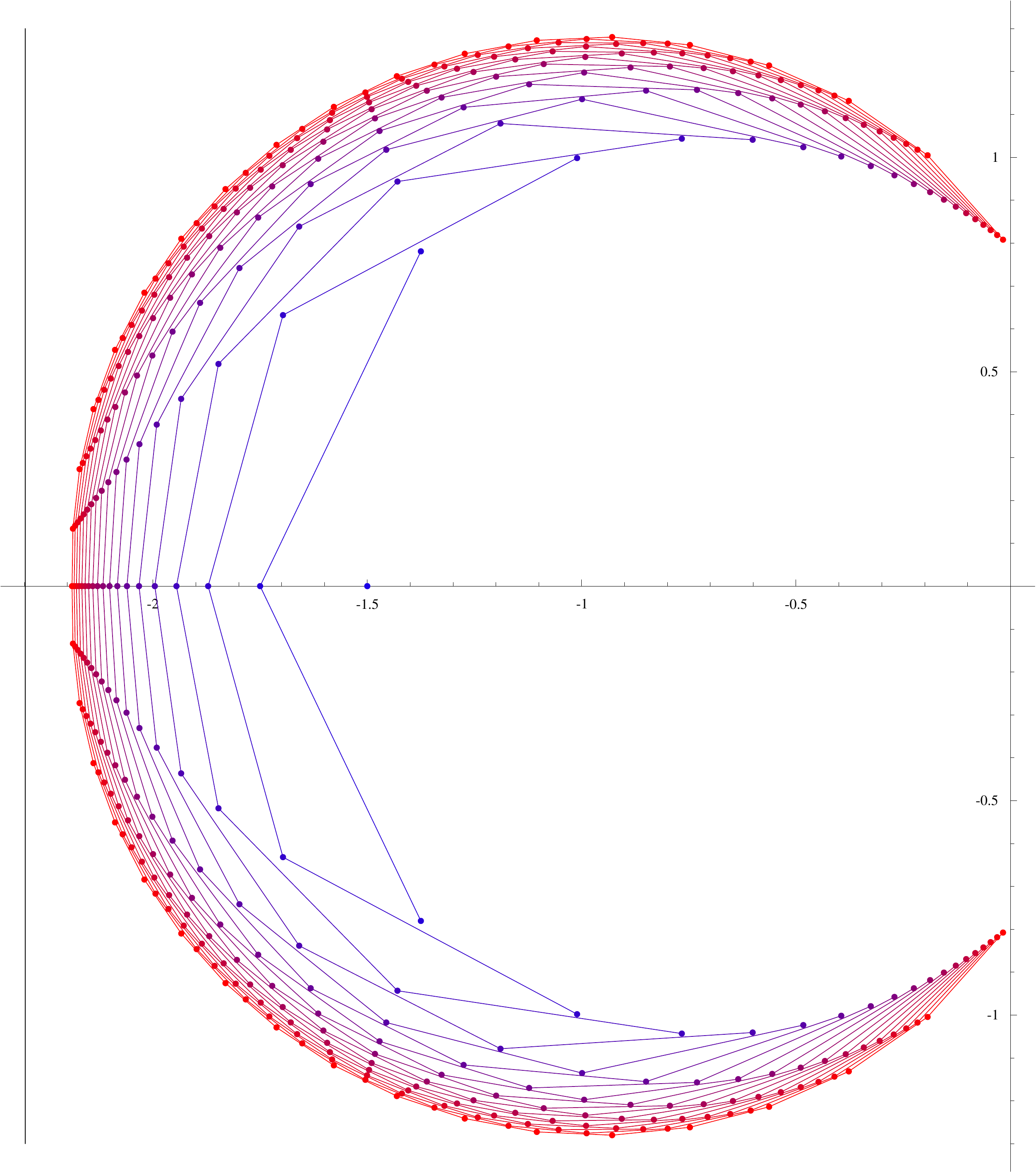}
\]
\caption{\label{fig.rooth1}Roots of the polynomials $Z_k(t)$ in the
  model at $h=1$, for $k \leq 20$. The vertical segment marks the
  limit of the real roots, $t_{\ast} = -\frac{\pi}{\sqrt{\pi^2-8}}=-2.2976\ldots$}
\end{figure}

\begin{figure}[tb]
\[
\includegraphics[scale=.4]{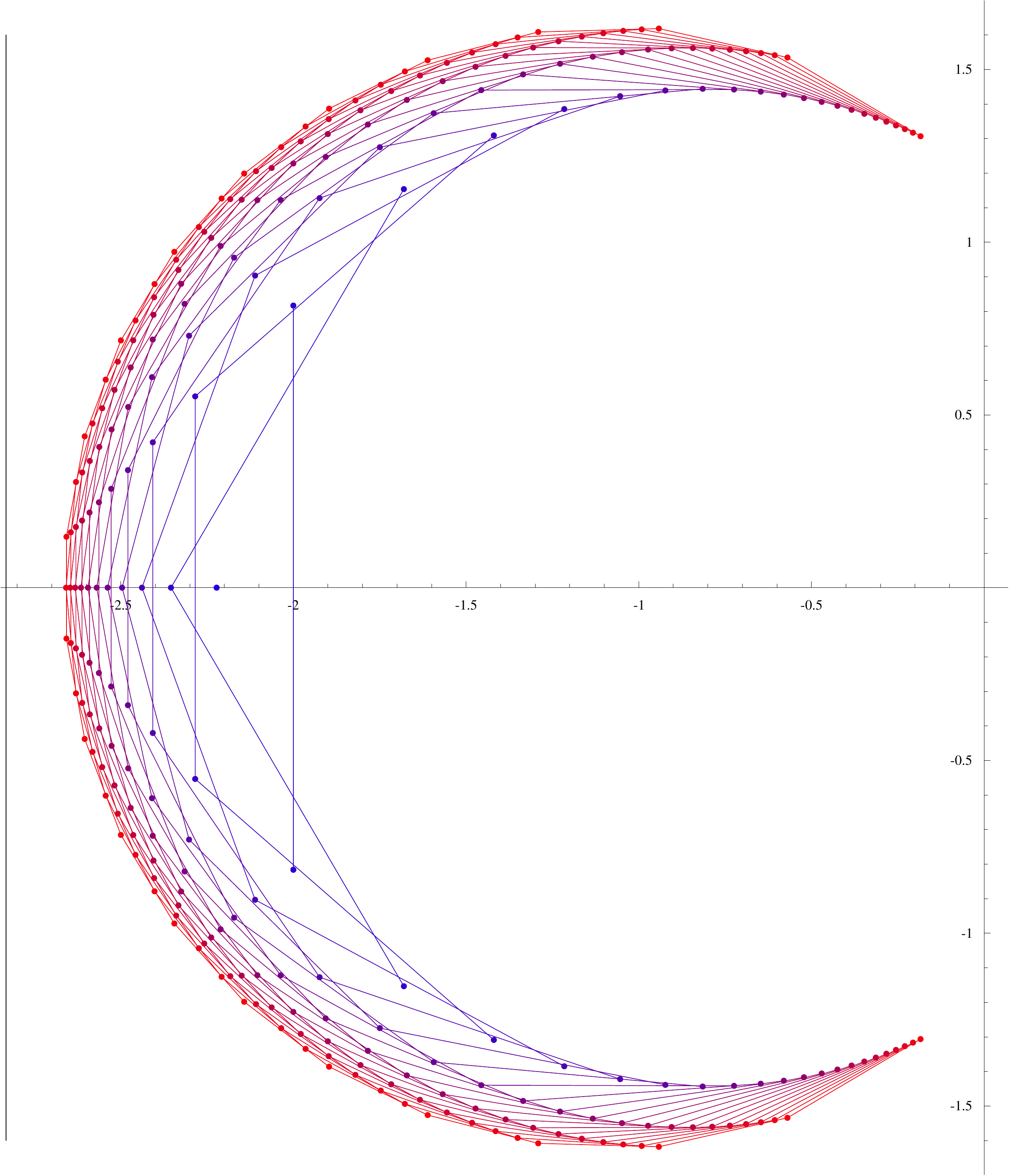}
\]
\caption{\label{fig.rooth2}Roots of the polynomials $Z_k(t)$ in the
  model at $h=2$, for $k \leq 25$. The vertical segment marks the
  limit of the real roots, $t_{\ast} = -2.8324\ldots$}
\end{figure}

Here we show the numerical results for the `true' quantities in our
model, for the cases $h=1$ (Figure \ref{fig.rooth1}) and $h=2$
(Figure \ref{fig.rooth2}).

From the discussions in Section \ref{ssec.anaquart} and
\ref{ssec.anacubi}, we know the point of intersection of the limit
curve with the horizontal axis (see equations (\ref{eq.valtast4}) and
(\ref{eq:t_ast})). It would be interesting to determine also the limit
curves, and in particular their endpoints.

%%%%%%%%%%%%%%%%%%%%%%%%%%%%%%%%%%%%%%%%%%%%%%%%%%%%%%%
\section[The Berker--Kadanoff phase at $t=-1$]
{The Berker--Kadanoff phase at \textbf{\textit{t}\,=\,-\!-1}}
\label{sec:comb_crit}

%-------------------------------------------------------
\subsection{The role of Bernardi embedding-activities}

The partition function \eqref{eq:ZForDef} of spanning forests over a
given graph $G$ corresponds to a (monovariate) specialization of the
(bivariate) Tutte polynomial $T_{G}(x,y)$ associated to a graph $G$
(see e.g.\ \cite{SokTutte}), namely
\be
\Zfor_G (t,1) = t\; T_{G}(t+1,1)
\ef.
\ee
The Tutte polynomial admits both a formulation in terms of spanning
subgraphs, and one as a sum over spanning trees weighted according to
their activities \cite{tuttebook}.
% \cite{Bollobas} 
The latter is particularly useful when studying $T_{G}(x,y)$ for 
$x, y \geq 0$, but not both $\geq 1$, so that the weights associated to
spanning subgraphs are not real-positive, but those associated to
trees are.

Activities are of two kinds. An edge in the tree may be
internally-active, while one not in the tree may be externally-active.
The $t=-1$ point of spanning forests is specially
simple. Externally-active edges have weight 1, so we do not need to
`count' them. Internally-active edges have weight 0, so they are just
forbidden. This simplifies the combinatorics: once an
internally-active edge has been identified in the tree, we can discard
the associated term, with no need to complete the exploration. The
partition function at fixed graph is an integer, and thus also the
generating function associated to random planar graphs has
integer-valued coefficients, provided we take an ensemble of suitably
`rooted' graphs, so that the automorphism group is trivialised.

The concrete determination of internally-active edges is not a natural
task. Tutte's definition of activities is associated to an arbitrary
but fixed labeling of the edges, which is not a canonical
notion. Alternatively, it could be defined as an average over all
possible labelings, but this would introduce a new set of variables,
which makes the treatment complicated.

Luckily enough, by a breakthrough result of Bernardi
\cite{Bernardi2008} we now know that, for graphs embedded on a
surface, one can define \emph{embedding-activities}, i.e.\ activities
determined in terms of a canonical labeling of the edges associated to
the tree. Remarkably, and for non-trivial reasons, the Tutte
polynomial based on `static' activities and the Bernardi polynomial
using embedding activities do coincide.

This result makes viable a probabilistic analysis of large random planar
graphs equipped with a ferromagnetically-critical Potts model at $0
\leq q \leq 4$, somewhat along the lines by which the same limit is
constructed for pure gravity,
% [....marckert, le gall, miermont....], 
an approach that has flourished in recent years 
\cite{sheffield2011quantum, chen2015basic, berestycki2015critical,
  Kassel2015, Gwynne2016}.
However, unfortunately, this probabilistic approach seems to be
confined to the ferromagnetic critical line, thus it is orthogonal to
the treatment of spanning forests, except that at the trivial
spanning-tree criticality.

%-------------------------------------------------------
\subsection{Reformulation in terms of spanning trees with internal activities}

Consider a graph $G$, embedded on a surface, and with a distinguished
half-edge (the \emph{root}). Given a spanning tree $T$ over it, we say
that an edge is \emph{internal} if it belongs to the tree, and
\emph{external} otherwise.  Define the \emph{fundamental cocycle} of
an internal edge $e$ as the set of external edges $f$ such that $T
\backslash \{e\} \cup \{f\}$ is a spanning tree. I.e., removing $e$
from the tree leaves with two components, and the cocycle is the set
of edges with endpoints on distinct components.  Analogously, define
the \emph{fundamental cycle} of an external edge $e$ as the set of
internal edges $f$ such that $T \backslash \{f\} \cup \{e\}$ is a
spanning tree. I.e., adding $e$ to the tree leaves with a unicyclic
subgraph, and the cycle is just the cycle of the subgraph.

% The notation $T \backslash \{e\}$ means the deletion of the edge $e$.  
Now we consider a walk on the surface, encircling the tree, and
starting from the root. Thus, we go along internal edges, while
crossing external ones.  Each internal edge happens to be adjacent to
this walk exactly twice, one per side, The same happens for external
edges, this time once per endpoint. Label internal and external edges
according to the order of the first (of the two) visit by the walk.

\begin{figure}[tb]
\[
\setlength{\unitlength}{65.625pt}
\begin{picture}(4,4)
\put(0,0){\includegraphics[scale=3.5]{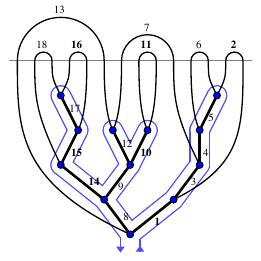}}
% \put(0,0){\line(1,0){4}}
\end{picture}
\]
\caption{\label{fig:activerpg}%
A rooted cubic random planar graph, equipped with a spanning tree.
The exploration path around the tree is shown in blue, and determines
the labeling of the edges (both internal and external). Active edges
are denoted by labels in boldface (this example has $4$ internally- and
$3$ externally-active edges).}
\end{figure}

This allows us to define the notion of activity: an external
(resp.\ internal) edge is \emph{active} if it is minimal in its
fundamental cycle (resp.\ cocycle).  Call
${\cal I}(T)$ and ${\cal E}(T)$ the number of internally- and
externally-active edges of $T$, according to embedding-activities
(this notation is somewhat elliptic, as this value depends on $G$ and
on the root position as well).  See Figure \ref{fig:activerpg} for an
example.

Thus, as a corollary of Bernardi result \cite{Bernardi2008},
the spanning-forest partition function
can be rewritten as
\begin{align}
  \Zfor_G (t,1) = 
  t\sum_{T \preceq G}(1+t)^{{\cal I}(T)}
\ef.
\end{align}
As we said, embedding activities are more canonical than the original
ones defined by Tutte. However, there is a tiny amount of
non-canonical degrees of freedom left. First, we decided to explore the
tree through a counter-clockwise contour. A clockwise contour would
have produced the same polynomial, although the terms associated to
the single trees would have been different. Second, we need to start
the contour at some point, so that the tree is in fact \emph{rooted}
on a half-edge.

As we know, this is not unusual for random planar graphs, also in
other models (and even just in pure gravity). The generating functions
of unrooted and of rooted configurations are normally easily related
one to another, while rooted ones have the advantage of avoiding graph
automorphisms.

So, similarly to what is done in equation \eqref{eq:ZSpForFixG}, to
facilitate the combinatorics it is convenient to re-weight the sum. In
this case we include a factor $|\Aut(G)|/(3|V(T)|)$, so that the
partition function for spanning forests on random planar graphs
becomes:
\begin{align}
  Z(t,g)
  &=
  \sum_{G \mathrm{~rooted}} \frac{g^{|V(T)|}}{3|V(T)|}
  \sum_{T \preceq G} (t+1)^{{\cal I}(T)} \nonumber \\
  &=
  \sum_{T' \mathrm{~rooted}} \frac{g^{|V_{\rm int}(T')|}}{3|V_{\rm int}(T')|}
  \sum_{\LL \in\LxL(T')} (t+1)^{{\cal I}(T')}  \, .
\end{align}
In the second line the sum is over cubic planar trees $T'$, and the
restriction of $T'$ to its non-leaf edges is the tree $T$. We also
denote by $V_{\rm int}(T')$ the set of non-leaf nodes of $T'$, which
thus coincides with all the nodes of $T$.
There are $n_{T'}=|V_{\rm int}(T')|+2$ leaves. The set $\LxL(T')$ is
the ensemble of link patterns composed of $n_{T'}/2$ arcs. Such a
pattern, used for pairing the leaves of $T'$, composes a graph $G$.
% Recall that a leaf of $T$ denotes an external half edge.
The activities are then computed in this graph $G$ (so that 
${\cal I}(T')$ is a shorthand for the true dependence 
${\cal I}(T',L)$).

In the next section we will study this function at $t=-1$, which
corresponds to counting configurations with no internally-active edge:
\begin{align}
  \label{eq:Z_activity}
  Z(-1,g)
  =
  \sum_{T} \frac{g^{|V(T)|}}{3|V(T)|}
  \sum_{\LL \in\LxL(T)} \delta_{{\cal I(T)},0}
  \, .  
\end{align}

%-------------------------------------------------------
\subsection{Recursion relations: resolving a branch}

Let us consider the ensemble of planar rooted trees, and let us
represent them as embedded in the lower half plane, and with leaves
attached to the boundary, as in the lower part of Figure
\ref{fig:activerpg}.  We denote by ${\cal S}_{k,\ell}$ the subset of
such trees, in which the vertices have all degree three, except
for the root vertex, which has degree $k+\ell$. We further
require that among these $k+\ell$ edges, the $\ell$ right-most
vertices are leaves.

The position of infinity in the half-plane breaks the rotational
symmetry among the edges adjacent to the root vertex, and we can
canonically identify the root half-edge with the right-most edge
adjacent to the root.

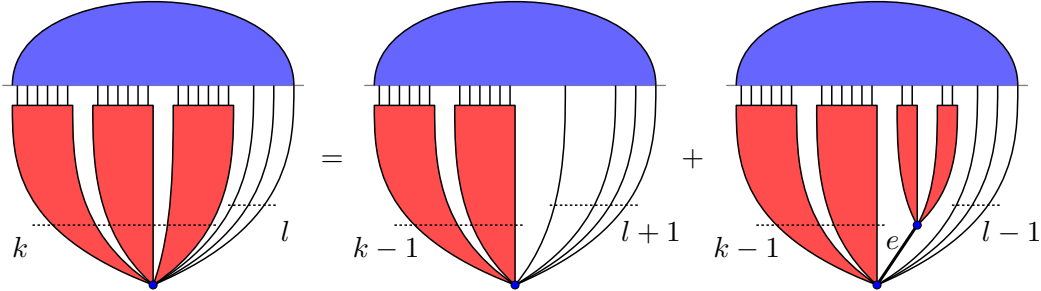
\begin{figure}[tb]
\[
% le tre figure distano 9 ul
\setlength{\unitlength}{15pt}
\begin{picture}(26,8)
\put(0,0){\includegraphics[scale=1.5]{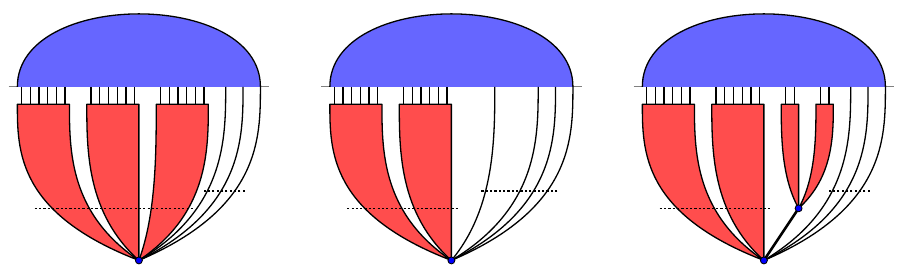}}
\put(.5,1.2){$k$}
\put(9,1.2){$k-1$}
\put(18,1.2){$k-1$}
\put(7.2,1.6){$l$}
\put(15.7,1.6){$l+1$}
\put(24.7,1.6){$l-1$}
\put(22.3,1.4){$e$}
\put(8.2,3.5){$=$}
\put(17.2,3.5){$+$}
\end{picture}
\]
  %% \centering
  %% \includegraphics[scale=1]{Rkl}
  \caption{Illustration of the recursion \eqref{eq:Rkl} satisfied by
    $T_{k,\ell}$.\label{fig:Rkl}}
%    The root is the rightmost leaf.
\end{figure}

Call $\LxL(T)$ the set of link patterns of size $n$, if $T$ has $2n$
leaves, and the empty set if it has an odd number of leaves. We
naturally associate to a pair $(T,L)$, with $T \in {\cal S}_{k,\ell}$
and $L \in \LxL(T)$, a planar graph, through the graphical
construction illustrated by Figure~\ref{fig:activerpg}. The left-most
drawing of Figure \ref{fig:Rkl} schematise such a structure, with red
triangoloids associated to the $k$ (possibly-)non-trivial branches of
$T$ attached to the root, and the blue half-disk associated to the
link pattern.

We will derive a recursion formula for the generating function of
these graphs, restricted to the set of graphs with no
internally-active edges
% following quantity:
\begin{align}
  \label{eq:Tkl}
  T_{k,\ell}(g) = \sum_{T\in{\cal S}_{k,\ell}} g^{|V(T)|-1}
  \sum_{\LL \in\LxL(T)} \delta_{{\cal I(T)},0}\, .
\end{align}
% The relation 
In the special case $(k,\ell)=(3,0)$, this coincides with the
partition function \eqref{eq:Z_activity} of the previous section, up
to a trivial differentiation
\begin{align}
  T_{3,0}(g) = 3\frac{\dd}{\dd g} Z(-1,g)
\ef,
\end{align}
so $T_{3,0}(g)$ is the object of our interest. Nonetheless, we are
induced to consider the more general class of functions
$T_{k,\ell}(g)$, in order to produce recursion relations (in the next
section, we get rid of the parameter $\ell$, while still being forced
to consider all values of $k$ at once).  Remark that, similarly to
what discussed above for general cubic graphs, also here we have a
parity constraint on $k$, $\ell$ and the number of vertices of $T$. In
particular, the even coefficients of $T_{3,0}$ are zero.

Our recursion is established by the following argument.  We start by
setting $T_{0,0}=1$, and $T_{k,\ell}=0$ if $k<0$ or $\ell<0$.  At
$k=0$ we have no internal structure in the tree (in particular, there
are no internal edges, and thus no internally-active ones!), and we
trivially have $T_{0,\ell} = \Cat_{\ell/2}$ for $\ell$ even, and zero
for $\ell$ odd (here $\Cat_n$ are the Catalan numbers). Furthermore,
$T_{k,\ell}$ is a power series in $g$ with constant term
$\Cat_{(k+\ell)/2}$.

Since embedding-activities are defined according to the edges visited
by the walk, starting at the right of the root as in
Figure~\ref{fig:activerpg}, the root edge is always active (either
internally or externally). Thus, in our case with no internally-active
edges, we shall have that the left-most branch of the tree must be a
leaf, i.e., that $T_{k,0}=T_{k-1,1}$ for $k>0$.

Having set the easy boundary conditions, we can now proceed to the
analysis of the interesting quantities, i.e.\ $T_{k,\ell}$ with both
$k$ and $\ell$ strictly-positive. The first, tautological recursion is
depicted in Figure \ref{fig:Rkl}, and it states that, in $T_{k,\ell}$,
the $k$-th branch of the root (counting from the left) is either a
leaf, or it starts with an internal edge. Calling $R_{k,\ell}$ the
generating function associated to the second case, we thus have
\begin{align}
  \label{eq:Rkl}
  T_{k,\ell}(g) = T_{k-1,\ell+1}(g) + R_{k,\ell}(g)\, .
\end{align}
We shall then analyse $R_{k,\ell}$ more closely (and express it as a
polynomial in the ($T_{k',\ell'}$)'s), in particular we shall enforce
the fact that the edge $e$ that we have singled out is not
internally-active. The only edges in the graph which have a smaller
label than $e$ are the external edges with an endpoint on one of the
$\ell$ leaves attached to the root, at the right of $e$.  Thus, for
the edge $e$ to be inactive, there must be at least one of these edges
such that its other endpoint is attached to a node $v$ of the tree
downstream to $e$.  We denote by $f$ the innermost of such arcs (i.e.,
the one such that the endpoint on the right is left-most among the
arcs with this property), and say that this endpoint is on the
$(2h+1)$-th of the $\ell$ leaves, counting from the left. This index
shall in fact be odd, because, by planarity and the constraint that
$f$ is the left-most arc with its defining property, the other
endpoints to its left shall be matched among themselves.

Now we analyse the path on the tree between $e$ and $v$.  This path
has other branches of the tree attached to it, some on the left, some
on the right. (say there are $i$ and $j$ such branches, respectively).
All the edges in the path are inactive, as they have an index higher
than the one of $e$. This makes immaterial the ordering among left and
right branches along the path.

The $j$ branches on the right reach the boundary of the half-plane
with a number of leaf terminations, which are then paired through the
link pattern. As we have already established that the leaf-edges $1$
to $2h$ are paired among themselves, and because of the presence of
edge $f$, and planarity, the full link pattern must factorise on this
interval of leaves, this being the crucial property of the analysis.

At this point, the structure associated to $R_{k,\ell}$ is completely
factorised into a simple data structure:
\begin{itemize}
\item an index $h \leq (\ell-1)/2$, for the position of the right
  endpoint of $f$;
\item a link pattern composed of $h$ arcs, matching the leaves from
  $1$ to $2h$, among the $\Cat_h$ possible ones;
\item two integers $i$ and $j$ for the number of left and right
  branches on the path (with $i+j \geq 1$, as we are guaranteed that
  the path from $e$ to $v$ has at least one edge);
\item a binary string for their ordering, among the $\binom{i+j}{i}$
  possible ones;
\item a structure of the form $T_{j,0}$, for the right branches;
\item a structure of the form $T_{k+i-1,\ell-2h-1}$, for the left
  branches, the $k-1$ remaining branches attached to the root, and the
  $\ell-2h-1$ remaining leaves on the far right.
\end{itemize}
Collecting all these contributions, and including an appropriate power
of $g$ for the counting of nodes in the path from $e$ to $v$, leads to
the equation
\begin{align}
  \label{eq:Rkl_zoom}
  R_{k,\ell} = \sum_{h=0}^{\lfloor\frac{\ell-1}{2} \rfloor}
  \Cat_{h} \sum_{\substack{ i,j\ge 0\\i+j\ge 1}} g^{i+j}
  \binom{i+j}{j} T_{k+i-1,\ell-2h-1} T_{j,0}
\end{align}
Combination of (\ref{eq:Rkl}) and (\ref{eq:Rkl_zoom}) gives
\be
\begin{split}
  T_{k,\ell} &= T_{k-1,\ell+1} + 
\sum_{h=0}^{\lfloor\frac{\ell-1}{2} \rfloor}
\sum_{\substack{ i,j\ge 0\\i+j\ge 1}} 
\frac{1}{h+1}\binom{2h}{h}
g^{i+j}
  \binom{i+j}{j} T_{k+i-1,\ell-2h-1} T_{j,0}
\\
&=
\sum_{h=0}^{\lfloor\frac{\ell-1}{2} \rfloor}
\sum_{i,j\ge 0}
\frac{1}{h+1}\binom{2h}{h}
g^{i+j}
  \binom{i+j}{j} T_{k+i-1,\ell-2h-1} T_{j,0}
\ef.
\end{split}
\ee
This system of equations is `only' quadratic (and somewhat analogous
to the structures emerging in Tutte's quadratic method). It is
triangular in the variable $\ell$, if solved w.r.t.\ $T_{k-1,\ell+1}$,
for example
\begin{align}
\label{eq.9887565476}
T_{k-1,2}
&= 
  T_{k,1} -
\sum_{\substack{ i,j\ge 0\\i+j\ge 1}} 
g^{i+j}
  \binom{i+j}{j} T_{k+i-2,1} T_{j-1,1}
\ef.
\end{align}
but unfortunately we cannot provide initial conditions for $T_{k,1}$
(as, in particular, we would content ourselves with $T_{2,1}$ alone).

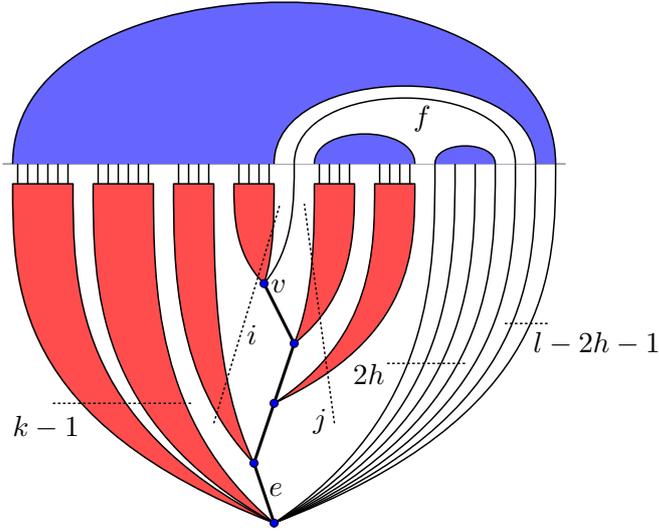
\begin{figure}[tb]
\[
\setlength{\unitlength}{15pt}
\begin{picture}(16,16)
\put(0,0){\includegraphics[scale=1.5]{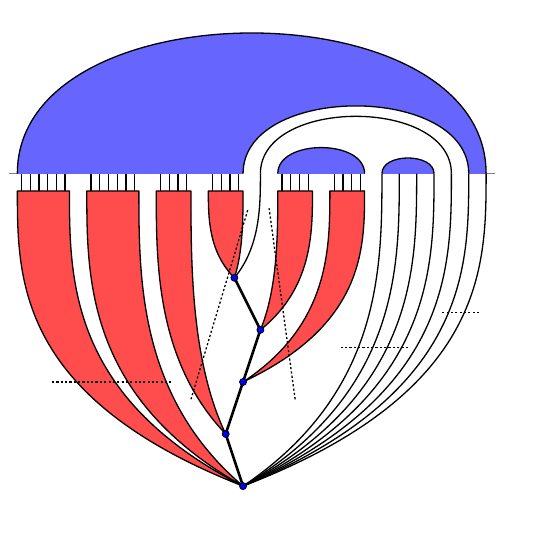}}
\put(.5,4.2){$k-1$}
\put(6.95,7.8){$v$}
\put(6.35,6.5){$i$}
\put(8,4.4){$j$}
\put(6.9,2.7){$e$}
\put(9,5.5){$2h$}
\put(13.5,6.3){$l-2h-1$}
\put(10.5,12){$f$}
\end{picture}
\]
%  \centering
%  \includegraphics[scale=1]{Rkl_zoom}
  \caption{Decomposition of a configuration entering $R_{k,\ell}$,
as described by equation~(\ref{eq:Rkl_zoom}).\label{fig:Rkl_zoom}}
\end{figure}
%

% -------------------------------------------------------
\subsection{Recursion relations: resolving a leaf}

We can produce another recursion relation for the quantities
$T_{k,\ell}$, in which, at difference with the previous section, we
investigate the local structure near to one of the $\ell$ leaves,
instead that near one of the $k$ branches.

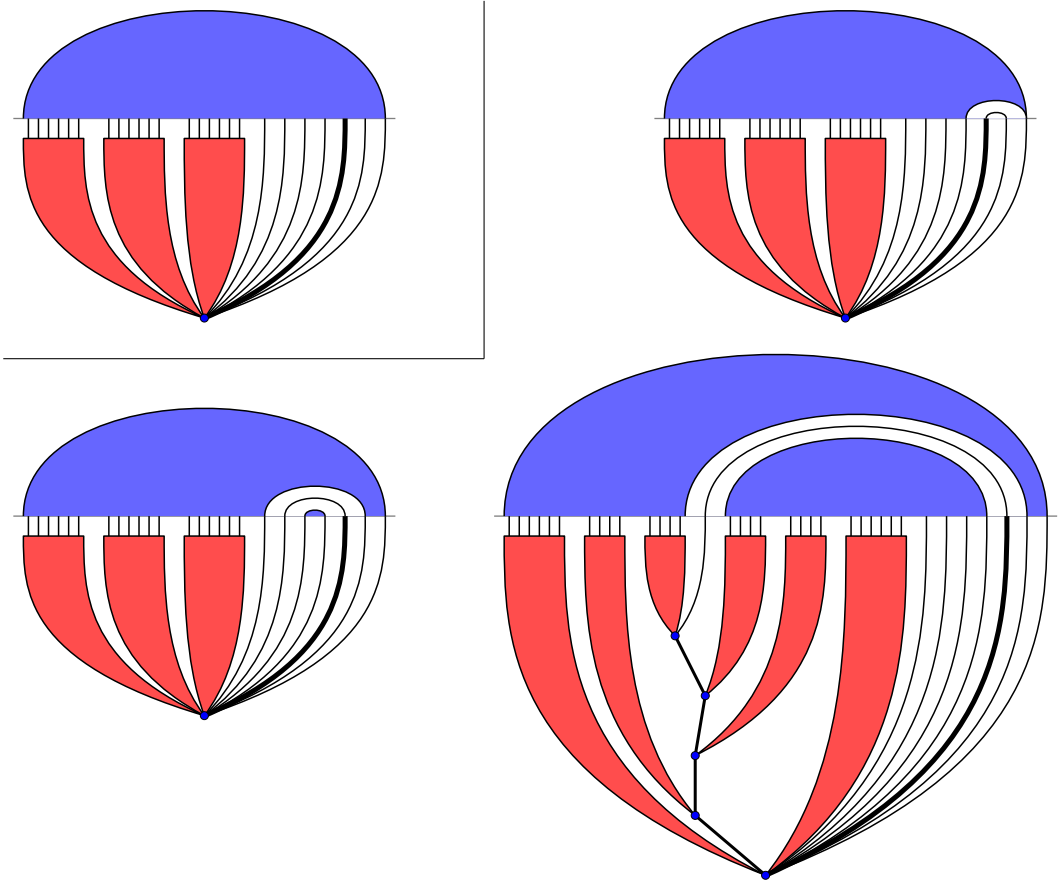
\begin{figure}[tb!]
\[
\setlength{\unitlength}{15pt}
\begin{picture}(28,24)
% \put(0,24){\line(1,0){5}}
\put(0,15){\line(1,0){12}}
\put(12,15){\line(0,1){9}}
\put(0,10){\includegraphics[scale=1.5]{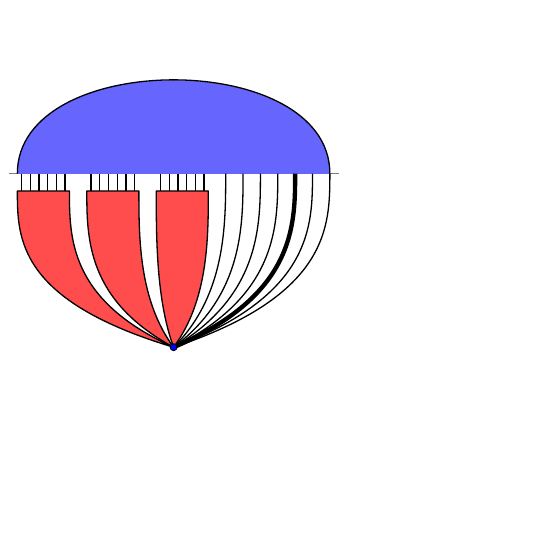}}
\put(16,10){\includegraphics[scale=1.5]{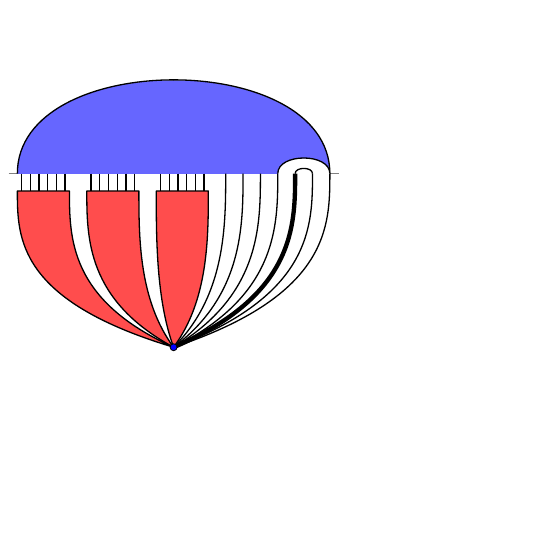}}
\put(0,0){\includegraphics[scale=1.5]{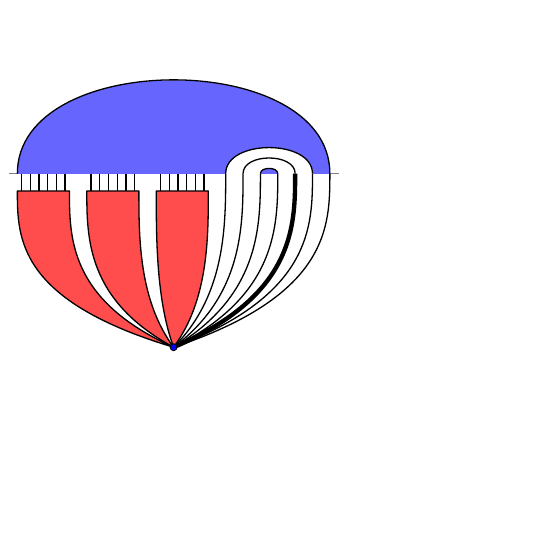}}
\put(12,0){\includegraphics[scale=1.5]{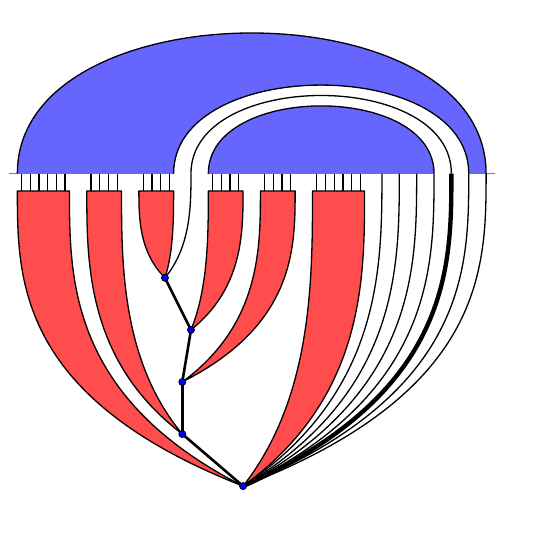}}
\end{picture}
\]
  \caption{In the top-left corner, a schematic representation of a
    configuration entering $T_{k,\ell}$. The three remaining images
    describe the three kinds of terms in
    equation~(\ref{eq.63545654}).\label{fig:leafrec}}
\end{figure}

Consider the $s$-th such leaf, counting from the right. Three events
may occur:
\begin{itemize}
\item The other endpoint of the edge is at its right, the arc covers
  $2h$ other endpoints. This gives a contribution 
  $C_h T_{k,\ell-2h-2}$, provided that $s-1 \geq 2h+1$.
\item The other endpoint of the edge is at its left, is one of the
  $\ell$ leaves, and the arc covers
  $2h$ other endpoints. This also gives a contribution 
  $C_h T_{k,\ell-2h-2}$, provided that $\ell-s \geq 2h+1$.
\item The other endpoint is on the $m$-th of the $k$ branches counting
  from the right. It reaches a vertex $v$, which is connected to the
  root of the tree by a path of length $i+j$, having attached to it
  $i$ branches on the right, and $j$ on the left. This gives a
  contribution $g^{i+j} \binom{i+j}{i} T_{m-1+i,\ell-s} T_{k-m+j,s-1}$
\end{itemize}
The reasoning by which we have the combinatorial factor 
$g^{i+j} \binom{i+j}{i}$, and we can identify the resulting structure
with $T_{k',\ell'}$ terms (because all and only the edges on the path
are certified non-active by the singled-out arc, and thus the path can
be shrunk to a point) is identical to the one analysed in the
previous section, and for this reason we omit to discuss it further
here (see however Figure \ref{fig:leafrec} for an illustration of the
three contributions).

This leads to the equation, valid for $1 \leq s \leq \ell$ (let
$\Theta_x$ denote the indicator function for event $x$)
\be
\label{eq.63545654}
\begin{split}
T_{k,\ell} 
&= \sum_{h\geq 0}
\big( \Theta_{s-1 \geq 2h+1} + \Theta_{\ell-s \geq 2h+1} \big)
C_h \; T_{k,\ell-2h-2}
\\
& \quad +
\sum_{m=1}^k
\sum_{i,j \geq 0}
g^{i+j} \binom{i+j}{i} 
\; T_{m-1+i,\ell-s} \; T_{k-m+j,s-1}
\ef.
\end{split}
\ee
Similarly to equation (\ref{eq.9887565476}), this is also triangular
in the parameter $\ell$. However, at difference with
(\ref{eq.9887565476}), now we can derive an equation involving only
$\ell \leq 1$ (recall that $T_{k,0} \equiv T_{k-1,1}$).  Calling
$T_k := T_{k,0} \equiv T_{k-1,1}$, we have
\be
\label{eq.2653453}
T_k =
\sum_{m=1}^{k-1}
\sum_{i,j \geq 0}
g^{i+j} \binom{i+j}{i} 
\; T_{m-1+i} \; T_{k-m+j-1}
\ee

% -------------------------------------------------------
\subsection{Numerical results}

In this section we will present a numerical study of the series $T_3$,
calculated by use of the recursion relation (\ref{eq.2653453}).  Say
that, for two power series
$f(x)=\sum_j f_j x^j$ and $g(x)=\sum_j g_j x^j$, 
$f \preceq g$ if $f_j \leq g_j$ for all $j$.  Let $f^{[k]}(x)$ the
truncation of the series $f$ at order $k$.  The relation
(\ref{eq.2653453}) can be seen as an iterated map, in which the
left-hand side at time $t+1$ is obtained by the expressions on the
right-hand side at time $t$. We fix a value $N$, and search for the
collection of $T_k^{[N-k]}$. This means that, when evaluating an
algebraic expression involving polynomials, we will restrict the
result at the prescribed order. We initialise $T_k$ to the
forementioned values $T_k(g=0) = C_{k/2}$ (if $k$ even, and $0$
otherwise). From the positivity of the involved coefficients, it is
easy to see that the forementioned iterated map is monotonic
w.r.t.\ the ordering $\preceq$, and reaches a fixed point in $\lfloor
N/2 \rfloor$ iterations.

From this map, it is easy to devise a program that produces the series
coefficients $(T_k)_j = [g^j] T_k(g)$, for $k+j \leq N$, and in
particular the desired values $(T_3)_j$.  We present in Table
\ref{tab:T30v} the first few such values.
\begin{table}[ht!]
\begin{center}
\begin{tabular}{ l l }
\hline
$j$ & $(T_{3})_j$\\
\hline
$1$ & $2$\\ 
$3$ & $18$\\ 
$5$ & $258$\\ 
$7$ & $4688$\\ 
$9$ & $98496$\\ 
$11$ & $2283372$\\ 
$13$ & $56838970$\\ 
$15$ & $1493264064$\\ 
$17$ & $40925517072$\\ 
$19$ & $1160500983808$\\ 
$21$ & $33843055573872$\\ 
$23$ & $1010383317486336$\\ 
$25$ & $30772480450675840$\\ 
$27$ & $953427414717730872$\\ 
$29$ & $29983902716549162970$\\ 
$31$ & $955366149199264460928$\\ 
$33$ & $30794714322277028617968$\\ 
$35$ & $1002901358278865429292960$\\ 
$37$ & $32964942669028735961001552$\\ 
$39$ & $1092605376996265409863119360$\\ 
% $41$ & $36488138498961852476189889792$\\ 
% $43$ & $1226940661506823065165727994496$\\ 
% $45$ & $41516699148649013379117203627568$\\ 
% $47$ & $1412944172847420605930972417246208$\\ 
% $49$ & $48342849001033872400052181169676416$\\ 
% $51$ & $1662146178496611076677384799206162432$\\ 
% $53$ & $57408984320404873797016845164048791680$\\ 
% $55$ & $1991244817212081032432718422131658764288$\\ 
% $57$ & $69339124398176919237234903929582149969920$\\ 
% $59$ & $2423416696155140295942186129121596152856816$\\ 
\hline
\end{tabular}
\label{tab:T30v}
\caption{Values of $(T_{3})_j$ for $j=1,3,\dots,39$. The coefficients
  for even $j$ are zero.}
\end{center}
\end{table}

As a technical remark, we note that the integers involved become soon
very big (they grow at an exponential rate), and the well-known
`Chinese remainder theorem strategy' has to be used in order to
improve on the na\"ive complexity. In a few words, this accounts to
perform the arithmetic of the recursion modulo a large prime number
$p$, and repeat the procedure for a collection of distinct primes
$p_i$'s. This allows to determine the congruence of $(T_k)_j$ modulo
each of these primes, and the Chinese remainder theorem allows to
efficiently reconstruct the unique integer value with such
congruences, within some rough upper-bound.

This allowed us to compute $(T_{3})_j$ up to order $j=403$, within a
fairly moderate running time.

It is our goal in this section to explicitly verify the prediction of
Bousquet-M\'elou and Courtiel \cite{BMCourt2015} on the asymptotic
behaviour of this series. In principle, there exist several quite
refined methods for handling numerical data of this form, most of them
resulting from variants of the classical method of Pad\'e
approximants. And, no telling, this is a well-known domain of
expertise of Tony Guttmann (see e.g.\ \cite{guttpade} for the
foundation paper on the \emph{method of Differential Approximants},
and \cite{guttpadeDG} for an extensive review). However, our series
has special properties: on one side we already have a number of
theoretical predictions (for example, the growth rate is predicted not
only by \cite{BMCourt2015}, but also from the solution of Kostov and
Staudacher \cite{KoStau}, and the exact mapping described here, the
two predictions being consistent); on the other side, if, as we
expect, the theoretical prediction is correct, we need to detect
logarithmic corrections, which are notoriously hard.

\begin{figure}
\[
\includegraphics[scale=.5]{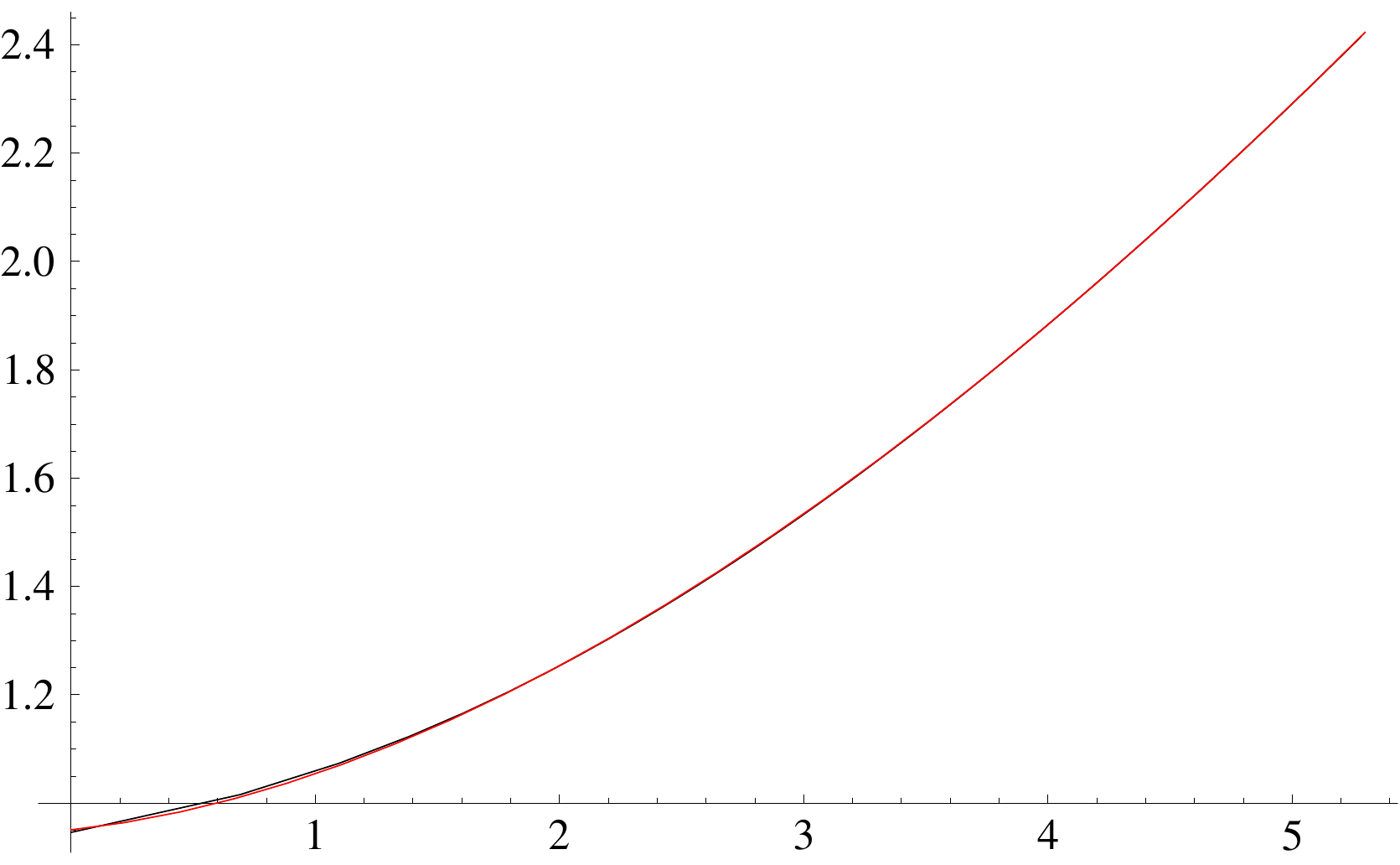}
\]
\caption{Fit of $B^{(3)}_n$ versus $x=\ln n$, using a function of the
  form (\ref{eq.277465}).
\label{fig.plotB3}}
\end{figure}

For these reasons, we perform instead a less orthodox analysis.
For brevity, call $A_n$ the series $(T_3)_{2n+1}$. We thus expect this
series to behave as
\be
A_n \sim K \left(\frac{\pi^2}{384}\right)^{-n} 
n^{-3} \big( \ln(n) \big)^{-\alpha}
\ee
and we want to check the hardest part of the theoretical prediction, $\alpha=2$,
at the best of our possibilities. Let us construct the series
\be
B_n := 
\frac{\left(\frac{\pi^2}{384}\right)^{-n} }{A_n}
\sim
\frac{1}{K}
n^{3} \big( \ln(n) \big)^{\alpha}
\ee
We expect the factor $n^3$ to be in fact affected by algebraic
corrections, thus, let us instead assume that $n^3$ is replaced by
some monic polynomial of degree 3. Similarly, we will assume that the
factor $\big( \ln(n) \big)^{\alpha}$ has indeed (more severe)
finite-size corrections of the form
\be
\big( \ln(n) \big)^{\alpha}
\quad \longrightarrow \quad
\big( \ln(n) \big)^{\alpha}
\left(
1 + \frac{a_1}{\ln n} + \frac{a_2}{(\ln n)^2} + \cdots
\right)
\label{eq.643564}
\ee
We may construct the series
associated to the lattice third-derivative,
\be
B^{(3)}_n := 
B_{n+3}-3B_{n+2}+3B_{n+1}-B_{n}
\ee
This manipulation cancels out the algebraic corrections in the $n^3$
factor, replacing the monic polynomial with the constant $6$, and lets
(\ref{eq.643564}) stable in form (while in fact shifting the
coefficients $a_i$). A first fit of the series 
$B^{(3)}_n$ as a function of $x=\ln n$, of the form
\be
f(x) = a_2 x^2 + a_1 x + a_0 + \frac{a_{-1}}{x+c}
\label{eq.277465}
\ee
(suggested by (\ref{eq.643564}) and the ansatz $\alpha=2$) is
encouraging (see Figure \ref{fig.plotB3}).

However, we are not satisfied. We want to \emph{fit} the value of
$\alpha$, instead of trying up the predicted value (and, on top of
this, with so many free parameters).

At this point we perform a further manipulation, and produce the
series
\be
C_n = \frac{B^{(3)}_{2n}}{B^{(3)}_n}
\ef.
\ee
Under the assumption that $B_n$ has no periodic behaviour (a fact that
can be consistently checked numerically on the data), this further
manipulation allows to further highlight the residual dependence from
$n$. That is, a function of the form (\ref{eq.643564}), in which we
keep as parameters $\alpha$ and the first $k$ of the $a_i$'s.

\begin{figure}
\[
\includegraphics[scale=.5]{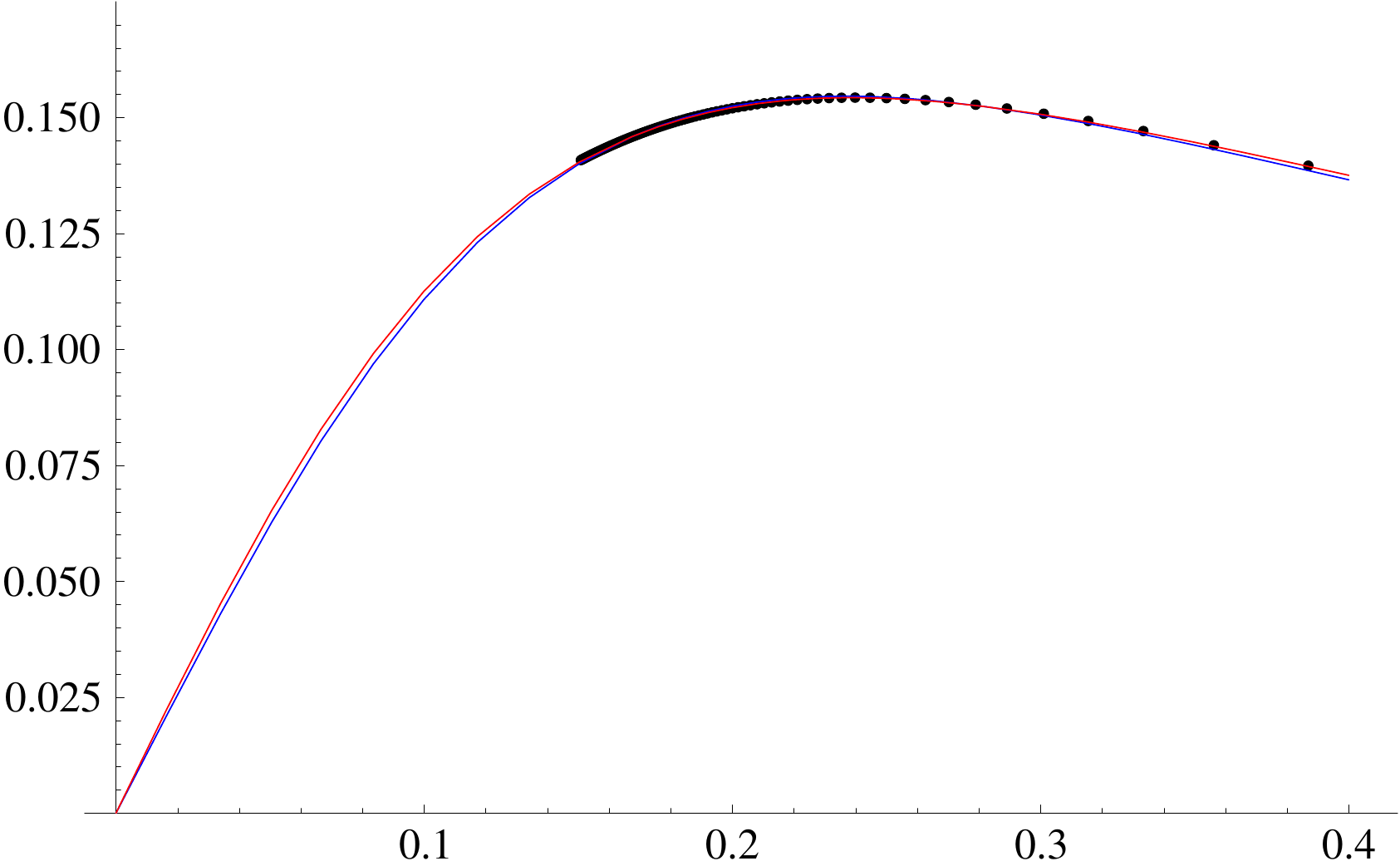}
\]
\caption{Fit of $C_n$ versus $x=\ln 2/\ln n$, using the ratio of functions of the
  form (\ref{eq.277465}), with $k$ correction terms, evaluated at $2n$
  (numerator) and at $n$ (denominator). Blue and red correspond to
  $k=2$ and $k=3$ respectively.\label{fig.fitC}}
\end{figure}

For $k\leq 1$ the fit does not converge. For $k=2$ and $k=3$ the
fitted values of $\alpha$ are $\alpha=1.84\ldots$ and
$\alpha=1.98\ldots$, respectively. For $k \geq 4$ the proliferation of
fit parameters, and their high covariance, make the fit less reliable.
We suppose that the incredibly good value at $k=3$ is merely
accidental, but the fact that these values are reasonably near to $2$
(and definitely off from 0) is a validation of the theoretical
prediction. In figure \ref{fig.fitC} we show the fit of the data with
functions at $k=2$ and $k=3$, as a function of $x=\ln 2 /\ln n$.

%%%%%%%%%%%%%%%%%%%%%%%%%%%%%%%%%%%%%%%%%%%%%%%%%%%%%%%%
\section{Conclusions}
%{Implications for criticality on regular lattices from KPZ}
\label{sec:flat_lattice}

As we have discussed above, there is an increasing evidence that the
model of spanning forests on random planar graphs of fixed degree
shows the KPZ counterpart of a Berker--Kadanoff phase on an interval
$(t_\ast,0]$ of fugacities, with $t_{\ast}$ being negative and
finite. In particular, at $t=0$ we have the well-known theory of
spanning trees, which has string susceptibility exponent $\gamma=-1$,
corresponding to a value $c=-2$.
For negative values of $t$, up to the values $t_\ast$ discussed in
Section \ref{sec.mainana}, the string susceptibility remains
$\gamma=-1$, while it develops logarithmic corrections, with
exponent~$\gamma'=-2$.  This is consistent with the value $c=-2$ found
for spanning forests on flat lattices in \cite{alan04}.
Therefore the picture we (and previous literature \cite{BMCourt2015})
have for spanning forests on random geometries, the KPZ relation, and
the understanding of the Berker--Kadanoff phase in flat geometry all combine
consistently.  Outside this range of $t$, the theory is massive, with
the possible exception of the theory at $t_{\ast}$, which may (and
shall) be a non-trivial CFT, either with vanishing central charge, as
follows from formula (\ref{eq:KPZ}), or a non-unitary CFT with a value
of $h_{\rm min}$ resulting into $\gamma=-1/2$, namely
$h_{\rm min} = c/25$.

So, is everything unraveled?

In our perspective, this is not the case. There are various aspects of
the phase diagram of this model --- and also of its counterpart in
regular lattices --- which are not clear, and deserve further
investigation.

One of these aspects is the origin of the logarithmic corrections
(also in their regular counterpart).  A satisfactory understanding
would require to elucidate the features of logarithmic CFTs coupled to
Liouville gravity, a topic that unfortunately, as far as we know, has
never been addressed in the literature.

A second aspect, discussed in the conclusions of
\cite{jacobsen2005arboreal}, is the combinatorial nature of the
criticality in the Berker--Kadanoff phase. It is conjectured there
that in this phase there is one tree occupying a finite fraction of
the lattice, although this fact is hard to investigate theoretically
or numerically, because of the lack of positivity in the Boltzmann
weights. This problem may be easier to approach in the gravitational
counterpart, especially in the cubic case, where we have an explicit
expression for the spectral density.

A third aspect is the nature of criticality at $t_{\ast}$. Indeed, at
discrepancy with \cite{jacobsen2005arboreal}, we do \emph{not} see the
counterpart of the theory with $c=-1$, consisting of symplectic
fermions and one non-compact boson (unless somehow this theory has an
unlikely field of lowest conformal dimension $h=-1/25$, compatible
with the string susceptibility we find at $t_{\ast}$). As a less
fundamental discrepancy, we do not see any relevant critical behaviour
occurring in the case of vanishing monomer weight, in the formulation
of the model as a gas of O$(-2)$ loops and dimers, as is the case in
\cite{jacobsen2005arboreal} for the regular square grid, so this may
be an accident of that lattice.

Another issue is the comprehension of the massive phase below
$t_{\ast}$ for quartic graphs, at the same level as done for cubic
graphs.  And an obvious issue is how to make rigorous the
present derivation, in the regime $t<-1$. 

Finally, it would be highly interesting to investigate the
antiferromagnetic regime of the Potts model on random planar graphs
beyond the case of spanning forests.  A particularly interesting
aspect is the behavior for $q$ a Beraha number, i.e.\ for
$q = 4 \cos^2(\pi/\delta)$, $\delta \in \mathbb{N}$. At these values
of $q$, the partition function of the $q$-state Potts model in the
Berker--Kadanoff phase on flat lattices develops severe singularities, whose
resolution is, to the best of our knowledge, only partially understood
\cite{Sal90,saleurNPB,SJAntiferro}.  (See also \cite{Jacobsen2014} for
a discussion of the Berker--Kadanoff phase for other graphs than the
triangular and square lattice.)  Nonetheless, it is conceivable that
at least the critical indices computed for generic $\delta\ge 2$
continue to hold even at the Beraha numbers, as it appears to be the
case for $\delta=2$, which corresponds to spanning forests.  The
quantum gravity counterpart of these and related issues may be
clarified by ongoing work of Bernardi and Bousquet-M{\'e}lou, as
continuation of \cite{Bernardi2011,Bernardi2015}.

We hope to address some of the questions outlined above in future work.

\end{document}